\documentclass[aps,prb,twocolumn,superscriptaddress,10pt,article,nofootinbib,longbibliography]{revtex4-2}
\usepackage{graphicx,amsfonts,amssymb,amsmath,hyperref,hypcap,enumerate}
\usepackage{color}
\usepackage{mathtools}

\begin{document}
\title{Hierarchy of quasi-symmetries and degeneracies in chiral crystal materials CoSi}
\author{Lun-Hui Hu}
\thanks{These two authors contributed equally}
\affiliation{Department of Physics, The Pennsylvania State University, University Park,  Pennsylvania 16802, USA}
\affiliation{Department of Physics and Astronomy, University of Tennessee, Knoxville, Tennessee 37996, USA}
\author{Chunyu Guo}
\thanks{These two authors contributed equally}
\affiliation{Laboratory of Quantum Materials (QMAT), Institute of Materials (IMX), \'{E}cole Polytechnique F\'{e}d\'{e}rale de Lausanne (EPFL), CH-1015 Lausanne, Switzerland}

\author{Yan Sun}
\affiliation{Max Planck Institute for Chemical Physics of Solids, 01187 Dresden, Germany}
\author{Claudia Felser}
\affiliation{Max Planck Institute for Chemical Physics of Solids, 01187 Dresden, Germany}

\author{Luis Elcoro}
\affiliation{Department of Condensed Matter Physics, University of the Basque Country UPV/EHU, Apartado 644, 48080 Bilbao, Spain}

\author{Philip J. W. Moll}
\affiliation{Laboratory of Quantum Materials (QMAT), Institute of Materials (IMX), \'{E}cole Polytechnique F\'{e}d\'{e}rale de Lausanne (EPFL), CH-1015 Lausanne, Switzerland}
\author{Chao-Xing Liu}
\email{cxl56@psu.edu}
\affiliation{Department of Physics, The Pennsylvania State University, University Park,  Pennsylvania 16802, USA}
\author{B. Andrei Bernevig}
\email{bernevig@princeton.edu}
\affiliation{Department of Physics, Princeton University, Princeton, New Jersey 08544, USA}
\affiliation{Donostia International Physics Center, P. Manuel de Lardizabal 4, 20018 Donostia-San Sebastian, Spain}
\affiliation{IKERBASQUE, Basque Foundation for Science, Bilbao, Spain}

\begin{abstract}
In materials, certain approximated symmetry operations can exist in a lower-order approximation of the effective model but are good enough to influence the physical responses of the system, and these approximated symmetries were recently dubbed ``quasi-symmetries''~\cite{guo_arxiv_2021}. In this work, we reveal a hierarchy structure of the quasi-symmetries and the corresponding nodal structures that they enforce via two different approaches of the perturbation expansions for the effective model in the chiral crystal material CoSi. In the first approach, we treat the spin-independent linear momentum (k) term as the zero-order Hamiltonian. Its energy bands are four-fold degenerate due to an SU(2)$\times$SU(2) quasi-symmetry. We next consider both the k-independent spin-orbit coupling (SOC) and full quadratic-k terms as the perturbation terms and find that the first-order perturbation leads to a model described by a self-commuting ``stabilizer code'' Hamiltonian with a U(1) quasi-symmetry that can protect nodal planes. In the second approach, we treat the SOC-free linear-k term and k-independent SOC term as the zero-order. They exhibit an SU(2) quasi-symmetry, which can be reduced to U(1) quasi-symmetry by a choice of quadratic terms. Correspondingly, a two-fold degeneracy for all the bands due to the SU(2) quasi-symmetry is reduced to two-fold nodal planes that are protected by the U(1) quasi-symmetry. For both approaches, including higher-order perturbation will break the U(1) quasi-symmetry and induce a small gap $\sim$ 1 meV for the nodal planes. These quasi-symmetry protected near degeneracies play an essential role in understanding recent quantum oscillation experiments in CoSi~\cite{guo_arxiv_2021}. 
\end{abstract}

\maketitle

{\it Introduction} -- 
Symmetry describes the invariance of a system under certain operations and plays a fundamental role in almost all branches of physics. In condensed matter physics, different quantum states of matter and the phase transition between them can be characterized via the principle of spontaneous symmetry breaking, as formulated in the Landau-Ginzburg theory~\cite{Goldenfeld_book_2018}. For example, the crystallization of a solid breaks continuous translation to discrete translation and the formation of ferromagnetism in a magnet breaks the full rotation symmetry, even though the microscopic interaction in these systems has full translation and rotation symmetries~\cite{mattis_book_2006}.

In the scenario of spontaneous symmetry breaking, the high-symmetry states appear at a high energy scale (or high temperature); when the energy scale is lowered, symmetry-breaking states start appearing. However, the opposite scenario also exists, and a high-symmetry state can emerge in the low-energy sector of a system~\cite{Volovik_book_2009}. For example, the Lorentz symmetry is accompanied by the emergence of the two-dimensional Dirac equation as a low-energy effective theory in graphene or at the surface of topological insulators, although both systems are non-relativistic~\cite{neto_rmp_2009,basov_rmp_2014,hasan_rmp_2010,qi_rmp_2011,bernevig_book_2013,franz_book_2013,shun_book_2018}. The existence of Dirac fermions and Lorentz symmetry leads to several exotic physical properties of graphene and topological insulators, making them appealing platforms to test quantum relativistic phenomena in table-top experiments~\cite{Geim_nm_2007}. Besides the space-time symmetry, emergent symmetries can also exist for the internal degree of freedom. For example, due to the spin and valley degrees, graphene has an additional SU(4) symmetry, which leads to intriguing physical phenomena, such as SU(4) quantum Hall ferromagnets\cite{yang2006collective,nomura2006quantum,young2012spin}.

Recently, we introduce the concept of ``quasi-symmetry'' to describe such emergent internal symmetry~\cite{guo_arxiv_2021}. More precisely, we refer to quasi-symmetry as a symmetry operator that only exists in a lower-order approximation of the effective Hamiltonian but is good enough to influence the physical responses of the material. The crystalline symmetry of solid material without magnetic order, described by 17 space groups in two dimensions (2D) and 230 space groups in 3D~\cite{dresselhaus_book_2007}, gives a strong constraint on the form of the low-energy effective model (in the spirit of ${\bf k\cdot p}$ type of Hamiltonian) around high symmetry momenta in the Brillouin zone~\cite{tang_prb_2021,Yu_scibull_2022,zhang_arxiv_2021,tang_arxiv_2022,Winkler_book_2003}. As a consequence, if one only keeps lower-order-k terms in the expansion, the effective Hamiltonian can generally possess additional symmetries, beyond the crystalline symmetry itself. With keeping further powers of ${\bf k}$ in the ${\bf k}\cdot{\bf p}$ expansion, these additional symmetries will gradually be broken by the higher-order-${\bf k}$ terms or other perturbations [e.g.,~spin-orbit coupling (SOC)], thus forming a hierarchy structure of quasi-symmetry groups.

\begin{figure}[t]
	\centering
	\includegraphics[width=\linewidth]{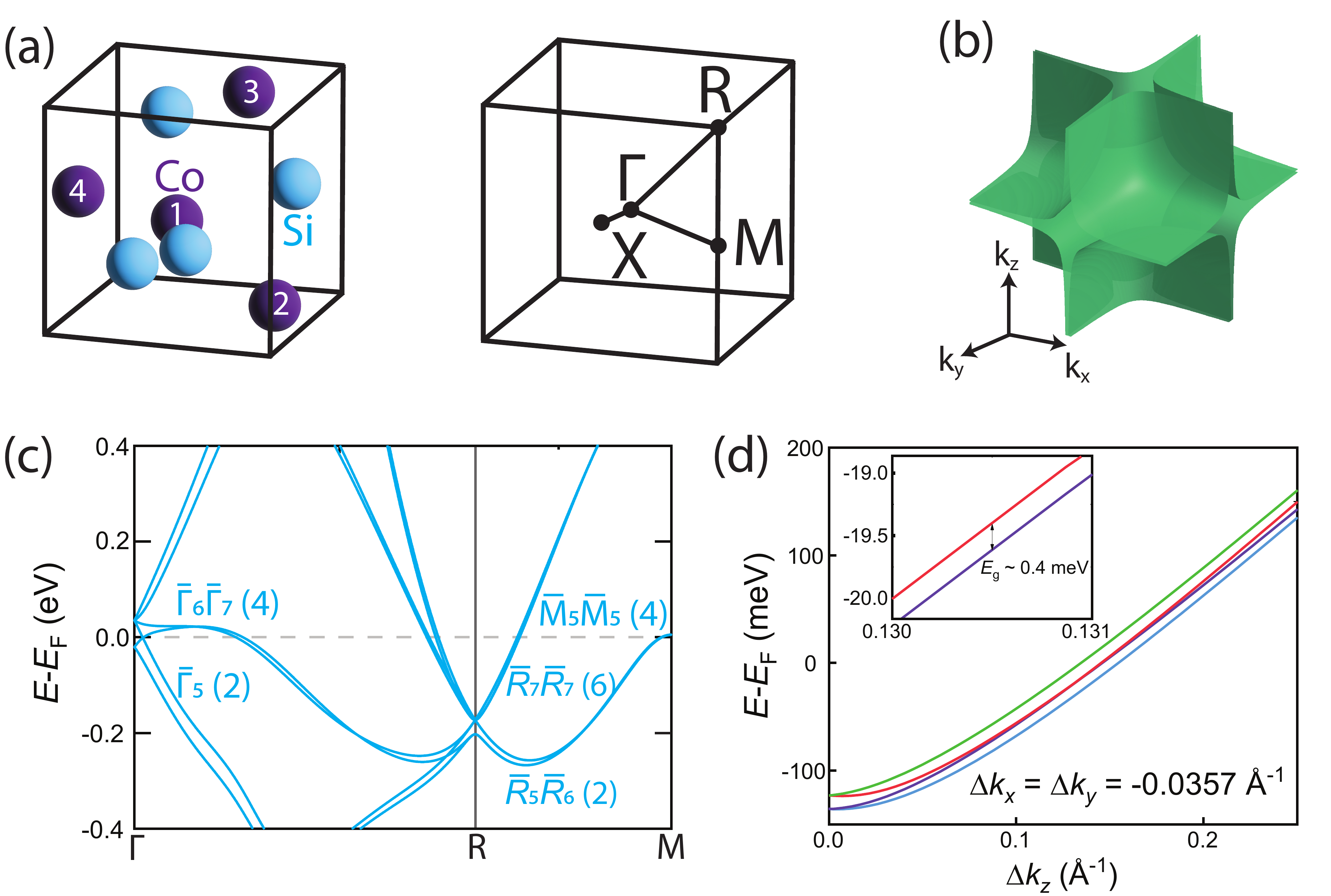}
	\caption{
	(a) One unit-cell with four Co and Si atoms, and the Brillouin zone with high-symmetry points ($\Gamma, R, X, M$).  
	(b) shows the quasi-symmetry protected nodal-planes. 
	(c) Electronic band structure of CoSi along $\Gamma-R-M$ lines. The irreps of the energy states at high-symmetry points are labelled.
	(d) The DFT bands along the $(-0.0357,-0.0357,k_z)$ direction and the inset shows the tiny gap $\sim 0.4$ meV.
	}	
	\label{fig1}
\end{figure}

\begin{figure}[t]
	\centering
	\includegraphics[width=\linewidth]{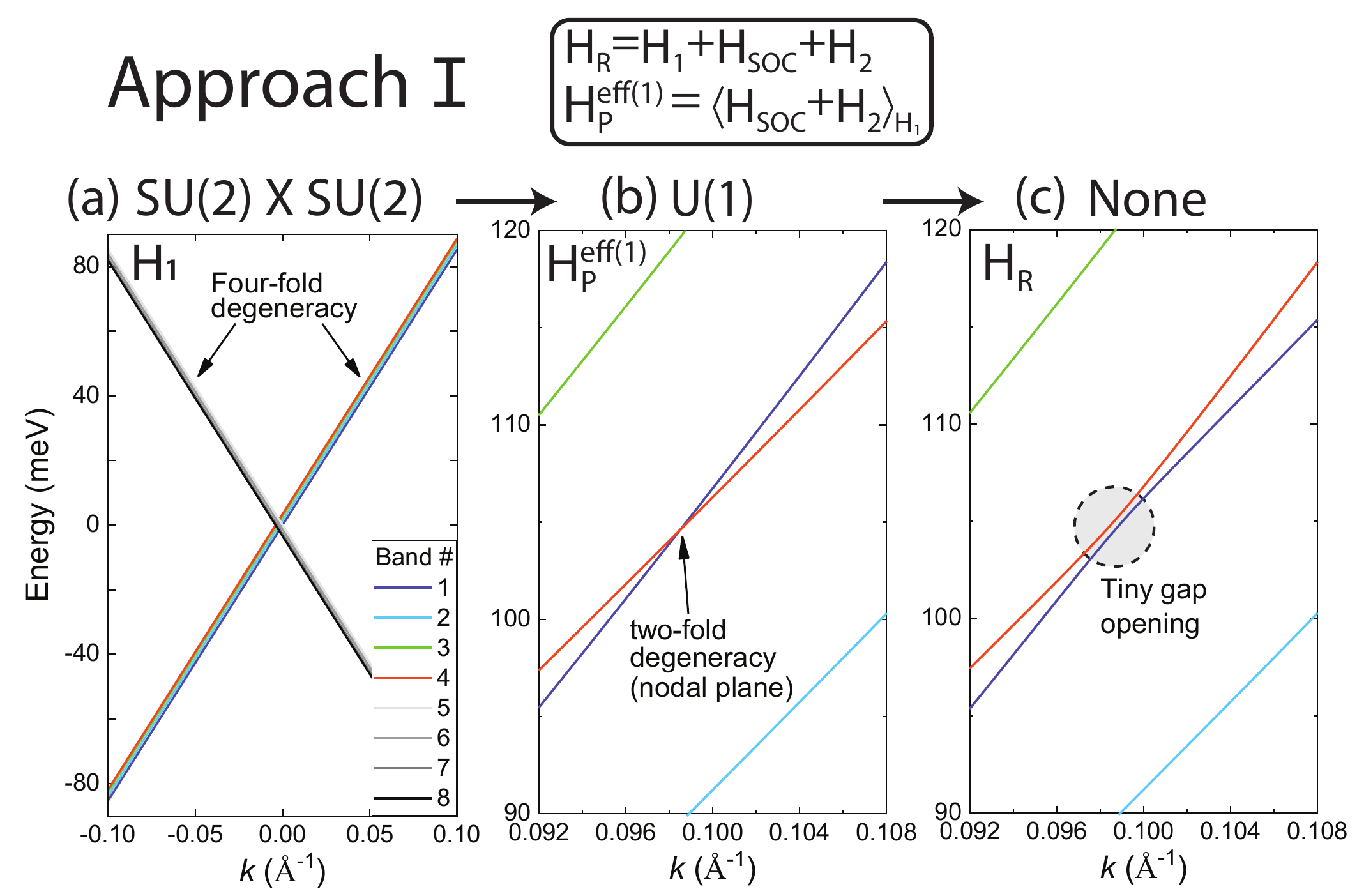}
	\caption{The summary of the hierarchy of quasi-symmetry with the approach I (self-commuting Hamiltonian). We plot the band splitting along the non-high-symmetry direction within the spherical coordinate ($\theta=\phi=\pi/3$). The left panel shows the SU(2)$\times$SU(2) quasi-symmetry protect the four-fold degeneracy. In the middle panel, all the bands split but there exists the U(1) quasi-symmetry protecting the nodal plane. The right panel shows a tiny gap appears once high-order perturbation corrections are involved. 
	}	
	\label{fig2}
\end{figure}

\begin{figure}[t]
	\centering
	\includegraphics[width=\linewidth]{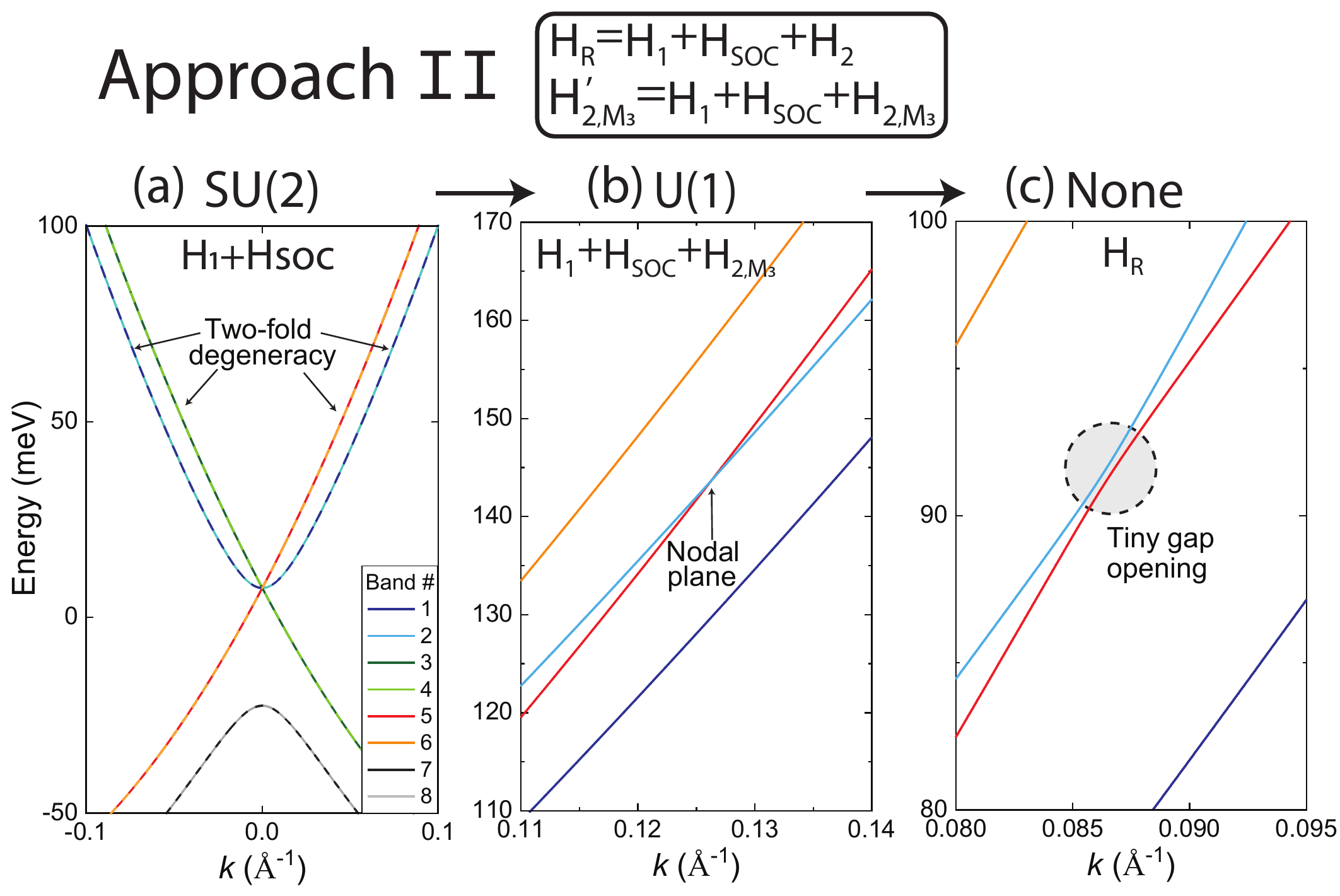}
	\caption{The hierarchy of quasi-symmetries by the Approach II. Here we plot the band splitting along a non-high-symmetry line with $\theta=\phi=\pi/3$. The left panel shows the band splitting of $\mathcal{H}_1+\mathcal{H}_{\text{soc}}$ is shown and each band is two-fold degeneracy required by the SU$_{\text{o}}$(2) quasi-symmetry. In the middle panel, all the bands split but there still exists a two-fold degeneracy which is protected by the U(1) quasi-symmetry. The right panel shows all the quasi-symmetries are broken and a tiny gap appears.  
	}	
	\label{fig3}
\end{figure}

This work aims in revealing such a hierarchy structure of quasi-symmetries in the low-energy effective Hamiltonian expansion for different orders of the momentum ${\bf k}$ for the material compound CoSi with a chiral crystal structure (space group No.~198). Recent experimental and theoretical work~\cite{guo_arxiv_2021} has shown that the quasi-symmetry exists in this compound and leads to the near-nodal-planes that are located at non-high-symmetry momenta, which are essential in understanding the transport measurement of quantum oscillations in CoSi. In this work, we will systematically discuss two approaches to constructing the effective Hamiltonian perturbatively for CoSi, and reveal the hierarchy structure of quasi-symmetries in different orders of the perturbation expansion. As discussed in Fig.~\ref{fig2}, our first approach treats the SOC-free linear-$k$ term as the zero order, which leads to a four-fold degeneracy protected by SU(2)$\times$SU(2) quasi-symmetry. Then we consider both the $k^2$ terms and SOC as the perturbation and project them into the subspace of these four-fold degenerate bands. The resulting effective Hamiltonian shows a striking ``self-commuting" feature that results in a U(1) quasi-symmetry for the protection of nodal planes in non-high-symmetric momenta (Fig.~\ref{fig1}(b)). Our second approach (Fig.~\ref{fig3}) treats both linear-$k$ term and SOC as the zero order Hamiltonian and shows all the bands are doubly degenerate due to the orbital SU(2) quasi-symmetry. We then consider $k^2$ terms as a perturbation and classify them into three different groups with each group selectively breaking the SU(2) quasi-symmetry into U(1) quasi-symmetry along a certain direction, which can also protect nodal planes. In both approaches, the second-order perturbation can induce a tiny gap ($\sim$ 1 meV for CoSi). In this sense, we dubbed these quasi-symmetry protected nodal planes to be near-nodal planes.

{\it Effective $\mathbf{k}\cdot\mathbf{p}$ model for CoSi} -- 
The crystal CoSi family crystallizes in a chiral cubic structure of space group (SG) $P2_{1}3$ (No.~198) without a center of inversion~\cite{tang_prl_2017}. Its cubic lattice with the lattice constant $a_0=4.433$ \AA, as shown in Fig.~\ref{fig1}(a), contains four Si atoms and four Co atoms in one unit cell. The corresponding  Brillouin zone (BZ) is shown in Fig.~\ref{fig1}(c), where high-symmetry points including $\Gamma$, $R$, and $M$ are marked. With the Seitz notation for the non-symmorphic symmetry operations, the three generators of SG 198 are $S_{2x} = \{ C_{2x} | \tfrac{1}{2} \tfrac{1}{2} 0 \}$,  $S_{2y} = \{ C_{2y} | 0 \tfrac{1}{2} \tfrac{1}{2} \}$, and  $C_{3} = \{ C_{3,(111)} | 0 0 0 \} $. The system has also time-reversal symmetry $\mathcal{T}$. With the density function theory (DFT) calculations, we obtain the electronic band structure with SOC, as shown in Fig.~\ref{fig1}(c) along the $\Gamma-R-M$ lines. Hole Fermi pockets are found around $\Gamma$ while electron Fermi pockets exist around $R$, $\mathbf{k}_R=(\pi,\pi,\pi)$. Here we focus on the electronic bands around $R$. For the SOC-free band structures, there are four-fold degenerate states (without spin degeneracy) at the $R$ point, which disperse linearly around $R$ and give rise to the electron Fermi pockets. The corresponding single-valued irreducible representation (irrep) is $R_1R_3$ based on the notations in the Bilbao Crystallographic Server~\cite{Bradlyn_nature_2017TQC,Xu_nature_2020,Elcoro_nc_2021}. Taking into account the spin degree of freedom, the eight-fold degenerate states at $R$ are split by SOC into higher-energy six-fold degenerate states~\cite{bradlyn_science_2016} (double-valued irrep $\bar{R}_7\bar{R}_7$) and lower-energy two-fold degenerate states (double-valued irrep $\bar{R}_5\bar{R}_6$) with a gap $\sim$ 30 meV.

As described in Ref.~\cite{guo_arxiv_2021} and Sec.~(A) of the Supplementary Material (SM)~\cite{sm2022}, the effective model to describe the energy bands around $R$ is constructed based on the little group at $R$ point generated by $S_{2x}$, $S_{2y}$, $C_{3}$ and $\mathcal{T}$. Up to $k^2$ order, the Hamiltonian contains three parts
\begin{eqnarray}\label{eq:H_R}
\mathcal{H}_R = \mathcal{H}_{1} + \mathcal{H}_{\text{soc}} + \mathcal{H}_2 ,
\end{eqnarray}
where $\mathcal{H}_{1}=C_0+ 2A_1 s_0  (\mathbf{k}\cdot\mathbf{L})$ includes a constant and linear-$k$ term, and $\mathcal{H}_{\text{soc}}=2\lambda_0 ( \mathbf{s} \cdot \mathbf{L})$. Here we define the operators $ L_x = \tfrac{1}{2} \sigma_y\tau_0, \; L_y = \tfrac{1}{2}\sigma_x\tau_y, \; L_z = -\tfrac{1}{2}\sigma_z\tau_y$, which satisfies the angular momentum commutation relation $[L_i,L_j] = i\epsilon_{ijk} L_k$ with Levi-Civita symbol $\epsilon_{ijk}$ and $i=x,y,z$. ${\bf s}$ represents the Pauli matrix in the spin space and both ${\bf \sigma, \tau}$ for the Pauli matrices in the orbital space. The basis for four orbitals of the $\sigma, \tau$ matrices are mainly ($>80\%$) composed of the mixing between the $t_{2g}$ and $e_g$ orbitals of the four Co atoms, as justified by the DFT calculations. And the detailed forms of the wave functions are shown in Sec.~(B) of the SM~\cite{sm2022}.

In addition, the $k^2$ order effective Hamiltonian $\mathcal{H}_2$ in Eq.~\eqref{eq:H_R} shows an intriguing structure and can be grouped into three classes, $\mathcal{H}_{2}=\mathcal{H}_{2, \mathcal{M}_1}+\mathcal{H}_{2, \mathcal{M}_2}+\mathcal{H}_{2, \mathcal{M}_3}$, where $\mathcal{H}_{2,\mathcal{M}_i}={\bf g}_i\cdot{\bf J}_i$ for i=1,2,3. Here we define ${\bf g}_1=(C_2k_xk_y, \\ -C_3k_xk_z,C_1k_yk_z)$,  ${\bf g}_2=(C_3k_xk_y,C_1k_xk_z, -C_2  k_yk_z)$,  ${\bf g}_3 \\ =  (C_1k_xk_y,  C_2k_xk_z, -C_3  k_y k_z)$, and ${\bf J}_1=( \sigma_x  \tau_x, -\sigma_z\tau_x, \\ \sigma_0 \tau_z)$, ${\bf J}_2=(\sigma_x\tau_z,\sigma_z\tau_z,\sigma_0\tau_x)$,  ${\bf J}_3=(\sigma_z\tau_0,\sigma_x\tau_0,\sigma_y\tau_y)$. The parameters for CoSi are obtained by fitting with the DFT bands~\cite{guo_arxiv_2021} and listed in Table.~(\ref{Table1}). It should be noted that all the bands at the $k_i=\pi$-planes ($i=x,y,z$) are doubly degenerate as a consequence of the anti-unitary symmetries $S_{2x}\mathcal{T}$, $S_{2y}\mathcal{T}$ and $S_{2z}\mathcal{T}$ in these planes~\cite{guo_arxiv_2021,huber_prl_2021,leonhardt_prm_2021}. Furthermore, the DFT calculations show near-nodal planes with tiny gaps $\sim 0.5$ meV at non-high-symmetry momenta shown in Fig.~\ref{fig1} (d), and we next discuss how to apply the perturbation theory to the model Hamiltonian $\mathcal{H}_R$ in Eq.~\eqref{eq:H_R} to understand these near-nodal planes, as well as the underlying quasi-symmetries.

\begin{table}[!htbp]
	\begin{tabular}{ c | c  c  c  c  c  c c}  
		\hline \hline
		Parameter & $C_0$ & $B_1$ & $A_1$ & $C_1$ & $C_2$ & $C_3$ & $\lambda_0$ \\    \hline    
		Value  & -0.18   & 2.123  & 0.853  & -0.042 & 0.546 & 3.345 & 0.0075   \\ 
		Unit & eV   & eV$\cdot\AA^2$  & eV$\cdot\AA$  & eV$\cdot\AA^2$ & eV$\cdot\AA^2$ & eV$\cdot\AA^2$ & eV  \\  \hline \hline
	\end{tabular}
	\caption{The parameters for the ${\bf k}\cdot {\bf p}$ Hamiltonian $\mathcal{H}_R$.}
	\label{Table1}
\end{table}

{\it Approach I: self-commuting Hamiltonian }--
We now precisely formulate the hidden quasi-symmetry that may appear at low-energy in the physics of the model Hamiltonian, and start with the linear-$k$-order Hamiltonian, SOC-free $\mathcal{H}_1$ in Eq.~\eqref{eq:H_R}, which is invariant under the spin SU$_{\text{s}}$(2) symmetry group. Moreover, an additional hidden SU$_{\text{o}}$(2) symmetry also exists for $\mathcal{H}_1$ in the orbital space, and can be generated by the operators
\begin{align}
\mathcal{M}_{1,2,3} = \tfrac{1}{2}\{ s_0\sigma_y\tau_z, s_0\sigma_y\tau_x, s_0\sigma_0\tau_y \},
\end{align}
which all commute with $\mathcal{H}_1$ and satisfy the commutation relations $[\mathcal{M}_{i}, \mathcal{M}_{j} ] = i\epsilon_{i,j,k} \mathcal{M}_k$. Thus, we refer to it as the SU$_{\text{o}}$(2) quasi-symmetry group for $\mathcal{H}_1$. As a result, the SU$_{\text{s}}$(2)$\times$SU$_{\text{o}}$(2) quasi-symmetry group protects the four-fold degeneracy for each band ($E_\pm(k)=\pm A_1 k + C_0$) at any nonzero $\mathbf{k}$ (see Fig.~\ref{fig2}(a)). Hereafter, we absorb the constant energy $C_0$ into Fermi energy $E_F$.

We now consider the perturbation from $\mathcal{H}_{\text{soc}}$ and $\mathcal{H}_2$. Without loss of generality, we choose four degenerate bands with positive energy ($E_+(k)=A_1 k$) as the basis,  $\{\vert\Psi_{+,s,i}\rangle\}$ with $s=\uparrow,\downarrow$ and i=1,2, and project $\mathcal{H}_{\text{soc}}$ and $\mathcal{H}_2$ into this subspace. Shown in Sec.~(C) of the SM~\cite{sm2022}, the projected four-band model is given by
\begin{align}\label{eq-eff-p-model}
\mathcal{H}_{P}^{\text{eff}(1)} = E_0 + \mathcal{H}_{\text{soc}}^{\text{eff}(1)}({\bf k}) + \mathcal{H}_{2}^{\text{eff}(1)}({\bf k}),
\end{align}
where $E_0=C_0+ A_1k +B_1k^2$, $\mathcal{H}_{\text{soc}}^{\text{eff}(1)}({\bf k})=\lambda_0 (\mathbf{k}\cdot \mathbf{s})\omega_0$, and  $\mathcal{H}_{2}^{\text{eff}(1)}({\bf k}) = \tilde{C}k^2 s_0 ( d_{x,{\bf k}} \omega_x + d_{y,{\bf k}}\omega_y + d_{z,{\bf k}}\omega_z )$ with $\tilde{C}=C_1-C_2+C_3$. Here $\omega_0$ is identity matrix and $\omega_i$ ($i=x,y,z$) are the Pauli matrices for the two spinless bands. The coefficients $d_{i,{\bf k}}$ depend on ${\bf k}$, and the detailed forms are given in Sec.~(C) of the SM~\cite{sm2022}. Such perturbation process can be well justified by satisfying both $A_1k \gg \lambda_0$ and $A_1k  \gg \tfrac{\sqrt{3}}{4} \tilde{C}k^2$, which results in the valid momentum range $0.01 < k < 1$ (\AA$^{-1}$), corresponding to a wide Fermi energy range $8.5 \ll E_F \ll 850$ (meV). Thus, the obtained effective model is relevant for the realistic experimental situations ($E_F$ in CoSi is 180 meV). Strikingly, we notice that the two terms in this effective Hamiltonian are self-commuting, namely
\begin{align}\label{eq-p-model-comm}
\left\lbrack \mathcal{H}_{\text{soc}}^{\text{eff}(1)}({\bf k}_1) , \mathcal{H}_{2}^{\text{eff}(1)}({\bf k}_2) \right\rbrack =0,  
\end{align}
which implies that Hamiltonian~\eqref{eq-eff-p-model} is a stabilizer code Hamiltonian~\cite{gottesman_pra_1996,gottesman_phd_1997,bernevig_prb_2021,lian_prb_2021}. The commutation relation in Eq.~\eqref{eq-p-model-comm} can be easily seen since $\mathcal{H}_{\text{soc}}^{\text{eff}(1)}$ ($\mathcal{H}_{2}^{\text{eff}(1)}$) contains an identity matrix in the $\omega$-space (spin ${\bf s}$-space). The self-commuting property implies the existence of a unitary symmetry operator $S_{\text{eff}}={\bf\frac{ k\cdot s}{|k|}}\omega_0$ that commutes with the whole Hamiltonian $\mathcal{H}_{P}^{\text{eff}(1)}$ for any momentum. $S_{\text{eff}}$ describes an internal symmetry and can be viewed as the generator of a U(1) group.

Due to the self-commuting nature, the eigen-state of the Hamiltonian in Eq.~\eqref{eq-eff-p-model} can be explicitly solved with its eigen-energy given by $E_{\alpha,\beta} = E_0 +\alpha\lambda_0 +\beta\sqrt{3}\tilde{C} \vert k_xk_yk_z\vert/k$ with $\alpha,\beta=\pm1$, where $\alpha$ labels the eigenvalues of $S_{\text{eff}}$. We notice that two eigen-energies $E_{+,-}$ and $E_{-,+}$ can be equal when the condition 
\begin{equation}\label{eq:nodalplane}
\lambda_0 = \sqrt{3}\tilde{C} \vert k_xk_yk_z\vert/k
\end{equation} 
is satisfied. It determines nodal planes of the effective model in Eq.~\eqref{eq-eff-p-model} in the whole momentum space. Fig.~\ref{fig2}(b) shows the U(1) quasi-symmetry protected two-fold degeneracy along a non-high-symmetry line ($\theta=\phi=\pi/3$), where the spherical coordinator $(k,\theta,\phi)$ is used with polar angle $\theta$ and azimuthal angle $\phi$. The Fermi sphere crosses the nodal planes to form nodal rings at the Fermi energy, which can be extracted by combining $E_F=E_{+,-}$ with Eq.~\eqref{eq:nodalplane}. Explicitly, the nodal rings at a fixed Fermi energy $E_F$ can be determined by 
\begin{align}\label{eq-curve-equ-nodal-line}
  f_{N}(\theta,\phi)  = \frac{2A_1^2\lambda_0}{\sqrt{3}\tilde{C}E_F^2} \lbrack 1 + \frac{2B_1E_F}{A_1^2}  
+ \sqrt{1+\frac{4B_1E_F}{A_1^2}}  \rbrack,
\end{align}
where $f_{N}(\theta,\phi)=\vert \sin2\phi\sin2\theta\sin\theta \vert$. We notice that Eq.~\eqref{eq-curve-equ-nodal-line} has solution only when $E_F\ge E_{W}$ with $E_{W} = \tfrac{2}{\sqrt{3}\tilde{C}}(A_1 \sqrt{\sqrt{3}\tilde{C}\lambda _0}+2B_1 \lambda _0)$. It coincides to the energy of Weyl point $E_{W}\approx 114.4$ meV, smaller than $E_F$ in CoSi. Thus, we expect the Fermi energy crosses the nodal plane in a ring form for CoSi. Therefore, up to the first order perturbation, we obtain a hierarchy of quasi-symmetry for CoSi, represented by (Fig.~\ref{fig2}) 
\begin{align}\label{eq-qs-hierarchy-2}
	\text{SU}_{\text{s}}(2)\times \text{SU}_{\text{o}}(2) 
	\stackrel{\left\langle\mathcal{H}_{\text{soc}}+\mathcal{H}_{2}\right\rangle_{\mathcal{H}_1}}{\xhookrightarrow{\quad\quad\quad\quad\quad\quad}} \text{U}(1),
\end{align}
and the corresponding energy bands are split from four-fold degeneracies at any momenta down to two-fold degeneracies that form nodal planes. Including further second-order perturbation corrections generate a tiny gap for the near-nodal planes, as shown in Fig.~\ref{fig2}(c). At $E_F=E_W$, the near-nodal rings at the Fermi energy shrink into nodal points, the Weyl points, which are stable to any order and do not rely on quasi-symmetries.

The quasi-symmetry $S_{\text{eff}}$ is essential in protecting the gapless nature of the near-nodal planes in the four-band effective model. To see that, we may consider a generic four-band Hamiltonian commuting with $S_{\text{eff}}$ for any momenta, which can only include the following terms $\mathcal{H}_{S} = \mu_0 s_0 \omega_0 + \mu_1 [({\bf k}\cdot{\bf s})\omega_0] + \mu_2 [s_0({\bf f}\cdot \boldsymbol{\omega})] + \mu_3 [({\bf k}\cdot{\bf s})({\bf g}\cdot\boldsymbol{\omega})]$. Here $\mu_{0,1,2,3}$ are all positive constants, ${\bf f}({\bf k})=(f_1({\bf k}),f_2({\bf k}),f_3({\bf k}))$ and ${\bf g}({\bf k})=(g_1({\bf k}),g_2({\bf k}),g_3({\bf k}))$ are two vectors of generic functions of ${\bf k}$. $\mathcal{H}_{S}$ contains all the terms in Eq.~\eqref{eq-eff-p-model}. The eigen-energies of $\mathcal{H}_{S}$ are $E_{\alpha,\beta}({\bf k})=\mu_0+\alpha (\mu_1k) + \beta\vert \mu_2{\bf f}({\bf k}) + \alpha \mu_3k{\bf g}({\bf k}) \vert$ with $\alpha,\beta=\pm$. Generally, all the bands are non-degenerate at generic momenta ${\bf k}$ once ${\bf f}({\bf k})$ and ${\bf g}({\bf k})$ are non-zero. Accidental degeneracy can occur when (1) $\mu_2{\bf f}({\bf k}) = \alpha \mu_3k{\bf g}({\bf k})$ for $E_{\alpha,+}=E_{\alpha,-}$, which gives a nodal point, and (2) $2\mu_1k - \sum_{\alpha=\pm} \vert \mu_2{\bf f}({\bf k}) + \alpha \mu_3k{\bf g}({\bf k}) \vert=0$ for $E_{+,-}({\bf k})=E_{-,+}({\bf k})$, which defines plane solutions for the degenerate subspace in 3D momentum space. In our model, the former corresponds to the degeneracy at ${\bf k}=0$, while the latter gives the nodal planes. It should be noted that a two-level degeneracy usually requires three constraint equations (codimension 3), and thus only Weyl nodes are stable in 3D momentum space~\cite{Wigner1967,murakami_prb_2007}. The presence of quasi-symmetry $S_{\text{eff}}$ reduces the number of the constraint equation to 1 (codimension 1), making the nodal planes stable. This can be viewed as a generalization of the Wigner-Von Neumann codimension theory.

{\it Approach II }--
In our second approach, $\mathcal{H}_1+\mathcal{H}_{\text{soc}}$ is treated as the zeroth order Hamiltonian and $\mathcal{H}_2$ as the perturbation. For $\mathcal{H}_1+\mathcal{H}_{\text{soc}}$, we find that the spin SU$_{\text{s}}$(2) symmetry is broken by SOC while the orbital SU$_{\text{o}}$(2) symmetry generated by $\mathcal{M}_{1,2,3}$ remains. The existence of SU$_{\text{o}}$(2) is due to the fact that spin ${\bf s}$ as a pseudo-vector behaves exactly the same as a vector due to the lack of inversion, mirror, {\it etc}. in chiral crystals, so $\mathcal{H}_{\text{soc}}$ can be obtained by replacing ${\bf k}$ by ${\bf s}$ in $\mathcal{H}_1$. The corresponding energy bands are given by $E_{1,\pm}({\bf k})=\pm A_1k+\lambda_0$ and $E_{2,\pm}({\bf k})=\pm \sqrt{A_1^2k^2 + 4\lambda_0^2}-\lambda_0$, and each band has two-fold degeneracy, as required by SU$_{\text{o}}$(2). The SOC-induced splitting between the $E_{1,\pm}(k)$ and $E_{2,\pm}(k)$ is $2\lambda_0$ for a large momentum $k$, which is depicted along the non-high-symmetry line ($\theta=\phi=\pi/3$) in Fig.~\ref{fig3}(a).

Generally, the $k^2$ terms of $\mathcal{H}_2$ break the SU$_{\text{o}}$(2) quasi-symmetry and lead to the splitting of all bands. One can show $[{\bf J}_i,\mathcal{M}_i]=0$ and $\{{\bf J}_i, \mathcal{M}_j\}=0$ for $i\neq j$, so that $[\mathcal{H}_{2,\mathcal{M}_i},\mathcal{M}_i]=0$. Without loss of generality, we can pick up one term, say $\mathcal{H}_{2, \mathcal{M}_3}$, which commutes with $\mathcal{M}_3$ but anti-commutes with $\mathcal{M}_{1}$ and $\mathcal{M}_{2}$. We show such choice of specific $k^2$-terms are general in Sec.~(D) of the SM~\cite{sm2022}. As a result, the term $\mathcal{H}_{2, \mathcal{M}_3}$ breaks the SU$_{\text{o}}$(2) quasi-symmetry group down into a U$_\text{o}$(1) group generated by $\mathcal{M}_3$. Thus, the two-fold degenerate bands $E_{1,+}({\bf k})$ and $E_{2,+}({\bf k})$ are split by $\mathcal{H}_{2, \mathcal{M}_3}$, as shown in Fig.~\ref{fig3}(b). The new eigen-states $\vert E_{i,\alpha,\beta}({\bf k})\rangle$ with $i=1,2$ are the common eigen-states of $\mathcal{H}_1+\mathcal{H}_{\text{soc}}+\mathcal{H}_{2, \mathcal{M}_3}$ and $\mathcal{M}_3$,
\begin{align}\label{eq-ek-quasi-nodal-bands}
\begin{split}
E_{1,\alpha,\beta} &= E_0 +\alpha \sqrt{f_{k^2} + 2\beta A_1\tilde{C}k_xk_yk_z} +\lambda_0, \\
E_{2,\alpha,\beta} &= E_0 +\alpha  \sqrt{f_{k^2} + 2\beta A_1\tilde{C}k_xk_yk_z + 4\lambda_0^2} -\lambda_0,
\end{split}
\end{align}
where $\alpha,\beta=\pm$, $E_0=B_1k^2 $ and $f_{k^2}=A_1^2k^2 + C_1^2k_x^2k_y^2 + C_2^2k_x^2k_z^2 + C_3^2k_y^2k_z^2$. The index $\beta$ labels the eigen-values of the $\mathcal{M}_3$ operators (see details in Sec.~(D) of the SM~\cite{sm2022}). When the splitting $2A_1\tilde{C}k_xk_yk_z/\sqrt{f_{k^2}}$ by $\mathcal{H}_{2, \mathcal{M}_3}$ reaches the SOC-induced splitting $2\lambda_0$, the condition
\begin{align}\label{eq-equ-nodal-plane-qsR-model}
E_{1,+,-} (\mathbf{k}) = E_{2,+,+} (\mathbf{k}), \; \forall \mathbf{k} \in k_xk_yk_z>0
\end{align}
is satisfied and leads to the band crossings that form nodal planes. Since $\mathcal{H}_{2, \mathcal{M}_3}$ increases with $k^2$ while the SOC-induced splitting $2\lambda_0$ is independent of ${\bf k}$, the condition~\eqref{eq-equ-nodal-plane-qsR-model} can always be satisfied at large enough $k$. As the two bands that form the nodal planes possess the opposite $\beta$ values ($\mathcal{M}_3$ parities), we expect the nodal planes are protected by quasi-symmetry. Turning on all the remaining $k^2$-terms break the U(1) quasi-symmetry and generates a tiny gap of nodal planes, as shown in Fig.~\ref{fig3}(c). The hierarchy structure of quasi-symmetries for the approach II is summarized in Fig.~\ref{fig3} as
\begin{align}
	\text{SU}_{\text{s}}(2)\times \text{SU}_{\text{o}}(2) \stackrel{\mathcal{H}_{\text{soc}}}{\xhookrightarrow{\quad\quad\;}} \text{SU}_{\text{o}}(2) 
	\stackrel{\mathcal{H}_{2,\mathcal{M}_i}}{\xhookrightarrow{\quad\quad\quad\;}} \text{U}_{\text{o}}(1).
\end{align}

{\it Conclusions and outlooks} -- 
In this work, we describe two different perturbation approaches to reveal the hierarchy structure of quasi-symmetry and near-degeneracy in electronic band structures of chiral crystal materials CoSi. Both approaches describe the physical consequence of near-nodal planes and thus are physically equivalent. The approach I reveals a self-commuting Hamiltonian in the first-order perturbation, while the approach II treats both the SOC and linear-k term as the zeroth order. 
We anticipate such a hierarchy structure of quasi-symmetry in the context of ${\bf k\cdot p}$ expansion of the effective models can generally appear in 230 space groups~\cite{zhang_arxiv_2021,tang_arxiv_2022}, which will be left for the future work. The hierarchy structure of quasi-symmetry also provides a natural starting point to discuss physical phenomena in different energy scales of the effective models. For example, in CoSi, the smallest energy scale $\sim 1$meV of the gap for near-nodal planes will easily be overcome by perturbations, e.g. disorder, and thus not be felt by electrons that take the cyclotron motion under magnetic fields, which is crucial in understanding the nearly angle-independent quantum oscillation spectrum in CoSi~\cite{guo_arxiv_2021}, as well as other experiments~\cite{xu_prb_2019,Wu_cpl_2019,wang_prb_2020,Ni_nc_2021,huber_prl_2021}. It is worth to note that the iso-structural compounds PtGa~\cite{Yao_nc_2020,Ma_nc_2021,Xu_cpl_2020}, PdAl~\cite{Schrter_np_2019}, PdGa~\cite{Schrter_science_2020,Sessi_nc_2020} and RhSi~\cite{chang_prl_2017,Rees_sa_2020} share the similar electronic band structure. Therefore, quasi-symmetry is also expected to play a major role in understanding their physical properties which require further experimental and theoretical attentions.

{\it Acknowledgement} --
We would like to acknowledge Carsten Putzke, Jonas Diaz,  Xiangwei Huang, Kaustuv Manna, Feng-Ren Fan, Chandra Shekhar, Zhen Bi, Kaijie Yang, Abhinava Chatterjee, Ruobing Mei and Rui-Xing Zhang for the helpful discussion.  LHH and CXL are supported by the Office of Naval Research (Grant No. N00014-18-1-2793).
BAB was supported by the Simons Investigator grant (No. 404513), the Office of Naval Research (ONR Grant No. N00014-20-1-2303),  the Schmidt Fund for Innovative Research, the BSF Israel US foundation (Grant No. 2018226), the Gordon and Betty Moore Foundation through Grant No. GBMF8685 towards the Princeton theory program and Grant No. GBMF11070 towards the EPiQS Initiative, the Schmidt DataX Fund at Princeton University from the Schmidt Futures Foundation and the European Research Council (ERC) under the European Union's Horizon 2020 research and innovation programme (grant agreement no. 101020833) and Princeton Global Network Funds. 
CXL and BAB also acknowledges the support from the NSF-MERSEC (Grant No. MERSEC DMR 2011750).
L.E. was supported by the Government of the Basque Country (Project IT1301-19) and the Spanish Ministry of Science and Innovation (PID2019-106644GB-I00).

\bibliographystyle{apsrev4-2}
\bibliography{ref}

\clearpage
\appendix
\begin{widetext}
\setcounter{page}{1}
\begin{center}
{\bf Supplementary materials for ``Hierarchy of quasi-symmetries and degeneracies in chiral crystal materials CoSi''}
\end{center}

\tableofcontents

\section{The effective ${\bf k}\cdot{\bf p}$ Hamiltonian around the $R$ point}	
\label{Appendix-A}

In this work, we mainly focus on the electronic bands of Cobalt Silicide (CoSi) around the $R$-point ${\bf k}_R=(\pi,\pi,\pi)$ (band structure is shown in Fig.~(1) in the main text), specifically, the four electron-type Fermi surfaces (FSs) around the Fermi energy. To understand the low-energy physics, we construct the effective Hamiltonian. For this purpose, in this appendix, we first discuss the space group 198 and its symmetry operators' matrix representations at the $R$ point. Then, we use the ${\bf k}\cdot{\bf p}$ theory to construct the effective model with the parameters fitting to the DFT bands.

Fig.~\ref{sm-fig1}(a) summarizes the general routines to construct the effective models with/without spin degree of freedom. First, we obtain the spin-orbit coupling (SOC)-free 4-band spinless model. Secondly, the spin degree of freedom is taken into account by considering the on-site atomic SOC, resulting a 8-band spinful model. This is valid when SOC is relatively weak. Based on this 8-band model, we then discuss the emergent internal quasi-symmetries and the corresponding hierarchy structure by using two approaches (discussed in the main text), illustrated in Fig.~\ref{sm-fig1}(b). The details of the approach I will be discussed in Sec.~\ref{Appendix-C}, and the details of the second approach will be discussed in Sec.~\ref{Appendix-D}.

\begin{figure}[!htbp]
	\centering
	\includegraphics[width=0.95\linewidth]{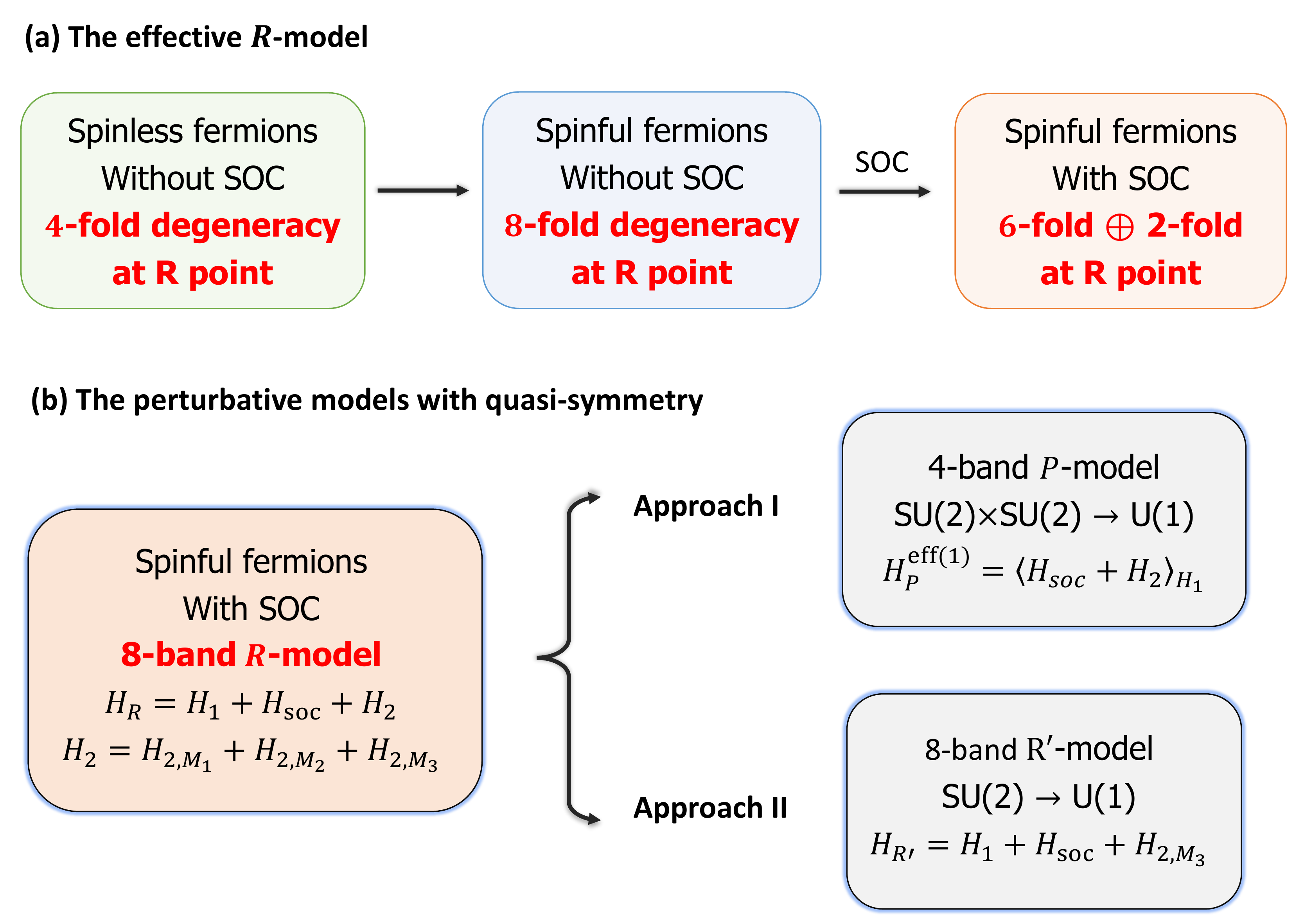}
	\caption{A brief summary of model analysis for the hierarchy of quasi-symmetry groups. 
	(a) shows the schematic process to construct the 8-band ${\bf k}\cdot {\bf p}$ effective Hamiltonian, labeled as $R$-model. At $R$ point, a 4-dimensional single-valued irreducible representation (Irrep) is our starting point, specifically, the basis for the 4-band spinless model. With the electron's spin degeneracy, it becomes a 8-fold degeneracy, which is split into a 6-fold degeneracy and a 2-fold degeneracy by the on-site atomic SOC.
	(b) shows the two approaches used in the main text to identify the quasi-symmetry with the perturbation theory. 
	In the ``Approach I'', we use first-order perturbation theory to project the effective low-energy 4-band $P$-model, and find the hierarchy of quasi-symmetry from SU(2)$\times$SU(2) down to U(1). 
	In the ``Approach II'', we add specially selected terms of the $k^2$-order Hamiltonian into the $\mathcal{H}_1+\mathcal{H}_{\text{soc}}$ to identify the hierarchy of quasi-symmetry from SU(2) down to U(1).
	}	
	\label{sm-fig1}
\end{figure}

\subsection{The Crystalline space group No.~198 and representations}
\label{Appendix-A-1}

As described in the main text, the CoSi crystallizes in a chiral cubic structure of space group (SG) $P2_{1}3$ (No. 198) without a center of inversion. Its lattice structure with lattice constant $a_x=a_y=a_z=4.433$ \AA, containing four Si atoms and four Co atoms in one unit cell. The corresponding  Brillouin zone (BZ) is also cubic. The SG 198 has 12 symmetry operations in addition to the translation sub-group. The three generators of SG 198 are: one threefold rotation symmetry along the ($111$) axis  and two twofold screw rotation symmetries along the $x$ and $y$ axis. Hereafter, the Seitz notation is taken for the non-symmorphic symmetry operations, i.e., a point group operation $\mathcal{O}$ followed by a translation $\mathbf{v}=v_i\mathbf{t}_i$, labeled as $ \hat{\mathcal{O}} = \{ \mathcal{O} | \mathbf{v} \} \text{ or } \hat{\mathcal{O}} = \{ \mathcal{O} | v_1v_2v_3\}$, with $\mathbf{t}_i$ ($i=1,2,3$) representing three basis vectors for a Bravais lattice in three dimensions. The rules for multiplication and inversion are defined as
\begin{align}
\begin{split}
	\{ \mathcal{O}_2 | \mathbf{v}_2 \} \{ \mathcal{O}_1 | \mathbf{v}_1 \} &= \{ \mathcal{O}_2\mathcal{O}_1 | \mathbf{v}_2 + \mathcal{O}_2 \mathbf{v}_1 \}, \\
	\{\mathcal{O}\vert \mathbf{v}  \}^{-1} &= \{ \mathcal{O}^{-1}\vert -\mathcal{O}^{-1}\mathbf{v} \}.
\end{split}
\end{align}
In addition to the translation operator $E_\mathbf{v}=\{E|\mathbf{v}\}$, the three symmetry generators of SG 198 are
\begin{align}\label{eq-little-SP198-cxl-operators}
	S_{2x} = \{ C_{2x} | \tfrac{1}{2} \tfrac{1}{2} 0 \}, 
	S_{2y} = \{ C_{2y} | 0 \tfrac{1}{2} \tfrac{1}{2}  \}, 
	C_{3} = \{ C_{3,(111)} | 0 0 0 \},
\end{align}
defined by $S_{2x}: (x,y,z)\to(x+\tfrac{1}{2} , -y +\tfrac{1}{2} , -z)$, $S_{2y}: (x,y,z)\to(-x,y+\tfrac{1}{2},-z+\tfrac{1}{2})$, and $C_{3}:(x,y,z)\to (y,z,x)$. Thus, one can check that $S_{2z}=\{C_{2z}|\tfrac{1}{2} 0 \tfrac{1}{2}  \}$ can be given by the combination of $S_{2x}$ and $S_{2y}$,
\begin{align} 
	S_{2x} S_{2y} = \{E|00\bar{1}\} S_{2z} \triangleq E_{00\bar{1}}S_{2z}.
\end{align}
In addition, the threefold rotation can also be along ($11\bar{1}$), ($1\bar{1}1$), and ($\bar{1}11$) axis. Therefore, the lattice of CoSi has three twofold and four threefold rotation or screw axes.

Next, we use the commutation relations of the symmetry group generators to directly construct the corresponding matrix representations. Alternatively, they can be found on the on the notations in the Bilbao Crystallographic Server~\cite{Bradlyn_nature_2017TQC,Xu_nature_2020,Elcoro_nc_2021}. The band calculation based on the density-functional theory (DFT) without SOC shows that all states are fourfold degenerate at $R$ point, which should belong to one 4D irreducible representation (Irrep). As mentioned in the main text, the 4D Irrep for the four-fold degenerate states close to the Fermi energy can be denoted as the single-valued Irrep $R_1R_3$ on the Bilbao~\cite{Bradlyn_nature_2017TQC,Xu_nature_2020,Elcoro_nc_2021}. Below, we discuss how this 4D Irrep can be established by considering the twofold screw rotations $S_{2x}$, $S_{2y}$, and time reversal (TR) symmetry $\mathcal{T}$. At $R$-point, we have
\begin{align}
	\begin{split}
		S_{2x}^2 = S_{2y}^2 =-1, \; S_{2x}S_{2y} = -S_{2y}S_{2x}, 
		[S_{2x},\mathcal{T}] = [S_{2y},\mathcal{T}] =0. 
	\end{split}
\end{align}
Without loss of generality, for the spinless fermions, the TR symmetry operator can be chosen as $\mathcal{T}=\mathcal{K}$. Based on the above commutation relations, we construct the matrix representations denoted as $G_{1,2,3}$ for the twofold screw rotations $S_{2x}$, $S_{2y}$, and the threefold rotation $C_3$, respectively. Let's choose $\Psi$ as the eigen-state of $G_1$ with eigenvalue $\lambda$, and the eigen-values for different states at the $R$ point constructed from $\Psi$ are given in the following table. 
\begin{center}
	\begin{tabular}{|c|c|c|c|c|}
		\hline 
		& $\Psi$ & $G_2\Psi$ & $\mathcal{T}\Psi$ & $G_2\mathcal{T}\Psi$ \\ 
		\hline 
		$G_1$ & $\lambda$ & $-\lambda$ & $\lambda^\ast = -\lambda$ & $-\lambda^\ast=\lambda$ \\ 
		\hline 
	\end{tabular} 
\end{center}
With $G_1^2=-1$, $\lambda$ is a purely imaginary number. By using $( G_2 \mathcal{T})^2=-1$, $G_2 \mathcal{T}$ is an anti-unitary symmetry operator, leading to the Kramer's degeneracy, namely $\langle \Psi \vert G_2 \mathcal{T} \Psi \rangle =0$.
This leads to two orthogonal states: $\Psi$ and $G_2T\Psi$. We now apply $G_2$ on these two states to generate the other two states, $G_2\Psi$ and $\mathcal{T}\Psi$, which are also eigen-states of $G_1$ with the eigenvalue $-\lambda$. The $G_2$-generated states have opposite $G_1$-eigenvalues compared with that of $\Psi$ or $G_2\mathcal{T}\Psi$. Therefore, $G_2\Psi$ and $\mathcal{T}\Psi$ are orthogonal to $\Psi$ and $G_2\mathcal{T}\Psi$, so
\begin{align}
	\langle \Psi \vert G_2 \Psi \rangle = \langle \Psi \vert \mathcal{T} \Psi \rangle  = \langle G_2\mathcal{T} \Psi \vert G_2 \Psi \rangle = \langle G_2\mathcal{T} \Psi \vert \mathcal{T} \Psi \rangle = 0.
\end{align}
Therefore, the four-fold degeneracy is formed by the four eigen-states at $R$,  
\begin{align}
	\{\Psi, G_2\Psi, \mathcal{T}\Psi,G_2\mathcal{T}\Psi \}.
\end{align}
In principle, a general basis of this 4D Irrep at $R$-point can be presented as 
\begin{align}\label{eq-general-4DIrrep-basis}
	\{\vert \Psi_1\rangle, \vert\Psi_2\rangle, \vert\Psi_3\rangle, \vert\Psi_4\rangle \}^T,
\end{align}
which serves as the basis for the effective SOC-free 4-band ${\bf k}\cdot{\bf p}$ Hamiltonian denoted as $\mathcal{H}_{\text{no-soc,R}}({\bf k})$. Hereafter, ${\bf k}$ is the relative momentum to $R$ point. In this work, we construct it up to $k^2$ order. Without loss of generality, we assume $\vert\Psi_i\rangle$ with $i=1,2,3,4$ are all real so that the representation for TR symmetry in this basis is given by
\begin{align}
	\mathcal{T}=I_{4\times4}\mathcal{K}, 
\end{align}
where $I_{4\times4}$ is the 4-by-4 identity matrix and $\mathcal{K}$ is the complex conjugate.

Next, we construct the matrix representations denoted as $G_1,G_2,G_3$ (4-by-4 matrices) for the twofold screw rotations $S_{2x}$, $S_{2y}$, and the threefold rotation $C_3$, respectively. For the spinless case, their commutation relations are summarized as 
\begin{align}\label{eq-cxl-g1g2g3-relation}
	G_3^3=1, G^2_1=G^2_2=-1, \;
	G_1G_2=-G_2G_1,\; 
	G_3^{-1}G_1G_3 = G_2, \;
	G_3^{-1}G_2G_3 = -G_1G_2,
\end{align}
because of $[\mathcal{T},G_1] = [\mathcal{T},G_2]=0$, we have $G_1=G_1^\ast$ and $G_2=G_2^\ast$. Moreover, $G_1^2=G_2^2=-1$, and then both $G_1$ and $G_2$ are anti-symmetric matrices, $G_{1,2}=-G_{1,2}^T$. Thus, $G_1 ,G_2$ can only be chosen from the following matrix set
\begin{align}
	 \{ i\sigma_y\tau_0, i\sigma_y\tau_x , i\sigma_y\tau_z, i\sigma_0\tau_y, i\sigma_x\tau_y, i\sigma_z\tau_y \},
\end{align}
where both $\sigma_{x,y,z}$ and $\tau_{x,y,z}$ represent the Pauli matrices, and $\sigma_0$, $\tau_0$ are two-by-two identity matrices. Considering $\{G_1,G_2\}=-1$, one can choose the representations as 
\begin{align}
	G_1 = i\sigma_y\tau_0 \text{ and } G_2 = i\sigma_x\tau_y.
\end{align}
Similarly, we now discuss how to construct $G_3$. According to the Bilbao~\cite{Bradlyn_nature_2017TQC,Xu_nature_2020,Elcoro_nc_2021}, there are two 4D single-valued Irreps: $R_1R_3$ and $R_2R_2$ of the little group at $R$ point. It also shows the trace of $G_3$ for the $R_1R_1$-Irrep ($R_2R_2$-Irrep) is $1$ ($-2$).  According to the DFT calculation without SOC (see the Sec.~\ref{Appendix-B-1} below), we find that the four states at $R$ point of CoSi near the Fermi energy are belonging to the $R_1R_3$-Irrep, because of 
\begin{align}
	\text{Tr}[C_3] = 1+1+e^{i\omega_0} + e^{-i\omega_0} = 1,
\end{align}
where $\omega_0=2\pi/3$, since we numerically check that these four states carry angular momentum $0,0,1,-1$ of $C_3$. To further satisfy both $\text{Tr}[C_3]=1$ and the commutation relations in Eq.~\eqref{eq-cxl-g1g2g3-relation}, we can choose 
\begin{align}\label{eq-spinless-C3-matrix}
	G_3 = \begin{pmatrix}
		1 & 0  & 0  &  0 \\ 
		0 & 0  & -1  & 0 \\ 
		0 & 0  & 0  & 1  \\ 
		0 & -1  & 0  & 0 
	\end{pmatrix}.
\end{align}
Please note that the choice of $G_3$ is not unique in the symmetry construction, while different choices of representation matrices just correspond to unitary transformation between different basis. For instance, the representation matrices $G_1,G_2, G_3$ and $\mathcal{T}$ are different form those on the Bilbao. This matrix chosen here is simple enough to make the construction of the effective ${\bf k}\cdot {\bf p}$ Hamiltonian become simper. Besides, in the Sec.~\ref{Appendix-B-2} below, we try to construct the basis made of the five $3d$-orbitals of the four Co atoms for the 4D Irrep $R_1R_3$. And the orbital basis can be explicitly shown by comparing to the Wannier functions from the DFT calculations.

\subsection{The spin-independent effective 4-band $R$-model}
\label{Appendix-A-2}

In this section, we construct the spin-independent 4-band ${\bf k}\cdot{\bf p}$ model Hamiltonian denoted as $\mathcal{H}_{\text{no-soc,R}}({\bf k})$ by using the matrix representations $G_1, G_2, G_3$ and TR symmetry $\mathcal{T}$ of the $R_1R_3$-Irrep. The general 4-by-4 SOC-free Hamiltonian $\mathcal{H}_{\text{no-soc,R}}({\bf k})$ is give by
\begin{align}
	\mathcal{H}_{\text{no-soc,R}}(\mathbf{k}) = \sum_{\mu\nu} h_{\mu\nu}(\mathbf{k}) \sigma_\mu\tau_\nu,
\end{align}
which should be invariant with any symmetry operators $g\in\{S_{2x},S_{2y},C_3,\mathcal{T}\}$ at the $R$-point. Here $\mathbf{k}$ is the momentum with reference to $\mathbf{k}_R=(\pi,\pi,\pi)$. Therefore, the Hamiltonian should satisfy 
\begin{align}
	\Delta^\dagger(g) \lbrack \mathcal{H}_{\text{no-soc,R}}(g\mathbf{k})  \rbrack  \Delta(g) = \mathcal{H}(\mathbf{k}),
\end{align}
where $\Delta(g)$ is the matrix representation for symmetry $g$, specifically, $G_1, G_2, G_3$ and $\mathcal{T}$. The classification of matrices ($\sigma_\mu\tau_\nu$ with $\mu,\nu=0,x,y,z$) and momentums ($k_i, k_ik_j$ with $i,j=x,y,z$) are summarized in Table.~\ref{tab-1} and Table.~\ref{tab-2}, respectively, from which the 4-by-4 Hamiltonian to the leading order becomes
\begin{align}\label{eq-ham0-k-order}
	\mathcal{H}_{1}(\mathbf{k}) = C_0 \sigma_0\tau_0 + A_1( k_x \sigma_y\tau_0 +  k_y \sigma_x\tau_y - k_z \sigma_z\tau_y).
\end{align}
The sign of $A_1$ is related to chirality of the crystal, which is chosen to be positive in this work. Moreover, we notice that the spin-independent Hamiltonian $\mathcal{H}_{1}(\mathbf{k})$ in Eq.~\eqref{eq-ham0-k-order} is isotropic with the full rotation symmetry. To see that, we could define the emergent angular momentum operators as
\begin{align}\label{eq-angular-mom-Lxyz}
	L_x = \tfrac{1}{2} \sigma_y\tau_0, \; L_y = \tfrac{1}{2}\sigma_x\tau_y, \; L_z = -\tfrac{1}{2}\sigma_z\tau_y,
\end{align}
which satisfies the commutation relation $[L_i,L_j] = i\epsilon_{ijk} L_k$ with Levi-Civita symbol $\epsilon_{ijk}$ and $i=x,y,z$.
Therefore, $\mathcal{H}_{1}(\mathbf{k})$ can be re-written as
\begin{align} \label{eq-ham0-k-order-Lk}
	\mathcal{H}_{1}(\mathbf{k}) &= C_0\sigma_0\tau_0 + 2A_1 (\mathbf{k}\cdot\mathbf{L}).
\end{align}
It is the linear-$k$ Hamiltonian presented in the main text (below Eq.~[1]). In addition, by similar symmetry analysis, the $k^2$ order effective Hamiltonian is given by
\begin{align}\label{eq-ham0-k2-order}
	\begin{split}
		\mathcal{H}_{2}(\mathbf{k}) &= B_1(k_x^2+k_y^2+k_z^2) 
		+ C_1(k_xk_y\sigma_z\tau_0 + k_yk_z\sigma_0\tau_z + k_xk_z\sigma_z\tau_z ) \\
		&+C_2(k_xk_y\sigma_x\tau_x -k_yk_z\sigma_0\tau_x + k_xk_z\sigma_x\tau_0 ) 
		+ C_3(k_xk_y\sigma_x\tau_z -k_yk_z\sigma_y\tau_y -k_xk_z\sigma_z\tau_x ),
	\end{split}
\end{align}
which can be re-organized into a compact way as presented in the main text, by noticing that the $k^2$ order effective Hamiltonian $\mathcal{H}_2$ shows an intriguing structure and can be grouped into three classes, 
\begin{align} \label{sm-eq-k2-ham-m1m2m3}
\mathcal{H}_{2}({\bf k})=\mathcal{H}_{2, \mathcal{M}_1}({\bf k}) +\mathcal{H}_{2, \mathcal{M}_2}({\bf k}) +\mathcal{H}_{2, \mathcal{M}_3}({\bf k}),
\end{align} 
where 
\begin{align}
	\mathcal{H}_{2,\mathcal{M}_i}({\bf k})={\bf g}_i({\bf k})\cdot{\bf J}_i,
\end{align}
for i=1,2,3. Here we define the parameter-momentum vectors 
\begin{subequations}
\begin{align}
	{\bf g}_1({\bf k}) &=(C_2k_xk_y,-C_3k_xk_z,C_1k_yk_z), \\
	{\bf g}_2({\bf k}) &=(C_3k_xk_y,C_1k_xk_z,-C_2k_yk_z), \\
	{\bf g}_3({\bf k}) &=(C_1k_xk_y,C_2k_xk_z, -C_3  k_y k_z).
\end{align}
\end{subequations}
And the operator vectors are
\begin{subequations}
\begin{align}
	{\bf J}_1 &=(\sigma_x\tau_x,-\sigma_z\tau_x,\sigma_0\tau_z), \\
	{\bf J}_2 &=(\sigma_x\tau_z,\sigma_z\tau_z,\sigma_0\tau_x), \\
	{\bf J}_3 &=(\sigma_z\tau_0,\sigma_x\tau_0,\sigma_y\tau_y).
\end{align}
\end{subequations}
The meaning of the subscript $\mathcal{M}_i$ is to be the role of quasi-symmetry operators, which has been explained in the main text and will be also discussed later in Sec.~\ref{Appendix-C} with details.

Combining Eq.~\eqref{eq-ham0-k-order} and Eq.~\eqref{eq-ham0-k2-order}, we finally get the effective 4-by-4 SOC-free Hamiltonian up to $k^2$-order,
\begin{align}\label{eq-ham-soc-free-tot}
	\mathcal{H}_{\text{no-soc,R}}({\bf k}) = \mathcal{H}_{1}(\mathbf{k}) + \mathcal{H}_{2}(\mathbf{k}),
\end{align}
which is called the SOC-free $R$-model Hamiltonian for short in the following discussions.

\begin{table}[!htbp]
	\begin{tabular}{c| c|c|c|c}
		\hline \hline
		$\sigma_\mu\tau_\nu$ & $S_{2x}=i\sigma_y\tau_0$  & $S_{2y}=i\sigma_x\tau_y$  & $\mathcal{T}=\mathcal{K}$  & $C_3$ in Eq.~\eqref{eq-spinless-C3-matrix} \\ 
		\hline 
		$\sigma_0\tau_0$ & $+$ & $+$  & $+$  & $\sigma_0\tau_0$  \\ 
		\hline 
		$\sigma_0\tau_x, \sigma_0\tau_z$ & $+$ & $-$  & $+$  & $-\sigma_x\tau_0, \sigma_z\tau_z$  \\ 
		\hline 
		$\sigma_0\tau_y$ & $+$ & $+$  & $-$  & $-\sigma_y\tau_z$  \\ 
		\hline 
		$\sigma_x\tau_0$ & $-$ & $+$  & $+$  & $\sigma_x\tau_x$  \\ 
		\hline 
		$\sigma_x\tau_x, \sigma_x\tau_z$ & $-$ & $-$  & $+$  & $-\sigma_0\tau_x, -\sigma_y\tau_y$  \\ 
		\hline 
		$\sigma_x\tau_y$ & $-$ & $+$  & $-$  & $-\sigma_z\tau_y$  \\ 
		\hline 
		$\sigma_y\tau_0$ & $+$ & $-$  & $-$  & $\sigma_x\tau_y$  \\ 
		\hline 
		$\sigma_y\tau_x, \sigma_y\tau_z$ & $+$ & $+$  & $-$  & $-\sigma_0\tau_y, \sigma_y\tau_x$  \\ 
		\hline 
		$\sigma_y\tau_y$ & $+$ & $-$  & $+$  & $\sigma_z\tau_x$  \\ 
		\hline 
		$\sigma_z\tau_0$ & $-$ & $-$  & $+$  & $\sigma_0\tau_z$  \\ 
		\hline 
		$\sigma_z\tau_x, \sigma_z\tau_z$ & $-$ & $+$  & $+$  & $-\sigma_x\tau_z, \sigma_z\tau_0$  \\ 
		\hline 
		$\sigma_z\tau_y$ & $-$ & $-$  & $-$  & $-\sigma_y\tau_0$  \\ 
		\hline\hline
	\end{tabular} 
	\caption{\label{tab-1}The classification of the Pauli matrices under the space group symmetry operators and time-reversal symmetry.}
\end{table}

Moreover, we notice that the linear-$k$ Hamiltonian $\mathcal{H}_1({\bf k})$ in Eq.~\eqref{eq-ham0-k-order-Lk} has a full rotational symmetry. This rotation is a combined rotation in both ${\bf k}$-space and orbital-space simultaneously. To show that, without loss of generality, we can first define a rotation in the momentum space to make 
\begin{align}
	R(\theta,\phi) (k_x, k_y, k_z)^T = k(0,0,1)^T,
\end{align}
where ${\bf k} = k(\sin\theta \cos \phi, \sin\theta\sin\phi, \cos\theta)$, and the rotation in the momentum space is
\begin{align}
	R(\theta,\phi) &\triangleq R_y(-\theta) R_z(-\phi) = \begin{pmatrix}
		\cos\theta & 0 & -\sin\theta \\
		0 & 1 & 0 \\
		\sin\theta & 0 & \cos\theta
	\end{pmatrix} \begin{pmatrix}
		\cos\phi & \sin\phi & 0 \\
		-\sin\phi & \cos\phi & 0 \\
		0 & 0 & 1
	\end{pmatrix}
	=\begin{pmatrix}
		\cos\theta\cos\phi & \cos\theta\sin\phi & -\sin\theta \\
		-\sin\phi & \cos\phi & 0 \\
		\sin\theta\cos\phi & \sin\theta\sin\phi & \cos\theta
	\end{pmatrix} .
\end{align}
To keep $\mathbf{k}\cdot\mathbf{L}$ invariant under rotations, we then define the associated rotation for the orbital subspace,
\begin{align}
	\mathcal{R}(\theta,\phi) \triangleq e^{-i\theta L_y} e^{-i\phi L_z} = \left(
	\begin{array}{cccc}
		\cos \left(\frac{\theta }{2}\right) \cos \left(\frac{\phi }{2}\right) & \cos \left(\frac{\theta }{2}\right) \sin \left(\frac{\phi }{2}\right) & \sin \left(\frac{\theta }{2}\right) \sin \left(\frac{\phi }{2}\right) & -\sin \left(\frac{\theta }{2}\right) \cos \left(\frac{\phi }{2}\right)  \\
		-\cos \left(\frac{\theta }{2}\right) \sin \left(\frac{\phi }{2}\right) & \cos \left(\frac{\theta }{2}\right) \cos \left(\frac{\phi }{2}\right) & \sin \left(\frac{\theta }{2}\right) \cos \left(\frac{\phi }{2}\right) & \sin \left(\frac{\theta }{2}\right) \sin \left(\frac{\phi }{2}\right) \\
		-\sin \left(\frac{\theta }{2}\right) \sin \left(\frac{\phi }{2}\right) & -\sin \left(\frac{\theta }{2}\right) \cos \left(\frac{\phi }{2}\right) & \cos \left(\frac{\theta }{2}\right) \cos \left(\frac{\phi }{2}\right) & -\cos \left(\frac{\theta }{2}\right) \sin \left(\frac{\phi }{2}\right) \\
		\sin \left(\frac{\theta }{2}\right) \cos \left(\frac{\phi }{2}\right) & -\sin \left(\frac{\theta }{2}\right) \sin \left(\frac{\phi }{2}\right) & \cos \left(\frac{\theta }{2}\right) \sin \left(\frac{\phi }{2}\right) & \cos \left(\frac{\theta }{2}\right) \cos \left(\frac{\phi }{2}\right) \\
	\end{array}
	\right),
\end{align}
Since $\mathbf{k}$ is rotated to the z-axis, we only need to compute the $L_z(\theta,\phi)$ after the rotation,
\begin{subequations}
\begin{align}
	L_z(\theta,\phi) &= \mathcal{R}(\theta,\phi) L_z \mathcal{R}^{-1}(\theta,\phi) = \frac{1}{2}\left(
	\begin{array}{cccc}
		0 & i \cos (\theta ) & -i \sin (\theta ) \cos (\phi ) & -i \sin (\theta ) \sin (\phi ) \\
		-i \cos (\theta ) & 0 & i \sin (\theta ) \sin (\phi ) & -i \sin (\theta ) \cos (\phi ) \\
		i \sin (\theta ) \cos (\phi ) & -i \sin (\theta ) \sin (\phi ) & 0 & -i \cos (\theta ) \\
		i \sin (\theta ) \sin (\phi ) & i \sin (\theta ) \cos (\phi ) & i \cos (\theta ) & 0 \\
	\end{array}
	\right), \\
	&= \left\lbrack  \sin(\theta)\cos(\phi)  L_x  + \sin(\theta)\sin(\phi) L_y + \cos(\theta) L_z \right\rbrack, \\
	&= {\bf L}\cdot \vec{n}_{{\bf k}}.
\end{align}
\end{subequations}
Here $\vec{n}_{{\bf k}}=\frac{{\bf k}}{k}$ is the direction of ${\bf k}$. Therefore, the linear-$k$ Hamiltonian becomes $\mathcal{H}_1(k,\theta,\phi) = C_0 + 2A_1 k L_z (\theta,\phi) $, which indicates the invariant of this Hamiltonian under the combined rotation $R(\theta,\phi)$ in the momentum space and $\mathcal{R}(\theta,\phi)$ in the orbital space keeps. Note that the helicity operator for low-energy Dirac fermions in spin-momentum coupled crystals is defined as $\mathcal{P}^{\bf S}_{\bf k}=\mathbf{S}\cdot \vec{n}_{\bf k}$ with $\mathbf{S}$ the spin matrix. For Hamiltonian that commutes with $\mathcal{P}^{\bf S}_{\bf k}$, such as, $\mathcal{H} \sim \mathbf{S}\cdot {\bf k}$, whose eigenstates at fixed ${\bf k}$ can be labeled by the eigenvalues $p=\pm 1/2$ of $\mathcal{P}^{\bf S}_{\bf k}$, $(\mathbf{S}\cdot {\bf k} ) \vert p\rangle = p k \vert p\rangle$, and these two states $\vert\pm 1/2\rangle$ represent  left-handed or right-handed states. Following this spirit, we define a similar helicity operator $\mathcal{P}^{\bf L}_{\bf k}={\bf L}\cdot \vec{n}_{\bf k}$ to reveal the angular momentum polarization along the moving direction in the absence of spin-orbit coupling. Thus, we obtain
\begin{align}
	\mathcal{H}_1(k,\theta,\phi)  \vert \pm \tfrac{1}{2} \rangle = (C_0 \pm A_1k) \vert  \pm \tfrac{1}{2} \rangle.
\end{align}
Here, $\pm \tfrac{1}{2}$ are the eigenvalues of $\mathcal{P}^{\bf L}_{\bf k}$. Each state has fourfold degeneracy if spin degeneracy is accounted. And, the explicit form of the eigen-wavefunctions will be given in Sec.~\ref{Appendix-C}.

\subsection{The spinful 8-band $R$-model with spin-orbit coupling}
\label{Appendix-A-3}

In this section, we further take spin degree of freedom into account and derive the effective 8-by-8 Hamiltonian with SOC. With the spin degree of freedom $\{\uparrow,\downarrow \}$, the spinful basis becomes
\begin{align}\label{eq-4dIR-basis-for-kp-ham-soc}
	\{ \vert \Psi_{1,\uparrow}\rangle,  \vert \Psi_{2,\uparrow}\rangle,  \vert \Psi_{3,\uparrow}\rangle, \vert \Psi_{4,\uparrow}\rangle \}^T 
	\oplus 
	\{\vert \Psi_{1,\downarrow}\rangle, \vert \Psi_{2,\downarrow}\rangle, \vert \Psi_{3,\downarrow}\rangle, \vert \Psi_{4,\downarrow}\rangle \}^T,
\end{align}
where $\vert \Psi_{i,\sigma}\rangle = \vert \Psi_i\rangle \otimes \vert \sigma\rangle$ with $i=1,2,3,4$ and $\sigma=\uparrow,\downarrow$. Thus, the spinful $R$-model consists of two parts,
\begin{align}
	\mathcal{H}_{R}(\mathbf{k}) = s_0\otimes \mathcal{H}_{\text{no-soc,R}}(\mathbf{k}) +  \mathcal{H}_{\text{soc}},
\end{align}
where the spin-independent part $\mathcal{H}_{\text{no-soc,R}}(\mathbf{k})$ is given by Eq.~\eqref{eq-ham-soc-free-tot} and $\mathcal{H}_{\text{soc}}$ represents the $k$-independent SOC Hamiltonian. Here $\mathcal{H}_{\text{soc}}$ is also constructed from the symmetry principle, and is in a similar form as $\mathcal{H}_1({\bf k})$. To show that, we need to consider the full rotation operators acting in both spin and orbital spaces,
\begin{align}
	&S_{2x} = (is_x)\otimes (i\sigma_y\tau_0), \\
	&S_{2y}=(is_y)\otimes (i\sigma_x\tau_y),\\ 
	&C_{3}=e^{i\frac{\pi}{3\sqrt{3}}(s_x+s_y+s_z) }\otimes \begin{pmatrix}
		1 & 0  & 0  &  0 \\ 
		0 & 0  & -1  & 0 \\ 
		0 & 0  & 0  & 1  \\ 
		0 & -1  & 0  & 0 
	\end{pmatrix},\\
	&\mathcal{T}=(is_y) \otimes \sigma_0\tau_0\mathcal{K}.
\end{align}
We use $s_{x,y,z}$ to be the Pauli matrices acting on the spin subspace, which obey
\begin{align}
	C_3: (s_x,s_y,s_z) \to (s_y,s_z,s_x),
\end{align}
which is the same as the transformation as momentum $(k_x,k_y,k_z)$ under $C_{3}$. It is because the material CoSi is a chiral crystal, so that the electron spin (pseudo-vector) behaves the same as the ${\bf k}$-vector (see their classifications in Table.~\ref{tab-2}). Therefore, the lowest-order SOC Hamiltonian reads
\begin{align}\label{eq-soc-ham}
	\begin{split}
		\mathcal{H}_{\text{soc}} &= \lambda_0 ( s_x \sigma_y\tau_0 + s_y \sigma_x\tau_y - s_z \sigma_z\tau_y )
		= 4 \lambda_0 ( \mathbf{S} \cdot \mathbf{L}),
	\end{split}
\end{align}
which is obtained by just replacing $(k_x,k_y,k_z)$ by $(s_x,s_y,s_z)$. Here we take the notation: the spin operators $\mathbf{S}=\tfrac{1}{2}(s_x,s_y,s_z)$ and the angular momentum operators $\mathbf{L}$ given by Eq.~\eqref{eq-angular-mom-Lxyz}. The SU(2) algebra for the angular momentum operators is represented as
\begin{subequations}
\begin{align}
	[S_i, S_j] &= i\epsilon_{ijk} S_k, \\
	[L_i, L_j] &= i\epsilon_{ijk} L_k. 
\end{align}
\end{subequations}
Moreover, in the next Sec.~\ref{Appendix-B}, we will use two approaches for the justification of the above SOC Hamiltonian.

In addition to the on-site SOC in Eq.~\eqref{eq-soc-ham}, the linear-$k$ SOC Hamiltonian generally reads
\begin{align}\label{eq-linear-k-soc-ham}
	\begin{split}
		\mathcal{H}_{k,\text{soc}}({\bf k}) &= \lambda_1(k_x s_x + k_y s_y + k_zs_z)\otimes \sigma_0 \tau_0 
		+ \lambda_2 ( k_x s_y\sigma_x\tau_x - k_y s_z\sigma_0\tau_x + k_z s_x\sigma_x\tau_0 ) \\
		&+\lambda_3 ( k_y s_x\sigma_x\tau_x - k_z s_y\sigma_0\tau_x + k_x s_z\sigma_x\tau_0 ) 
		+ \lambda_4(k_x s_y\sigma_x\tau_z - k_ys_z\sigma_y\tau_y -k_zs_x\sigma_z\tau_x) \\
		&+\lambda_5(k_ys_x\sigma_x\tau_z - k_z s_y\sigma_y\tau_y - k_xs_z\sigma_z\tau_x) 
		+\lambda_6(k_x s_y\sigma_z\tau_0 + k_y s_z\sigma_0\tau_z + k_z s_x\sigma_z\tau_z) \\
		&+\lambda_7 (k_ys_x\sigma_z\tau_0 + k_z s_y\sigma_0\tau_z + k_x s_z\sigma_z\tau_z).
	\end{split}
\end{align}
Combining the SOC-free $R$-model in Eq.~\eqref{eq-ham-soc-free-tot} with the SOC Hamiltonians in Eq.~\eqref{eq-soc-ham} and Eq.~\eqref{eq-linear-k-soc-ham}, we finally get the effective 8-band ${\bf k}\cdot{\bf p}$ Hamiltonian with SOC as
\begin{align}\label{eq-R-model-H0-soc-k2-ksoc}
	\mathcal{H}_{R}(\mathbf{k}) =  s_0\otimes \lbrack \mathcal{H}_{1}(\mathbf{k}) + \mathcal{H}_{2}(\mathbf{k}) \rbrack +  \mathcal{H}_{\text{soc}} + \mathcal{H}_{k,\text{soc}}({\bf k}) ,
\end{align}
which is called the 8-band $R$-model with SOC for short in the following discussions. The basis for the $R$-model is made of the five $d$-orbitals of the four Co atoms, whose details are shown in Sec.~\ref{Appendix-B}. Since the SOC is relatively weak in CoSi, the linear-$k$ SOC terms are neglected in the main text. Nevertheless, their influence on the quasi-nodal planes will be addressed in the following Sec.~\ref{Appendix-E}.

\begin{table}[!htbp]
	\begin{tabular}{c| c|c|c|c}
		\hline \hline
		Momentums & $C_{2x}$  & $C_{2y}$  & $\mathcal{T}$  & $C_{3,(111)}$ \\ 
		\hline 
		$k_x$ & $+$ & $-$  & $-$  & $k_y$  \\ 
		\hline 
		$k_y$ & $-$ & $+$  & $-$  & $k_z$  \\ 
		\hline 
		$k_z$ & $-$ & $-$  & $-$  & $k_x$  \\ 
		\hline 
		$k_x^2,k_y^2,k_z^2$ & $+$ & $+$  & $+$  & $k_y^2,k_z^2,k_x^2$  \\ 
		\hline 
		$k_xk_y$ & $-$ & $-$  & $+$  & $k_yk_z$  \\ 
		\hline 
		$k_xk_z$ & $-$ & $+$  & $+$  & $k_yk_x$   \\ 
		\hline
		$k_yk_z$ & $+$ & $-$  & $+$  & $k_zk_x$   \\ 
		\hline
		$s_x$ & $+$ & $-$  & $-$  & $s_y$  \\ 
		\hline 
		$s_y$ & $-$ & $+$  & $-$  & $s_z$  \\ 
		\hline 
		$s_z$ & $-$ & $-$  & $-$  & $s_x$  \\ 
		\hline \hline
	\end{tabular} 
	\caption{\label{tab-2}The momentums and spin operators under the space group symmetry operators and time-reversal symmetry. }	
\end{table}

\subsection{The sixfold degenerate states at $R$ point for spin-1/2 fermions}
\label{Appendix-A-4}

Next we briefly discuss the energy level splitting of the $R$-model in Eq.~\eqref{eq-R-model-H0-soc-k2-ksoc} at $R$ point (i.e.,~${\bf k}=0$) due to the presence of the on-site SOC Hamiltonian $\mathcal{H}_{\text{soc}}$. Solving $\mathcal{H}_R({\bf k}=0)$ gives rise to a six-fold degeneracy (energy $\lambda_0$) and a two-fold degeneracy (energy $-3\lambda_0$). The DFT calculation also implies that $\lambda_0>0$, so that the sixfold degenerate states have higher energy than the twofold states. And the sixfold is the double-valued $\bar{R}_6\bar{R}_7$ irrep while the twofold is the $\bar{R}_5\bar{R}_6$ irrep based on the irrep notations on the Bilbao~\cite{Bradlyn_nature_2017TQC,Xu_nature_2020,Elcoro_nc_2021}.
Here we analytically solve the on-site SOC Hamiltonian. To do that, we first apply a unitary transformation 
\begin{align}
	U=s_0\otimes\sigma_0\otimes \left(\begin{array}{cc}
		i & 1 \\	-i & 1 \\ \end{array} \right),
\end{align}
which transforms $\mathcal{H}_{\text{soc}}$ into a block diagonal form
\begin{align}
	U \lbrack \mathcal{H}_{\text{soc}} \rbrack U^\dagger = \begin{pmatrix}
		\mathcal{H}_+(0)	&  0 \\ 
		0	& \mathcal{H}_-(0)
	\end{pmatrix}, 
\end{align}
where the subscript $\pm$ labels the eigenvalues of $\tau_y$ and two blocks are given by
\begin{align}
	\mathcal{H}_+(0) &= C_0  + \lambda_0\left( s_x\sigma_y + s_y\sigma_x - s_z\sigma_z \right) = C_0 + \lambda_0 (\mathbf{s}\cdot \mathbf{\sigma}'), \\
	\mathcal{H}_-(0) &= C_0  + \lambda_0\left( s_x\sigma_y - s_y\sigma_x + s_z\sigma_z \right) = C_0 + \lambda_0 (\mathbf{s}\cdot \mathbf{\sigma}''),
\end{align} 
where $\mathbf{\sigma}'$ and $\mathbf{\sigma}''$  are defined as $(\sigma_y,\sigma_x,-\sigma_z)$ and $(\sigma_y,-\sigma_x,\sigma_z)$, respectively. In fact, these two blocks, $\mathcal{H}_+(0)$ and $\mathcal{H}_-(0)$, are related by TR symmetry. To show it, please notice that the TR symmetry is presented as $\mathcal{T}=is_y\sigma_0\tau_0\mathcal{K}$ in the original basis. After the unitary transformation, it becomes $\mathcal{T}_U=is_y\sigma_0\tau_x\mathcal{K}$.

We take $\mathcal{H}_+(0)$ as an example, where $\mathbf{\sigma}'$ can be treated as pseudo-spin, so it preserves $\mathbf{J}=\tfrac{1}{2}\mathbf{s} + \tfrac{1}{2}\mathbf{\sigma}'$. Therefore, the addition of two spin-1/2 naturally leads to one singlet state and three degenerate triplet states as $\frac{1}{2}\otimes \frac{1}{2} = 1\oplus 0$. By using the identity
\begin{equation}
	(\mathbf{s}\cdot \mathbf{\sigma}') = 2\left\lbrack  \mathbf{J}^2 -(\tfrac{1}{2}\mathbf{s})^2 - (\tfrac{1}{2}\mathbf{\sigma}')^2 \right\rbrack = 2(j(j+1) - \tfrac{1}{2}\times \tfrac{3}{2}\times2),  
\end{equation}
where $j=0$ for singlet state and $j=1$ for triplet states, we can solve the eigen-energies as $E_{s}=C_0-3\lambda_0$ for the singlet state and $E_t=C_0+\lambda_0$ for the three-fold triplet states. Similarly, the $\mathcal{H}_-(0)$ block also has one singlet state with energy $E_{s}=C_0-3\lambda_0$ and three triplet state with energy $E_t=C_0+\lambda_0$. Therefore, the on-site SOC Hamiltonian splits the eight states at $R$ point into a six-fold degeneracy and another two-fold degeneracy.

Alternatively, the sixfold degeneracy can be viewed from the spin-1 excitation with its time-reversal (TR)-related partner. And we examine the sixfold degenerate states by symmetry arguments~\cite{bradlyn_science_2016}. At the $R$ point, we have the following commutation relations for spin-$1/2$ fermions,
\begin{align}\label{eq-G1G2G3-relation-spinful}
	\begin{split}
		G_3^3=-1, G^2_1=G^2_2=1, 
		G_1G_2=G_2G_1, 
		G_3^{-1}G_1G_3 = G_2, 
		G_3^{-1}G_2G_3 = G_1G_2,
	\end{split}
\end{align}
which provides the sufficient condition for a 3D Irrep at the $R$ point. With $G_1 \Psi = \lambda_1 \Psi $ and $G_2 \Psi = \lambda_2 \Psi $, we have
\begin{align}
	\begin{split}
		&G_1 ( G_3 \Psi ) = \lambda_2 ( G_3 \Psi ),    \quad
		G_2 (G_3 \Psi ) = \lambda_1 \lambda_2 ( G_3 \Psi ), \\
		&G_1 (G_3^2 \Psi) = \lambda_1\lambda_2 (G_3^2 \Psi),  \quad
		G_2 (G_3^2 \Psi) = \lambda_1 (G_3^2 \Psi).
	\end{split}
\end{align}
If either $\lambda_1\neq1$ or $\lambda_2\neq 1$, $\Psi$, $G_3\Psi$ and $G_3^2\Psi$ all carry different eigenvalues under $G_1$ and $G_2$. Thus, it can lead to the basis for a 3D Irrep represented by $\{ \Psi, G_3\Psi, G_3^2\Psi \}$, as proved in Ref.~\cite{bradlyn_science_2016}. Next, let us discuss the effect of TR symmetry for the spin-$1/2$ system, and consider the eigenvalues of the states $\mathcal{T}\Psi_i$. Because $[\mathcal{T},G_1]=[\mathcal{T},G_2]=0$ and the eigenvalues of $G_1$ and $G_2$ are real, $\mathcal{T}\Psi_i$ has the same eigenvalues of $G_1$ and $G_2$ as $\Psi_i$ with $\Psi_{i}\in \{\Psi, G_3 \Psi, G_3^2 \Psi \}$. So, all these six basis functions are orthogonal with each other
\begin{align}
	\{\Psi, G_3 \Psi, G_3^2 \Psi \} \oplus \{\mathcal{T}\Psi, G_3 \mathcal{T}\Psi, G_3^2 \mathcal{T}\Psi \},
\end{align}
which generally forms the sixfold degeneracy~\cite{bradlyn_science_2016}.

\section{The justification of the SOC Hamiltonian and The $3d$-orbital basis of the $R$-model}	
\label{Appendix-B}

In this section, we use two approaches to justify the on-site SOC Hamiltonian obtained by the symmetry argument (see Eq.~\eqref{eq-soc-ham}). One is the full tight-binding (TB) model based on the Wannier function method~\cite{marzari_rmp_2012}, from which we can obtain the exact wave functions. The other is the construction of the 4D Irrep by using the five $3d$-orbitals of the four Co atoms, providing an alternative understanding for the single-valued $R_1R_3$ Irrep and the SOC Hamiltonian.

\subsection{The Justification of SOC Hamiltonian from the full tight-binding model}
\label{Appendix-B-1}

In this section, we discuss the four spinless degenerate wave functions (i.e.,~$R_1R_3$-Irrep) via a full TB model without SOC based on the Wannier function method from the DFT calculations. The model Hamiltonian (a 52-by-52 matrix) includes four Co atoms ($4s,4p,3d$-orbitals) and four Si atoms ($3s, 3p$-orbitals) in one uni-cell. In each simple cubic unit cell, these Co and Si atoms are located at 
\begin{align}
		\begin{split}
			\mathbf{R}_{\text{Co}_1} &= ( 0.14,  0.14,  0.14),  \; \; 
			\mathbf{R}_{\text{Co}_2} = (-0.14, -0.36,  0.36), \\
			\mathbf{R}_{\text{Co}_3} &= (-0.36,  0.36, -0.14), \; \; 
			\mathbf{R}_{\text{Co}_4} = ( 0.36, -0.14, -0.36), \\
			\mathbf{R}_{\text{Si}_1} &= (-0.157, -0.157, -0.157), \; \; 
			\mathbf{R}_{\text{Si}_2} = ( 0.157,  0.343, -0.343), \\
			\mathbf{R}_{\text{Si}_3} &= ( 0.343, -0.343,  0.157), \; \; 
			\mathbf{R}_{\text{Si}_4} = (-0.343,  0.157,  0.343),
		\end{split}
\end{align}
which are in the unit of the lattice constant $a_0=4.45$ \AA. The atomic orbitals of the full TB model are
\begin{align} \label{sm-eq-full-tb-basis}
		\begin{split}
			\Psi_{\text{TB}} =\{ 
			&\phi_{\text{Co}_1,4s}, \phi_{\text{Co}_2,4s},  \phi_{\text{Co}_3,4s}, \phi_{\text{Co}_4,4s},  
			\phi_{\text{Co}_1,4p_y}, \phi_{\text{Co}_1,4p_z},  \phi_{\text{Co}_1,4p_x}, 
			\phi_{\text{Co}_2,4p_y}, \phi_{\text{Co}_2,4p_z},  \phi_{\text{Co}_2,4p_x}, 
			\phi_{\text{Co}_3,4p_y}, \\
			&\phi_{\text{Co}_3,4p_z}, \phi_{\text{Co}_3,4p_x}, \phi_{\text{Co}_4,4p_y}, \phi_{\text{Co}_4,4p_z},  \phi_{\text{Co}_4,4p_x}, 
			\phi_{\text{Co}_1,3d_{xy}}, \phi_{\text{Co}_1,3d_{yz}},  \phi_{\text{Co}_1,3d_{3z^2-1}},  \phi_{\text{Co}_1,3d_{xz}},  
			\phi_{\text{Co}_1,3d_{x^2-y^2}}, \\
			&\phi_{\text{Co}_2,3d_{xy}}, \phi_{\text{Co}_2,3d_{yz}},  \phi_{\text{Co}_2,3d_{3z^2-1}},  \phi_{\text{Co}_2,3d_{xz}},  
			\phi_{\text{Co}_2,3d_{x^2-y^2}}, 
			\phi_{\text{Co}_3,3d_{xy}}, \phi_{\text{Co}_3,3d_{yz}},  \phi_{\text{Co}_3,3d_{3z^2-1}},  
			\phi_{\text{Co}_3,3d_{xz}},   \\
			&\phi_{\text{Co}_3,3d_{x^2-y^2}}, \phi_{\text{Co}_4,3d_{xy}}, \phi_{\text{Co}_4,3d_{yz}},  \phi_{\text{Co}_4,3d_{3z^2-1}},  
			\phi_{\text{Co}_4,3d_{xz}},   	\phi_{\text{Co}_4,3d_{x^2-y^2}}, 
			\phi_{\text{Si}_1,3s}, \phi_{\text{Si}_2,3s},  \phi_{\text{Si}_3,3s}, \phi_{\text{Si}_4,3s},  \\
			&\phi_{\text{Si}_1,4p_y}, \phi_{\text{Si}_1,4p_z},  \phi_{\text{Si}_1,4p_x}, 
			\phi_{\text{Si}_2,4p_y}, \phi_{\text{Si}_2,4p_z},  \phi_{\text{Si}_2,4p_x}, 
			\phi_{\text{Si}_3,4p_y}, \phi_{\text{Si}_3,4p_z},  \phi_{\text{Si}_3,4p_x}, 
			\phi_{\text{Si}_4,4p_y}, \phi_{\text{Si}_4,4p_z},  \phi_{\text{Si}_4,4p_x}
			\}.
		\end{split}
\end{align}
Here $s,p,d$ are the real atomic orbitals. In the atomic orbital basis, we now discuss the crystal symmetries at $R$ point. For instance, the spinless symmetry operator $C_3$ is constructed as
\begin{align}\label{eq-C3-52-by-52}
		C_{3} = \left\lbrack C_{3,\text{Co}}  \right\rbrack \oplus  \left\lbrack C_{3,\text{Co}} \otimes C_{3,p} \right\rbrack \oplus 
		\left\lbrack C_{3,\text{Co}} \otimes C_{3,d} \right\rbrack \oplus \left\lbrack C_{3,\text{Si}}  \right\rbrack
		\oplus \left\lbrack C_{3,\text{Si}} \otimes C_{3,p} \right\rbrack,
\end{align}
where 
\begin{align}
		\begin{split}
			C_{3,\text{Co}} =  C_{3,\text{Si}} = \begin{pmatrix}
				1 & 0 & 0 & 0 \\ 
				0 & 0 & 0 & 1 \\ 
				0 & 1 & 0 & 0 \\ 
				0 & 0 & 1 & 0
			\end{pmatrix} , \; \; 
			C_{3,p}=\begin{pmatrix}
				0 & 1 & 0 \\ 
				0 & 0 & 1 \\ 
				1 & 0 & 0
			\end{pmatrix} , \; \; 
			C_{3,d}=\begin{pmatrix}
				0 & 1 & 0 & 0 & 0 \\ 
				0 & 0 & 0 & 1 & 0 \\ 
				0 & 0 & -\tfrac{1}{2} & 0 & \tfrac{\sqrt{3}}{2}  \\ 
				1 & 0 & 0 & 0 & 0 \\ 
				0 & 0 & -\tfrac{\sqrt{3}}{2} & 0 & -\tfrac{1}{2}
			\end{pmatrix} .
		\end{split}
\end{align}
Similarly, at $R$ point, the symmetry operator $S_{2x}$ is represented by
\begin{align}\label{eq-s2x-52-by-52}
		S_{2x} = \left\lbrack C_{2x,\text{Co}}  \right\rbrack \oplus  \left\lbrack C_{2x,\text{Co}} \otimes C_{2x,p} \right\rbrack \oplus 
		\left\lbrack C_{2x,\text{Co}} \otimes C_{2x,d} \right\rbrack \oplus \left\lbrack C_{2x,\text{Si}}  \right\rbrack
		\oplus \left\lbrack C_{2x,\text{Si}} \otimes C_{2x,p} \right\rbrack,
\end{align}
where 
\begin{align}
	C_{2x,\text{Co}} = C_{2x,\text{Si}} = \begin{pmatrix}
			0 & 0 & 1 & 0 \\ 
			0 & 0 & 0 & 1 \\ 
			-1 & 0 & 0 & 0 \\ 
			0 & -1 & 0 & 0
		\end{pmatrix}, \; \; 
		C_{2x,p} = \begin{pmatrix}
			-1 & 0 & 0 \\ 
			0 & -1 & 0 \\ 
			0 & 0 & 1
		\end{pmatrix} , \; \; 
		C_{2x,d} = \begin{pmatrix}
			-1 & 0 & 0 & 0 & 0 \\ 
			0 & 1 & 0 & 0 & 0 \\ 
			0 & 0 & 1 & 0 & 0  \\ 
			0 & 0 & 0 & -1 & 0 \\ 
			0 & 0 & 0 & 0 & 1
		\end{pmatrix} .
\end{align}

\subsubsection{Atomic $d$-orbital basis: the complex wavefunctions for the single-valued $R_1R_3$ Irrep}	

We then numerically solve the full SOC-free $52\times52$ TB model for the wavefunctions.  At the $R$ point, all states are fourfold degenerate. In this work, we only focus on those four degenerate wave functions, whose energy is closest to the Fermi energy, as labeled by $\vert \Psi_{TB} \rangle$
\begin{align}
	\vert \Psi_{TB} \rangle =\left\{\vert \Psi_{R,1}^{TB} \rangle , \vert \Psi_{R,2}^{TB} \rangle , \vert \Psi_{R,3}^{TB}\rangle,\vert \Psi_{R,4}^{TB}\rangle \right\}. 
\end{align} 
Other bands that are far away from the Fermi energy are neglected. Even though the dimension of each spinless wavefunction ($\vert \Psi^{TB}_{R,i}\rangle$) is $52$, we notice that the $3d$-orbitals of Co contribute to 80\% of total density of states. Then, it is reasonable to ignore the other contributions ($4s$ and $4p$-orbitals of the four Co atoms and all orbitals of the four Si atoms). The dimension is reduced to $20$ since we only keep the five $3d$-orbitals of the four Co atoms. Therefore, the dimension-reduced subspace at the $R$ point	 is expanded by  
\begin{align} \label{sm-eq-tb-basis-d-orbitals-1}
\{ \text{Co}_1, \text{Co}_2, \text{Co}_3, \text{Co}_4 \}\otimes\{d_{xy} , d_{yz}, d_{3z^2-1}, d_{xz}, d_{x^2-y^2} \}.
\end{align}
The three-dimensional Cartesian coordinates are used for defining the $d$-orbitals of the cubic lattice CoSi, whose definitions are given in Sec.~\ref{Appendix-B-2}. This convention is used throughout this work.

This set of basis for these five $d$-orbitals is used throughout this section. Therefore, the spinless basis in Eq.~\eqref{sm-eq-full-tb-basis} is reduced to 
\begin{align}
\begin{split}
\{ 
& \phi_{\text{Co}_1,3d_{xy}}, \phi_{\text{Co}_1,3d_{yz}},  \phi_{\text{Co}_1,3d_{3z^2-1}},  \phi_{\text{Co}_1,3d_{xz}},  
		\phi_{\text{Co}_1,3d_{x^2-y^2}}, \phi_{\text{Co}_2,3d_{xy}}, \phi_{\text{Co}_2,3d_{yz}},  \phi_{\text{Co}_2,3d_{3z^2-1}},  \phi_{\text{Co}_2,3d_{xz}}, \phi_{\text{Co}_2,3d_{x^2-y^2}},  \\ 
& \phi_{\text{Co}_3,3d_{xy}}, \phi_{\text{Co}_3,3d_{yz}},  \phi_{\text{Co}_3,3d_{3z^2-1}},  
		\phi_{\text{Co}_3,3d_{xz}},   \phi_{\text{Co}_3,3d_{x^2-y^2}}, \phi_{\text{Co}_4,3d_{xy}}, \phi_{\text{Co}_4,3d_{yz}},  \phi_{\text{Co}_4,3d_{3z^2-1}},  
		\phi_{\text{Co}_4,3d_{xz}},   	\phi_{\text{Co}_4,3d_{x^2-y^2}}, 
\}.
\end{split}
\end{align}
In this dimension-reduced basis, the symmetry operator $C_3$ in Eq.~\eqref{eq-C3-52-by-52}  and  $S_{2x}$ in Eq.~\eqref{eq-s2x-52-by-52}  become
\begin{align}
	C_{3} = C_{3,\text{Co}} \otimes C_{3,d}, \; \text{and } S_{2x} =  C_{2x,\text{Co}} \otimes C_{2x,d}.
\end{align}
These two are now $20\times 20$ matrices. Correspondingly, the four degenerate states at the $R$ point, $\vert \Psi^{TB}_{R,i}\rangle$ with $i=1,2,3,4$ are correspondingly reduced and renormalized (i.e.,~dimension 20). By symmetry principle, these four degenerate states form a 4D single-valued irrep. Thus, they can be used to construct the matrix representations of the symmetry operators of SG 198 and time-reversal. This can be helpful to check these fourfold degeneracy belongs to $R_1R_3$ Irrep or $R_2R_2$ Irrep. As we mentioned in Sec.~\ref{Appendix-A-1}, the trace of $G_3$ for the $R_1R_3$-Irrep ($R_2R_2$-Irrep) is $1$ ($-2$). For this purpose, the matrix representation $G_3$ for symmetry $C_3$ is calculated via
\begin{align}
	\left \lbrack G_3 \right\rbrack_{i,j} = \langle \Psi^{TB}_{R,i} \vert C_{3}   \vert \Psi^{TB}_{R,j} \rangle,
\end{align}
with $i,j=1,2,3,4$. Similarly, the matrix representation $G_1$ for $S_{2x}$ is given by $\langle \Psi_{TB} \vert S_{2x}\vert \Psi_{TB}\rangle$. And the numerical results for the wavefunctions are listed in Table.~\ref{sm-fig2}, which are orthogonal to each other
\begin{align}
	\langle \Psi^{TB}_{R,i} | \Psi^{TB}_{R,j} \rangle  = \delta_{i,j},
\end{align}
here $\delta_{i,j}$ is the Kronecker delta function. This also leads to the following relations
\begin{align}\label{eq-C3TB-trace}
	\text{Tr}[G_3]=1 \text{ and } \text{Tr}[G_1] =0.
\end{align}
This confirms these states form the $R_1R_3$ Irrep, as mentioned in the main text and Sec.~\ref{Appendix-A-1}. Furthermore, it is also ready to construct a basis under which the matrix representations of symmetry operators ($C_3$, $S_{2x}$, $S_{2y}$ and $\mathcal{T}$) are exactly given in Sec.~\ref{Appendix-A-3}. Because the SOC Hamiltonian can be justified once the $d$-orbital basis is constructed. To do that, we first take a proper unitary transformation, i.e., linear combination of these states $\vert \Psi^{TB}_{R,i}\rangle$ with $i=1,2,3,4$, to make these states $\vert \Psi^{TB}_{R,i}\rangle$ are eigen-states of $C_3$. Under this $d$-orbital basis, the matrix representations are correspondingly constructed as follows. The $G_3$ is given by
\begin{align} \label{sm-eq-g3-complex-basis}
	G_3 = \text{Diag}\left\lbrack 1, e^{i\omega_0}, 1, e^{-i\omega_0} \right\rbrack,
\end{align}
where $\omega_0=2\pi/3$. These are complex wave functions and the TR symmetry requires  $\Psi_{R,1}^{TB}=(\Psi_{R,3}^{TB})^\ast$ and $\Psi_{R,2}^{TB}=(\Psi_{R,4}^{TB})^\ast$. The time-reversal symmetry operator $\mathcal{T}$ is given by
\begin{align} \label{sm-eq-trs-complex-basis}
	\mathcal{T} &=  \left(
	\begin{array}{cccc}
		0 & 0 & 1 & 0 \\
		0 & 0 & 0 & 1 \\
		1 & 0 & 0 & 0 \\
		0 & 1 & 0 & 0 \\
	\end{array}
	\right)\mathcal{K},
\end{align}
where $\mathcal{K}$ is complex conjugate operator. Moreover, the matrix representation $G_1$ for $S_{2x}$ is given by
\begin{align} \label{sm-eq-g1-complex-basis} 
G_1 = \left( \begin{array}{cccc}
		-i\sqrt{\frac{1}{3}} & -\sqrt{\frac{2}{3}}  & 0 & 0 \\
		\sqrt{\frac{2}{3}}  & i\sqrt{\frac{1}{3}}   & 0 & 0 \\
		0 & 0 & i\sqrt{\frac{1}{3}}  & -\sqrt{\frac{2}{3}} \\
		0 & 0 & \sqrt{\frac{2}{3}}   & -i\sqrt{\frac{1}{3}} 
	\end{array}	\right).
\end{align}
In addition, one can also check that the matrix representation $G_2$ for the symmetry $S_{2y}$ is obtained to obey the commutation relations in Eq.~\eqref{eq-cxl-g1g2g3-relation}.

\begin{table}[t]
	\centering
	\includegraphics[width=\linewidth]{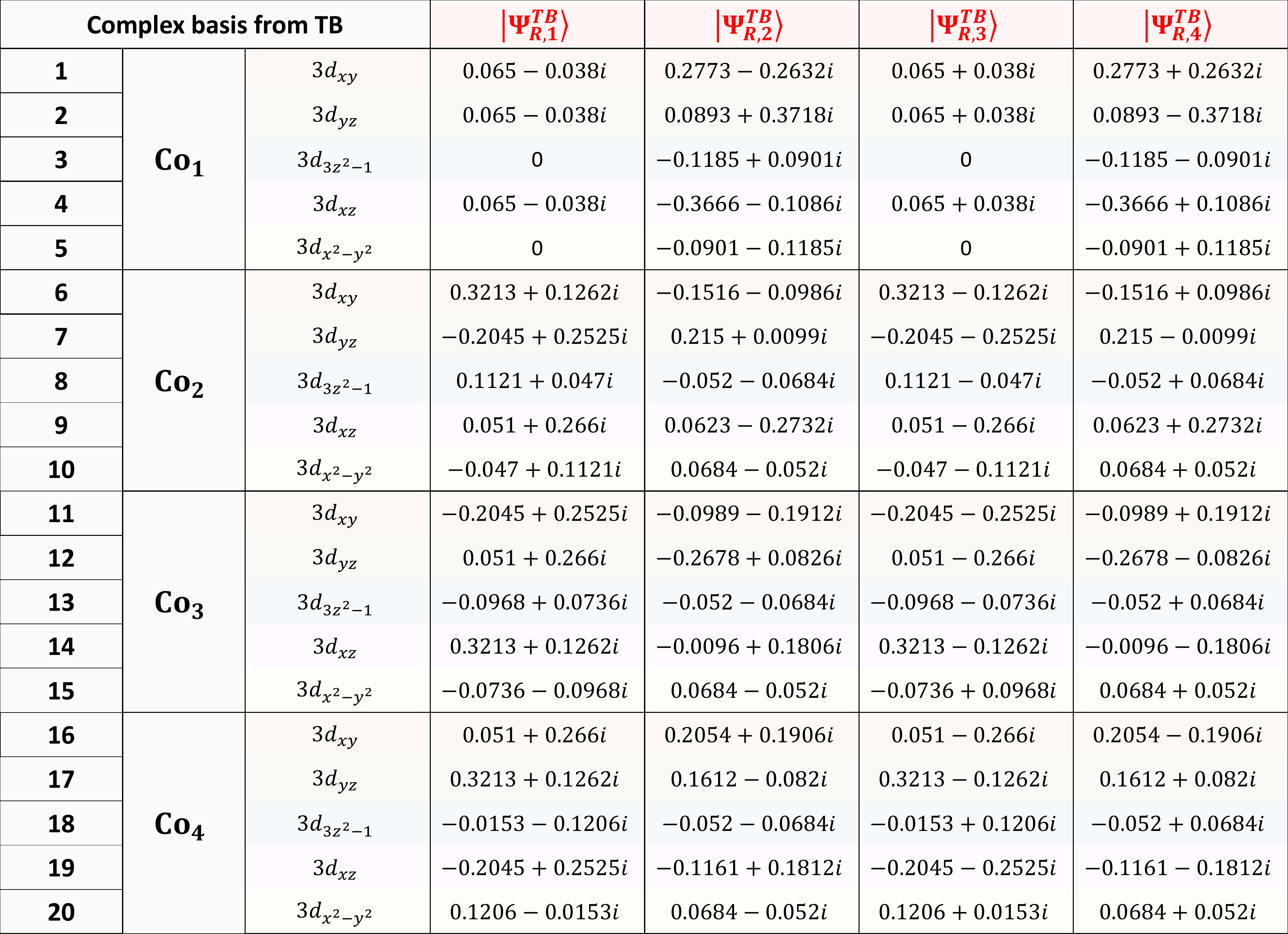}
	\caption{The numerically calculated {\bf complex} wavefunctions at the $R$ point, where the first column represents the component number of the wavefunctions. Since these wave functions are made of the five $3d$-orbitals of the four Co atoms, shown in the second and third columns. These four wavefunctions $\vert \Psi_{R,i}^{TB}\rangle$ with $i=1,2,3,4$ are the common eigenstates of the Hamiltonian and the $C_3^{TB}$ operators with the $C_3^{TB}$-eigenvalues $1,e^{i\omega_0},1,e^{-i\omega_0}$, respectively. These wavefunctions also show that time-reversal symmetry relates $\Psi_{R,1}^{TB}$ ($\Psi_{R,2}^{TB}$) and $\Psi_{R,3}^{TB}$ ($\Psi_{R,4}^{TB}$) by applying a complex conjugate. 
	}	
	\label{sm-fig2}
\end{table}

We now consider the band splitting induced by the atomic SOC. For the full TB model at the $R$ point, one can add the on-site atomic SOC Hamiltonian for the Co atoms,
\begin{align}
	\mathcal{H}_{\text{3d,soc}}= \lambda_{\text{soc}} \, \sum_{i=1}^4   \mathbf{s}_i \cdot \mathbf{L}_{3d}^i,
\end{align}
where $i$ labels the four Co atoms, $\mathbf{L}_{3d}$ represents the three angular momentum operators based on the five $d$-orbitals (i.e.,~$\{d_{xy} , d_{yz}, d_{3z^2-1}, d_{xz}, d_{x^2-y^2} \}$), and $\mathbf{s}$ are Pauli matrices acting on the spin subspace. We assume the SOC terms from other orbitals are small enough to be neglected. The explicit forms of operators  $\mathbf{s}$ and $\mathbf{L}_{3d}$ are given by
\begin{align}
	s_+ &= \frac{s_x + i s_y}{2} = \begin{pmatrix}
		0 & 1 \\ 
		0 & 0
	\end{pmatrix} , \; 
	s_- = \frac{s_x - i s_y}{2} =  \begin{pmatrix}
		0 & 0 \\ 
		1 & 0
	\end{pmatrix} , \; 
	s_z = \begin{pmatrix}
		1/2 & 0 \\ 
		0 & -1/2
	\end{pmatrix} , \\
	L_{3d,+} & = \begin{pmatrix}
		0 & -1 & 0 & -i & 0 \\ 
		1 & 0 & -i\sqrt{3} & 0 & -i \\ 
		0 & i\sqrt{3} & 0 & \sqrt{3}  & 0 \\ 
		i & 0 & -\sqrt{3} & 0 & 1 \\ 
		0 & i & 0 & -1 & 0
	\end{pmatrix} , \; 
	L_{3d,-} = \begin{pmatrix}
		0 & 1 & 0 & -i & 0 \\ 
		-1 & 0 & -i\sqrt{3} & 0 & -i \\ 
		0 & i\sqrt{3} & 0 & -\sqrt{3}  & 0 \\ 
		i & 0 & \sqrt{3} & 0 & -1 \\ 
		0 & i & 0 & 1 & 0
	\end{pmatrix} , \; 
	L_{3d,z} = \begin{pmatrix}
		0 & 0 & 0 & 0 & 2i \\ 
		0 & 0 & 0 & i & 0 \\ 
		0 & 0 & 0 & 0  & 0 \\ 
		0 & -i & 0 & 0 & 0 \\ 
		-2i & 0 & 0 & 0 & 0
	\end{pmatrix}.
\end{align}
Here we have defined $L_{3d,\pm} = (L_{3d,x}\pm i L_{3d,y} )/2$. With this $d$-orbital basis ($R_1R_3$ Irrep) shown in Table.~\ref{sm-fig2}, we next project the atomic SOC terms into the effective ${\bf k}\cdot {\bf p}$ Hamiltonian to justify the $k$-independent SOC Hamiltonian in Eq.~\eqref{eq-soc-ham}. To do that, we consider the basis at the $R$ point with spin degree of freedom, 
\begin{align} \label{sm-eq-basis-complex-r1r3-basis}
\begin{split}
&\{\vert\uparrow\rangle  \otimes \vert \Psi_{TB}\rangle ,  \vert\downarrow\rangle \otimes \vert \Psi_{TB}\rangle \} \\
=&\left\{ \vert \uparrow\rangle \otimes \vert \Psi^{TB}_{R,1} \rangle , 
 \vert \uparrow\rangle \otimes \vert \Psi^{TB}_{R,2} \rangle ,
 \vert \uparrow\rangle \otimes \vert \Psi^{TB}_{R,3} \rangle, 
 \vert \uparrow\rangle \otimes \vert \Psi^{TB}_{R,4} \rangle,
 \vert \downarrow\rangle \otimes \vert \Psi^{TB}_{R,1} \rangle , 
 \vert \downarrow\rangle \otimes \vert \Psi^{TB}_{R,2} \rangle ,
 \vert \downarrow\rangle \otimes \vert \Psi^{TB}_{R,3} \rangle, 
 \vert \downarrow\rangle \otimes \vert \Psi^{TB}_{R,4} \rangle
\right\}
\end{split}
\end{align} 
where $\vert \Psi^{TB}_{R,i} \rangle$ with $i=1,2,3,4$ have been shown in Table.~\ref{sm-fig2}. After projecting $\mathcal{H}_{\text{3d,soc}}$ onto the spinfull basis at the $R$ point in Eq.~\eqref{sm-eq-basis-complex-r1r3-basis}, the effective SOC Hamiltonian at the $R$ point reads
\begin{align}\label{eq-soc-TB-model}
	\mathcal{H}_{\text{eff,soc}} = \lambda_{\text{eff,soc}}  \mathbf{s} \cdot \tilde{\mathbf{L}}.
\end{align}
Here, the effective angular momentum operators, denoted as $\tilde{L}_x , \tilde{L}_y ,\tilde{L}_z $, are given by
\begin{align}
	[ \tilde{L}_{a} ]_{ij} =  \left\langle \Psi^{TB}_{R,i} \right\vert  I_{4\times4} \otimes L_{3d,a}  \left\vert  \Psi^{TB}_{R,j} \right\rangle,
\end{align}
where the index $a=\{x,y,z\}$, $i,j=1,2,3,4$, and $I_{4\times4}$ is a 4-by-4 identity matrix. And the numerical results of these 4-by-4 matrices are given by
\begin{subequations}
\begin{align}\label{eq-angular-momentum-lxlylz}
\tilde{L}_x &= \left( \begin{array}{cccc}
			-a_1 & ia_2 & 0 & 0 \\
			-ia_2 & a_1 & 0 & 0 \\
			0 & 0 & a_1 & ia_2 \\
			0 & 0 & -ia_2 & -a_1 \\
\end{array}	\right), \\
\tilde{L}_y &= \left(\begin{array}{cccc}
			-a_1 & -a_2e^{i\omega_0'} & 0 & 0 \\
			-a_2e^{-i\omega_0'} & a_1 & 0 & 0 \\
			0 & 0 & a_1 & a_2e^{-i\omega_0'} \\
			0 & 0 & a_2e^{i\omega_0'} & -a_1 \\
\end{array}	\right),  \\
\tilde{L}_z  &= \left(\begin{array}{cccc}
			-a_1 & a_2e^{-i\omega_0'} & 0 & 0 \\
			a_2e^{i\omega_0'} & a_1 & 0 & 0 \\
			0 & 0 & a_1 & -a_2e^{i\omega_0'} \\
			0 & 0 & -a_2e^{-i\omega_0'} & -a_1 \\
\end{array}\right), 
\end{align}
\end{subequations}
where $\omega_0'=\pi/6$, $a_1=1/(2\sqrt{3})$ and $a_2= 1/\sqrt{6}$. Moreover, the above effective angular momentum operators satisfy the standard (anti-)commutation relations
\begin{subequations}
\begin{align}\label{eq-lxlylz-TB}
	[\tilde{L}_i , \tilde{L}_j] &= i \epsilon_{ijk} \tilde{L}_k, \\
	\{ \tilde{L}_i , \tilde{L}_j\} & =\tfrac{1}{2}\delta_{ij} ,
\end{align}
\end{subequations}
where $i,j,k=\{x,y,z\}$, $\delta_{ij}$ is the Kronecker delta function, and $\epsilon_{ijk}$ is the Levi-Civita symbol. Moreover, we also notice that the eigenvalues of $\mathcal{H}_{\text{eff,soc}}$ are $-3\lambda_{\text{eff,soc}}$ (twofold)  and $\lambda_{\text{eff,soc}}$ (sixfold), by diagonalizing the SOC Hamiltonian~\eqref{eq-soc-TB-model}. This explains the energy splitting at the $R$ due to the presence of SOC, indicting that the important role of SOC in the low-energy physics. This is consistent with the analysis in Sec.~\ref{Appendix-A-4}.

\subsubsection{The real wavefunctions: atomic $d$-orbital basis for the 8-band $R$-model}

In this section, we construct the $d$-orbital basis for the $R$-model in the main text (details are shown in Sec.~\ref{Appendix-A-2} and Sec.~\ref{Appendix-A-3}). Recall that the spinful 8-band $R$-model in Eq.~\eqref{eq-R-model-H0-soc-k2-ksoc} is constructed only by symmetry arguments, but the corresponding {\bf physical} basis has not been derived. To solve this issue, we take the following unitary transformation $\mathcal{U}$, which connects these two representations,
\begin{align}
	\text{Matrix representations in the {\bf Complex} basis } \longleftrightarrow \text{ Matrix representation for the 8-band } R \text{ model}.
\end{align}
Once this is proved, the important on-site SOC Hamiltonian $\mathcal{H}_{\text{soc}}$ can be easily justified. Practically, $\mathcal{U}$ needs to be mathematically constructed to make the matrix representations ($G_1$ for $S_{2x}$ and $G_3$ for $C_3$) be transferred correspondingly as
\begin{align}
	\mathcal{U}^{-1}  G_3 \mathcal{U}  = \begin{pmatrix}
		1 & 0  & 0  &  0 \\ 
		0 & 0  & -1  & 0 \\ 
		0 & 0  & 0  & 1  \\ 
		0 & -1  & 0  & 0 
	\end{pmatrix},  \\
	\mathcal{U}^{-1}  G_1 \mathcal{U} = \left(
	\begin{array}{cccc}
		0 & 0 & 1 & 0 \\
		0 & 0 & 0 & 1 \\
		-1 & 0 & 0 & 0 \\
		0 & -1 & 0 & 0 \\
	\end{array}
	\right).
\end{align}
where the matrices $G_3$ and $G_1$ are given by Eq.~\eqref{sm-eq-g3-complex-basis} and Eq.~\eqref{sm-eq-g1-complex-basis}, respectively. The right side are the matrix representations used for the 8-band $R$ model in Sec.~\ref{Appendix-A-1}. Solving this problem is equivalently to define the following basis transformation based on the complex wavefunctions given by Table.~\ref{sm-fig2},
\begin{subequations}
\begin{align}
  \vert \Psi_{R,1}^{TB}\rangle_{\text{real}}  &= \frac{1}{\sqrt{2}} \left\lbrack \vert \Psi_{R,1}^{TB}\rangle + \vert \Psi_{R,3}^{TB}\rangle \right\rbrack, \\
  \vert \Psi_{R,2}^{TB}\rangle_{\text{real}}  &= -i\frac{1}{\sqrt{6}} \left\lbrack  \vert \Psi_{R,1}^{TB}\rangle - \vert \Psi_{R,3}^{TB}\rangle \right\rbrack  - \frac{1}{\sqrt{3}} \left\lbrack a_p^\ast \vert \Psi_{R,2}^{TB}\rangle + a_p \vert \Psi_{R,4}^{TB}\rangle \right\rbrack, \\
  \vert \Psi_{R,3}^{TB}\rangle_{\text{real}}  &=  i\frac{1}{\sqrt{6}} \left\lbrack  \vert \Psi_{R,1}^{TB}\rangle - \vert \Psi_{R,3}^{TB}\rangle \right\rbrack - \frac{1}{\sqrt{3}} \left\lbrack \vert \Psi_{R,2}^{TB}\rangle + \vert \Psi_{R,4}^{TB}\rangle \right\rbrack , \\
  \vert \Psi_{R,4}^{TB}\rangle_{\text{real}}  &= i\frac{1}{\sqrt{6}} \left\lbrack  \vert \Psi_{R,1}^{TB}\rangle - \vert \Psi_{R,3}^{TB}\rangle \right\rbrack + \frac{1}{\sqrt{3}} \left\lbrack a_p \vert \Psi_{R,2}^{TB}\rangle + a_p^\ast \vert \Psi_{R,4}^{TB}\rangle \right\rbrack,
\end{align}
\end{subequations}
where $a_p=\exp(i\tfrac{\pi}{3})$. Please notice that time-reversal is just a complex conjugate used for the construction of the 8-band $R$ model in Sec.~\ref{Appendix-A-1}, we have to define the real wavefunctions by the above basis transformation. In other words, the time-reversal symmetry in the complex basis is given by Eq.~\eqref{sm-eq-trs-complex-basis}, which needs to be transformed to 
\begin{align}
	\mathcal{T}=I_{4\times4}\mathcal{K},
\end{align}
with $I_{4\times4}$ a 4-by-4 identity matrix and $\mathcal{K}$ the complex conjugate. It is only possible for real basis wavefunctions. Therefore, the corresponding unitary transformation matrix $\mathcal{U}$ is represented as 
\begin{align}\label{sm-eq-com-real-U-trans}
	\mathcal{U} = \left(\begin{array}{cccc}
		\frac{1}{\sqrt{2}} & -\frac{i}{\sqrt{6}} & \frac{i}{\sqrt{6}} & \frac{i}{\sqrt{6}} \\
		0 & -\frac{a_p^\ast}{\sqrt{3}} & -\frac{1}{\sqrt{3}} & \frac{a_p}{\sqrt{3}} \\
		\frac{1}{\sqrt{2}} & \frac{i}{\sqrt{6}} & -\frac{i}{\sqrt{6}} & -\frac{i}{\sqrt{6}} \\
		0 & -\frac{a_p}{\sqrt{3}} & -\frac{1}{\sqrt{3}} & \frac{a_p^\ast}{\sqrt{3}} \\
	\end{array}	\right),
\end{align}
which is unitary transformation because of $\mathcal{U}\mathcal{U}^\dagger = \mathcal{U}\mathcal{U}^{-1} = I_{4\times4}$ and it gives rise to
\begin{align}
	\left( \vert \Psi_{R,1}^{TB}\rangle,  \vert \Psi_{R,2}^{TB}\rangle, \vert \Psi_{R,3}^{TB}\rangle, 
	\vert \Psi_{R,4}^{TB}\rangle  \right) \mathcal{U}  = 
	\left( \vert \Psi_{R,1}^{TB}\rangle_{\text{real}},  \vert \Psi_{R,2}^{TB}\rangle_{\text{real}}, \vert \Psi_{R,3}^{TB}\rangle_{\text{real}},  \vert \Psi_{R,4}^{TB}\rangle_{\text{real}}  \right) ,	
\end{align}
Therefore, the new wavefunctions are also orthogonal to each other, $_{\text{real}}\langle \Psi_{R,i}^{TB} \vert \Psi_{R,j}^{TB} \rangle_{\text{real}}=\delta_{i,j}$. Each component of the real wavefunctions are shown in Table.~\ref{sm-fig3}. Fo the $R$ point, we can use this real $d$-orbital basis in Table.~\ref{sm-fig3} to construct the on-site SOC Hamiltonian in Sec.~\ref{Appendix-A-3}. To show that, after the transformation, the projected angular momentum operators in Eq.~\eqref{eq-angular-momentum-lxlylz} become
\begin{subequations}
\begin{align}
	\mathcal{U}^{-1}  \tilde{L}_x \mathcal{U}  = \frac{1}{2}\left(
		\begin{array}{cccc}
			0 & 0 & -i & 0 \\
			0 & 0 & 0 & -i \\
			i & 0 & 0 & 0 \\
			0 & i & 0 & 0 \\
		\end{array}	\right) \triangleq L_x, \\
	\mathcal{U}^{-1}  \tilde{L}_y \mathcal{U}  = \frac{1}{2}\left(
		\begin{array}{cccc}
			0 & 0 & 0 & -i \\
			0 & 0 & i & 0 \\
			0 & -i & 0 & 0 \\
			i & 0 & 0 & 0 \\
		\end{array}	\right)  \triangleq L_y, \\
	\mathcal{U}^{-1}  \tilde{L}_z \mathcal{U}  = -\frac{1}{2}\left(
		\begin{array}{cccc}
			0 & -i & 0 & 0 \\
			i & 0 & 0 & 0 \\
			0 & 0 & 0 & i \\
			0 & 0 & -i & 0 \\
		\end{array}	\right)\triangleq L_z.
\end{align}
\end{subequations}
These matrices in the right side are $L_x = \tfrac{1}{2} \sigma_y\tau_0, \; L_y = \tfrac{1}{2}\sigma_x\tau_y, \; L_z = -\tfrac{1}{2}\sigma_z\tau_y$, which are just the angular momentum operators defined in Eq.~\eqref{eq-angular-mom-Lxyz}. According to Eq.~\eqref{eq-soc-TB-model}, the on-site SOC Hamiltonian becomes $\mathcal{U}^{-1}\mathcal{H}_{\text{eff,soc}} \mathcal{U} = 4\lambda_0 ({\bf S}\cdot {\bf L})$, with ${\bf S}=\tfrac{1}{2}(s_x,s_y,s_z)$ and ${\bf L}=(L_x, L_y, L_z)$. The SU(2) algebra for the angular momentum is given by $[S_i,S_j]=i\epsilon_{ijk}S_k$, $\{S_i,S_j\}=\tfrac{1}{2}\delta_{ij}$ and $[L_i,L_j]=i\epsilon_{ijk}L_k$, $\{L_i,L_j\}=\tfrac{1}{2}\delta_{ij}$. As a brief conclusion, we have numerically found the real basis function for the 8-band $R$-model, and also confirmed the form of the angular momentum operators, which are required for the construction of the on-site SOC Hamiltonian. The wavefunctions in Table.~\ref{sm-fig3} shows that the mixing of $t_{2g}$ and $e_g$ orbitals are essential for the on-site SOC Hamiltonian in the spinful $R$ model, as mentioned in the main text.

\begin{table}[t]
	\centering
	\includegraphics[width=\linewidth]{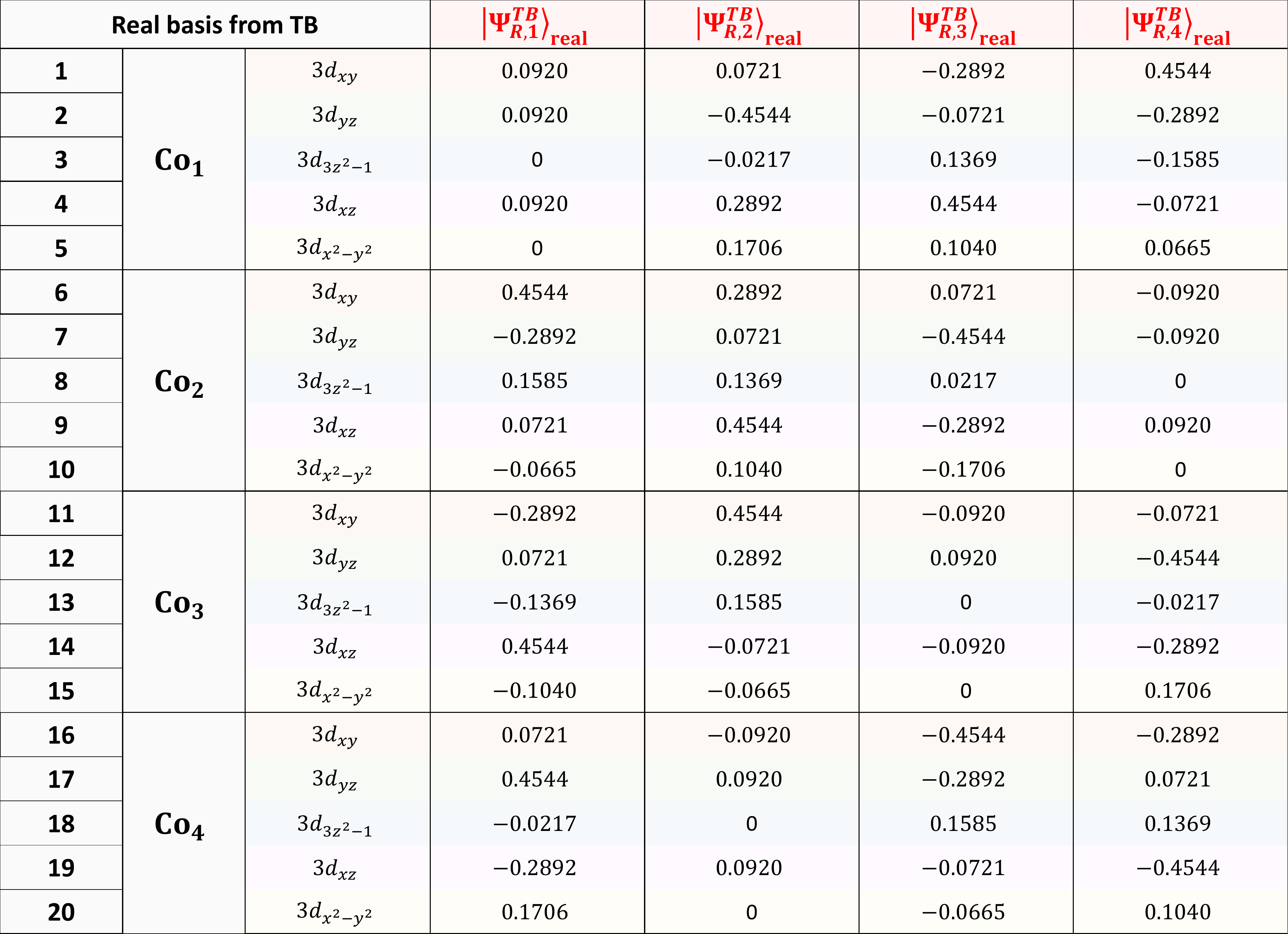}
	\caption{The numerically calculated {\bf real} wavefunctions at the $R$ point, the first column represents the component number of the wavefunctions. Since these wave functions are made of the five $3d$-orbitals of the four Co atoms, shown in the second and third columns. These real wavefunctions are obtained via a unitary transformation in Eq.~\eqref{sm-eq-com-real-U-trans} on the complex wavefunctions in Table.~\ref{sm-fig2}. In these new spinless basis, time-reversal symmetry is just a complex conjugate. This basis becomes the spinless basis for the 8-band $R$ model after taking the spin degree of freedom into account, represented by $\vert \uparrow\rangle \otimes \{ \vert \Psi_{R,1}^{TB}\rangle_{\text{real}},  \vert \Psi_{R,2}^{TB}\rangle_{\text{real}}, \vert \Psi_{R,3}^{TB}\rangle_{\text{real}},  \vert \Psi_{R,4}^{TB}\rangle_{\text{real}}   \} , \vert \downarrow \rangle \otimes \{ \vert \Psi_{R,1}^{TB}\rangle_{\text{real}},  \vert \Psi_{R,2}^{TB}\rangle_{\text{real}}, \vert \Psi_{R,3}^{TB}\rangle_{\text{real}},  \vert \Psi_{R,4}^{TB}\rangle_{\text{real}}   \} $.
	}	
	\label{sm-fig3}
\end{table}

\subsection{The 4D irreducible representations for the $3d$-orbitals of Co atoms}
\label{Appendix-B-2}

In this section, we use the atomic $3d$-orbitals to mathematically construct the Irreps for the eigen-states at the $R$ point and then justify the physical basis and the SOC Hamiltonian. This is equivalent to the numerical construction from the full TB model in the last two sections, but it will provide more theoretical understanding of the atomic basis of our model. We start from the spinless case with the dimension $4\times 5=20$ (four Co atoms and five $3d$-orbitals on each atom) and will show that these 20 atomic orbitals can consist of five 4D Irreps for the group generated by $\{S_{2x},S_{2y},C_3,\mathcal{T}\}$. In a unit cell, the positions of four Co atoms are (labelled by $l=1,2,3,4$)
\begin{align}
\begin{split}
	\mathbf{r}_1 &= (0.1451, 0.1451, 0.1451), \;\; 
	\mathbf{r}_2 = (0.3549, 0.8549, 0.6451), \\
	\mathbf{r}_3 &= (0.6451, 0.3549, 0.8549), \;\; 
	\mathbf{r}_4 = (0.8549, 0.6451, 0.3549).
\end{split}
\end{align}
The general basis consists four Co atoms and the relevant atomic orbitals, represented by $\Psi_{\alpha,l,\mathbf{k}_R}(\mathbf{r})$ with $\alpha$ for the five $d$-orbitals and $l$ for the four Co atoms.

Next, we focus on the little group at the $R$ point with $\mathbf{k}_R=(\pi,\pi,\pi)$. For a general symmetry operators $\mathcal{O} = \{ \mathcal{O} | \mathbf{v} \} $, the corresponding matrix representation is given by
\begin{align}
	\mathcal{O}  \Psi_{\alpha,l,\mathbf{k}_R}(\mathbf{r}) = e^{i\mathbf{k}_\mathcal{O} \cdot \mathbf{r}_{l'}  }  e^{-i(\mathbf{k}_R + \mathbf{k}_\mathcal{O})\cdot\mathbf{v}}
	\sum_{\beta} D_{\beta\alpha} \Psi_{\beta,l',\mathcal{O}\mathbf{k}_R}(\mathbf{r}),
\end{align}
where $\mathcal{O}\mathbf{k}_R = \mathbf{k}_R + \mathbf{k}_{\mathcal{O}}$ with $\mathbf{k}_\mathcal{O}$ a reciprocal lattice vector, $\mathcal{O}$ transform the Co atom $\mathbf{r}_l$ to the Co atom $\mathbf{r}_{l'}$, and $D_{\beta\alpha}$ is the matrix element for the transformation of the atomic $d$-orbitals. To see the different roles of the three $t_{2g}$ orbitals and the two $e_g$ orbitals, we use $ t_{2g}\oplus e_g \equiv \{ d_{xy},  d_{yz},  d_{xz},  d_{3z^2-1}, d_{x^2-y^2}  \}$ throughout this section. 
Note that the sequence of the five $d$-orbitals for this basis is dfferent from that used in Eq.~\eqref{sm-eq-tb-basis-d-orbitals-1}. We use the real $d$-orbitals wave functions,
\begin{align}
	\phi_{xy} = \sqrt{15} \frac{x y}{r^2}, \; 
	\phi_{yz} = \sqrt{15} \frac{y z}{r^2}, \;
	\phi_{xz} = \sqrt{15} \frac{x z}{r^2}, \; 
	\phi_{3z^2-1} = \frac{\sqrt{5}}{2} \frac{-x^2-y^2+2 z^2}{r^2}, \;
	\phi_{x^2-y^2} = \frac{\sqrt{15}}{2} \frac{x^2-y^2}{r^2}.
\end{align}
These real $d$ orbitals are orthogonal with each other and normalized by 
\begin{align}
	\frac{1}{4\pi}\int_{0}^{\pi} \sin\theta \, d\theta \int_{0}^{2\pi} d\phi\; \left( \phi_{i} [\theta,\phi] \phi_{j}[\theta,\phi] \right)= \delta_{i,j},
\end{align}
with $i,j=\{xy, yz, xz, 3z^2-1, x^2-y^2\}$ and spherical coordinator $(x,y,z)=r(\sin\theta\cos\phi,\sin\theta\sin\phi,\cos\theta)$. Here $r$ is the radial distance, $\theta$ is polar angle, and $\phi$ is the azimuthal angle.  Based on this basis ($\{ d_{xy},  d_{yz},  d_{xz},  d_{3z^2-1}, d_{x^2-y^2}  \}$), the matrix representations $D_x$, $D_y$ and $D_3$ for symmetries $S_{2x}$, $S_{2y}$ and $C_3$, respectively, are give by
\begin{align} \label{sm-eq-dx-dy-d3}
	D_{x} =  \begin{pmatrix}
		-1 & 0 & 0 & 0 & 0 \\ 
		0 & 1 & 0 & 0 & 0 \\ 
		0 & 0 & -1 & 0 & 0 \\ 
		0 & 0 & 0 & 1 & 0 \\ 
		0 & 0 & 0 & 0 & 1
	\end{pmatrix}, \;\;\;\;   
	D_y = \begin{pmatrix}
		-1 & 0 & 0 & 0 & 0 \\ 
		0 & -1 & 0 & 0 & 0 \\ 
		0 & 0 & 1 & 0 & 0 \\ 
		0 & 0 & 0 & 1 & 0 \\ 
		0 & 0 & 0 & 0 & 1
	\end{pmatrix}, \;\;\;\;   
	D_3 = \begin{pmatrix}
		0 & 1 & 0 & 0 & 0 \\ 
		0 & 0 & 1 & 0 & 0 \\ 
		1 & 0 & 0 & 0 & 0 \\ 
		0 & 0 & 0 & -\frac{1}{2} & \frac{\sqrt{3}}{2} \\ 
		0 & 0 & 0 & -\frac{\sqrt{3}}{2} & -\frac{1}{2}
	\end{pmatrix}.
\end{align}

Therefore, the $S_{2x}$ operator transforms the eigen-wave functions at the $R$ point as 
\begin{align}\label{eq-matrix-S2x}
	S_{2x} (\Psi_{1},\Psi_{2},\Psi_{3},\Psi_{4}) = (\Psi_{1} D_{x},\Psi_{2} D_{x},\Psi_{3} D_{x},\Psi_{4} D_{x} ) \times \begin{pmatrix}
			0 & 0  & e^{i\vec{\kappa}_x\cdot\mathbf{r}_1}  &  0 \\ 
			0 & 0  & 0  & e^{i\vec{\kappa}_x\cdot\mathbf{r}_2} \\ 
			e^{i\vec{\kappa}_x\cdot\mathbf{r}_3} & 0  & 0  & 0  \\ 
			0 & e^{i\vec{\kappa}_x\cdot\mathbf{r}_4}  & 0  & 0 
	\end{pmatrix} ,
\end{align}
where we use $\Psi_{l} = \{\Psi_{\alpha,l,\mathbf{k}_R}(\mathbf{r})\}$ for short, $S_{2x}\mathbf{k}_R = \mathbf{k}_R + \vec{\kappa}_x$ with $\vec{\kappa}_x=(0,-2\pi,-2\pi)$. The matrix $D_x$ is given by Eq.~\eqref{sm-eq-dx-dy-d3}. Also, please note that $e^{-i(\mathbf{k}_R + \vec{\kappa}_x)\cdot(\tfrac{1}{2},\tfrac{1}{2},0)}=1$. From Eq.~\eqref{eq-matrix-S2x}, we learn that
\begin{align}
	S_{2x} \Psi_1 = e^{i\vec{\kappa}_x\cdot\mathbf{r}_3} \Psi_{3} D_{x}, \;
	S_{2x} \Psi_2 = e^{i\vec{\kappa}_x\cdot\mathbf{r}_4} \Psi_{4} D_{x}, \;
	S_{2x} \Psi_3 = e^{i\vec{\kappa}_x\cdot\mathbf{r}_1} \Psi_{1} D_{x}, \;
	S_{2x} \Psi_4 = e^{i\vec{\kappa}_x\cdot\mathbf{r}_2} \Psi_{2} D_{x}.
\end{align}
Similarly, the matrix representation of $S_{2y}$ is 
\begin{align}\label{eq-matrix-S2y}
	S_{2y} (\Psi_{1},\Psi_{2},\Psi_{3},\Psi_{4}) = (\Psi_{1} D_{y},\Psi_{2} D_{y},\Psi_{3} D_{y},\Psi_{4} D_{y} ) \times \begin{pmatrix}
			0 & 0  & 0  & e^{i\vec{\kappa}_y\cdot\mathbf{r}_1} \\ 
			0 & 0  & e^{i\vec{\kappa}_y\cdot\mathbf{r}_2}  & 0 \\ 
			0 & e^{i\vec{\kappa}_y\cdot\mathbf{r}_3}  & 0  & 0  \\ 
			e^{i\vec{\kappa}_y\cdot\mathbf{r}_4} & 0  & 0  & 0 
	\end{pmatrix} ,
\end{align}
where $S_{2y}\mathbf{k}_R = \mathbf{k}_R + \vec{\kappa}_y$ with $\vec{\kappa}_y=(-2\pi,0,-2\pi)$. The matrix $D_y$ is given by Eq.~\eqref{sm-eq-dx-dy-d3}. Note that $e^{-i(\mathbf{k}_R + \vec{\kappa}_y)\cdot(0,\tfrac{1}{2},\tfrac{1}{2})}=1$. From Eq.~\eqref{eq-matrix-S2y}, we have
\begin{align}
	S_{2y} \Psi_1 = e^{i\vec{\kappa}_y\cdot\mathbf{r}_4} \Psi_{4} D_{y}, \;
	S_{2y} \Psi_2 = e^{i\vec{\kappa}_y\cdot\mathbf{r}_3} \Psi_{3} D_{y}, \;
	S_{2y} \Psi_3 = e^{i\vec{\kappa}_y\cdot\mathbf{r}_2} \Psi_{2} D_{y}, \;
	S_{2y} \Psi_4 = e^{i\vec{\kappa}_y\cdot\mathbf{r}_1} \Psi_{1} D_{y}.
\end{align}
Moreover, the matrix representation of $C_3$ is,
\begin{align}\label{eq-matrix-C3}
	C_3 (\Psi_{1},\Psi_{2},\Psi_{3},\Psi_{4}) = (\Psi_{1} D_{3},\Psi_{2} D_{3},\Psi_{3} D_{3},\Psi_{4} D_{3} ) \times \begin{pmatrix}
			1 & 0  & 0  & 0 \\ 
			0 & 0  & 1  & 0 \\ 
			0 & 0  & 0  & 1  \\ 
			0 & 1  & 0  & 0 
	\end{pmatrix} ,
\end{align}
where the matrix $D_3$ is given by Eq.~\eqref{sm-eq-dx-dy-d3}. And this gives rise to
\begin{align}
	C_3 \Psi_1 =  \Psi_{1} D_{3}, \;
	C_3 \Psi_2 =  \Psi_{4} D_{3}, \;
	C_3 \Psi_3 =  \Psi_{2} D_{3}, \;
	C_3 \Psi_4 =  \Psi_{3} D_{3}.
\end{align}

To remove the phases $e^{i\vec{\kappa}_x\cdot\mathbf{r}_l}$ in Eq.~\eqref{eq-matrix-S2x} and $e^{i\vec{\kappa}_y\cdot\mathbf{r}_l}$ in Eq.~\eqref{eq-matrix-S2y}, we take the following basis functions,
\begin{align}\label{eq-new-basis-psi-1234}
	\tilde{\Psi}_1 = e^{i\vec{\kappa}_x\cdot\mathbf{r}_1} \Psi_1 , \;
	\tilde{\Psi}_2 = e^{i\vec{\kappa}_x\cdot\mathbf{r}_2} \Psi_2 , \;
	\tilde{\Psi}_3 =  \Psi_3 , \;
	\tilde{\Psi}_4 =  \Psi_4 .
\end{align}
which provide the basis functions for the effective $\mathbf{k}\cdot\mathbf{p}$ Hamiltonian. Based on this new basis, the matrix representations of $S_{2x}$, $S_{2y}$ and $C_{3}$ become
\begin{align}
	S_{2x} (\tilde{\Psi}_1,\tilde{\Psi}_2,\tilde{\Psi}_3,\tilde{\Psi}_4) 
		&= (\tilde{\Psi}_1 D_{x},\tilde{\Psi}_2 D_{x},\tilde{\Psi}_3 D_{x},\tilde{\Psi}_4 D_{x} ) \times \begin{pmatrix}
			0 & 0  & 1  &  0 \\ 
			0 & 0  & 0  & 1 \\ 
			-1 & 0  & 0  & 0  \\ 
			0 & -1  & 0  & 0 
		\end{pmatrix} , \\
	S_{2y} (\tilde{\Psi}_1,\tilde{\Psi}_2,\tilde{\Psi}_3,\tilde{\Psi}_4) 
		&=(\tilde{\Psi}_1 D_{y},\tilde{\Psi}_2 D_{y},\tilde{\Psi}_3 D_{y},\tilde{\Psi}_4 D_{y} ) \times \begin{pmatrix}
			0 & 0  & 0  &  1 \\ 
			0 & 0  & -1  & 0 \\ 
			0 & 1  & 0  & 0  \\ 
			-1 & 0  & 0  & 0 
		\end{pmatrix} , \\
	C_3(\tilde{\Psi}_1,\tilde{\Psi}_2,\tilde{\Psi}_3,\tilde{\Psi}_4) 
		&=(\tilde{\Psi}_1 D_{3},\tilde{\Psi}_2 D_{3},\tilde{\Psi}_3 D_{3},\tilde{\Psi}_4 D_{3} ) \times \begin{pmatrix}
			1 & 0  & 0  &  0 \\ 
			0 & 0  & -1  & 0 \\ 
			0 & 0  & 0  & 1  \\ 
			0 & -1  & 0  & 0 
		\end{pmatrix}. 
\end{align}
One can easily check that the above matrix representations based on the atomic orbitals satisfy the commutation relations in Eq.~\eqref{eq-cxl-g1g2g3-relation}. The matrix representation for TR operator $\mathcal{T}=\mathcal{K}$ is
\begin{align}
	\mathcal{T}(\tilde{\Psi}_1,\tilde{\Psi}_2,\tilde{\Psi}_3,\tilde{\Psi}_4) 
	&=(\tilde{\Psi}_1 ,\tilde{\Psi}_2 ,\tilde{\Psi}_3 ,\tilde{\Psi}_4  ) \times (e^{i\theta_0} \mathcal{K}) \to (\tilde{\Psi}_1 ,\tilde{\Psi}_2 ,\tilde{\Psi}_3 ,\tilde{\Psi}_4  ) \times (\mathcal{K}),
\end{align}
where $\theta_0$ is a constant and can be absorbed into the definition of the wavefunctions.

Based on the basis $ \{ e^{i\vec{\kappa}_x\cdot \mathbf{r}_1} \phi_{\text{Co}_1}, e^{i\vec{\kappa}_x\cdot \mathbf{r}_2} \phi_{\text{Co}_2},  \phi_{\text{Co}_3},  \phi_{\text{Co}_4} \} \otimes \{ d_{xy},  d_{yz},  d_{xz},  d_{3z^2-1}, d_{x^2-y^2} \}$, these matrix representations for $S_{2x}$, $S_{2y}$ and $C_3$ are reducible and our next task is to find an appropriate set of basis to decompose these $20\times20$ matrices into a direct sum of several matrices. To show that, we first consider the $C_3$ rotation with the representation matrix represented as
\begin{align}
	C_3 = \begin{pmatrix}
			1 & 0  & 0  &  0 \\ 
			0 & 0  & -1  & 0 \\ 
			0 & 0  & 0  & 1  \\ 
			0 & -1  & 0  & 0 
	\end{pmatrix} \otimes D_3 \triangleq C_{3,\text{Co}}\otimes D_3,
\end{align}
Similarly, 
\begin{align}
\begin{split}
	S_{2x} &=  \begin{pmatrix}
		0 & 0  & 1  &  0 \\ 
		0 & 0  & 0  & 1 \\ 
		-1 & 0  & 0  & 0  \\ 
		0 & -1  & 0  & 0 
	\end{pmatrix} \otimes D_x \triangleq C_{2x,\text{Co}}\otimes D_x, \\
	S_{2y} &=   \begin{pmatrix}
		0 & 0  & 0  &  1 \\ 
		0 & 0  & -1  & 0 \\ 
		0 & 1  & 0  & 0  \\ 
		-1 & 0  & 0  & 0 
	\end{pmatrix} \otimes D_y \triangleq C_{2y,\text{Co}}\otimes D_y,
\end{split}
\end{align} 
where $D_3$, $D_x$ and $D_y$ are given by Eq.~\eqref{sm-eq-dx-dy-d3}.

We now use the eigen-states of $C_3$ to reduce these $20\times 20$ matrices. The eigenvalues of $C_{3,\text{Co}}$ are given by
\begin{align}
	\{ 1, 1, e^{i\omega_0} , e^{-i\omega_0} \}
\end{align}
with $\omega_0=2\pi/3$. And note that the matrix $D_3$, $D_x$ and $D_y$ are block-diagonalized as
\begin{subequations} \label{sm-eq-d3-dx-dy-t2g-eg}
\begin{align}
	D_3&=[\text{Matrix}_{t_{2g}}]\oplus[\text{Matrix}_{e_g}] = \begin{pmatrix}
		0 & 1 & 0 \\
		0 & 0 & 1 \\
		1 & 0 & 0 
	\end{pmatrix}_{t_{2g}} \oplus \begin{pmatrix}
		 -\frac{1}{2} & \frac{\sqrt{3}}{2} \\ 
		 -\frac{\sqrt{3}}{2} & -\frac{1}{2}
	\end{pmatrix}_{e_g}, \\
	D_{x} &=[\text{Matrix}_{t_{2g}}]\oplus[\text{Matrix}_{e_g}] = \begin{pmatrix}
		-1 & 0 & 0 \\
		0  & 1 & 0 \\
		0  & 0 & -1 
	\end{pmatrix}_{t_{2g}} \oplus \begin{pmatrix}
		1 & 0 \\ 
		0 & 1
	\end{pmatrix}_{e_g}, \\
	D_{y} &=[\text{Matrix}_{t_{2g}}]\oplus[\text{Matrix}_{e_g}] = \begin{pmatrix}
		-1 & 0  & 0 \\
		0  & -1 & 0 \\
		0  & 0  & 1 
	\end{pmatrix}_{t_{2g}} \oplus \begin{pmatrix}
		1 & 0 \\ 
		0 & 1
	\end{pmatrix}_{e_g}.
\end{align}
\end{subequations}
Then, the eigenvalues of $D_3$ are classified into two types,
\begin{align}
	t_{2g}-\text{type: } \{1,e^{i\omega_0} , e^{-i\omega_0}\}   \text{ and } e_g-\text{type: } \{e^{i\omega_0} , e^{-i\omega_0}\}.
\end{align}
As a result, the eigenvalues of $C_3$ are given by 
\begin{align}
\text{eig}[C_3] = \text{eig}[C_{3,\text{Co}}]\times \text{eig}[D_3]  = \{ 1, 1, e^{i\omega_0} , e^{-i\omega_0} \} \times \left\lbrack \{1,e^{i\omega_0} , e^{-i\omega_0}\}  \oplus  \{e^{i\omega_0} , e^{-i\omega_0}\} \right\rbrack.
\end{align} 
Therefore, there are 20 numbers of eigenvalues in total for the matrix $C_3$ with three different values, i.e., $1$, $e^{i\omega_0} $ and $e^{-i\omega_0}$. There are {\bf 6} states with the $C_3$-eigenvalue $1$, {\bf 7} states with the $C_3$-eigenvalue $e^{i\omega_0}$, and {\bf 7} states with the $C_3$-eigenvalue $e^{-i\omega_0}$. These eigen-states of $C_3$ can be classified by time-reversal symmetry (i.e.,~complex conjugate), and there are two scenarios for the Irreps, which are labelled by $A$-type-Irrep and $B$-type-Irrep, 
\begin{subequations} \label{sm-eq-A-B-irrep-C3-eig}
\begin{align}
	&A-\text{type: } \{ 1, 1, e^{i\omega_0} , e^{-i\omega_0} \}, \; \text{with the trace of }C_3 = 1. \\
	&B-\text{type: } \{ e^{i\omega_0} , e^{-i\omega_0} ,e^{i\omega_0} , e^{-i\omega_0} \}, \; \text{with the trace of }C_3= -2.
\end{align}
\end{subequations}
As mentioned in Sec.~\ref{Appendix-A-1}, there are only two 4D single-valued Irreps at the $R$ point based on the Bilbao~\cite{Bradlyn_nature_2017TQC,Xu_nature_2020,Elcoro_nc_2021}: $R_1R_3$ (trace of $C_3$ is 1) and $R_2R_2$ (trace of $C_3$ is -2) of the little group at $R$ point. Thus, the $A-$type Irrep is $R_1R_3$ and the $B-$type Irrep is $R_2R_2$. On the other hand, we have
\begin{align}
	\text{Tr}[C_3] = \text{Tr}[C_{3,\text{Co}}]\times \text{Tr}[D_3] =-1 \triangleq 1 + 1 +  1 - 2 - 2,
\end{align}
which indicates that there are 3 sets of basis for the A-type Irrep and 2 sets of basis for the B-type Irrep. 
Thus, the trace of $C_3$ operators suggest the full representation matrix should be reduced into
\begin{subequations} \label{sm-eq-irrep-total}
\begin{align}
	&\text{The } e_g-\text{type: }  \{ 1, 1, e^{i\omega_0} , e^{-i\omega_0} \} \otimes \{e^{i\omega_0} , e^{-i\omega_0} \} = A_1 \text{ Irrep} \oplus B_1 \text{ Irrep}. \\
	&\text{The } t_{2g}-\text{type: } \{ 1, 1, e^{i\omega_0} , e^{-i\omega_0} \} \otimes  \{1,e^{i\omega_0} , e^{-i\omega_0}\}   =  A_2 \text{ Irrep} \oplus A_3 \text{ Irrep} \oplus  B_2 \text{ Irrep},  
\end{align}
\end{subequations}
where we label $A_{1,2,3}$ for the three A-type Irrep and $B_{1,2}$ for the two B-type Irrep. As $t_{2g}$ and $e_g$ orbitals both contain $A$ and $B$-type Irreps, 
the eigen-states of the Hamiltonian should be their mixture, as already seen from our numerical results in Tab.~\ref{sm-fig3} in Sec.~\ref{Appendix-B}. Below we will provide a detailed process to construct the explicit forms of these Irreps for $t_{2g}$ and $e_g$ orbitals, separately.

\subsubsection{The four eigen-states of $C_{3,\text{Co}}$}

To construct the Irreps in the following two sub-sections (Sec.~\ref{sm-sub-sub-sec-eg} and Sec.~\ref{sm-sub-sub-sec-t2g}), we first consider the eigen-states of the matrix $C_{3,\text{Co}}$ and find that 
\begin{subequations} \label{sm-eq-basis-wf-C3co}
\begin{align}
	\text{Eigenvalue}-1: \;\; \phi_{a0} &= \tilde{\Psi}_1 = (1,0,0,0) ^T, \\
	\text{Eigenvalue}-1: \;\; \phi_{b0} &= \frac{1}{\sqrt{3}} \left( -\tilde{\Psi}_2  + \tilde{\Psi}_3 + \tilde{\Psi}_4  \right) = \frac{1}{\sqrt{3}} (0,-1,1,1)^T, \\
	\text{Eigenvalue}-e^{i\omega_0}: \;\;  \phi_{b+} &=\frac{1}{\sqrt{3}}  \left( -e^{i\omega_0}\tilde{\Psi}_2  + e^{-i\omega_0} \tilde{\Psi}_3 + \tilde{\Psi}_4  \right) =\frac{1}{\sqrt{3}}\left( 0,  -e^{i\omega_0}, e^{-i\omega_0} ,1  \right)^T, \\
	\text{Eigenvalue}-e^{-i\omega_0}: \;\;  \phi_{b-} &= \frac{1}{\sqrt{3}}  \left( -e^{-i\omega_0}\tilde{\Psi}_2  + e^{i\omega_0} \tilde{\Psi}_3 + \tilde{\Psi}_4  \right) =\frac{1}{\sqrt{3}}\left( 0,  -e^{-i\omega_0}, e^{i\omega_0} ,1  \right)^T = (\phi_{b+})^\ast, 
\end{align}
\end{subequations}
where the left side before the colon is the eigenvalue of the matrix $C_{3,\text{Co}}$ and  the right side corresponds to the eigen-state. Besides, it means $C_{3,\text{Co}} \vert \phi_{a0}\rangle = \vert \phi_{a0}\rangle$, $C_{3,\text{Co}} \vert \phi_{b0}\rangle = \vert \phi_{b0}\rangle$, $C_{3,\text{Co}} \vert \phi_{b+}\rangle = e^{i\omega_0} \vert \phi_{b+}\rangle$ and $C_{3,\text{Co}} \vert \phi_{b-}\rangle = e^{-i\omega_0} \vert \phi_{b-}\rangle$. Therefore, a unitary transformation ($\mathcal{U}_{3,\text{Co}}$) to diagonalize $C_{3,\text{Co}}$ can be defined as 
\begin{align} \label{sm-eq-c3-diagonal-wf}
\begin{split}
	&\mathcal{U}_{3,\text{Co}}  = \frac{1}{\sqrt{3}}\begin{pmatrix}
		\sqrt{3} & 0 & 0 & 0 \\
		0 & -1 & -e^{i\omega_0} & -e^{-i\omega_0} \\
		0 & 1 & e^{-i\omega_0} & e^{i\omega_0} \\
		0 & 1 & 1 & 1 
		\end{pmatrix}, \\
\Rightarrow \;\;  &(\mathcal{U}_{3,\text{Co}})^{\dagger}  [ C_{3,\text{Co}} ] \mathcal{U}_{3,\text{Co}} = \begin{pmatrix}  
		1 & 0 & 0  & 0 \\
		0 & 1 & 0  & 0 \\
		0 & 0 & e^{i\omega_0}  & 0 \\
		0 & 0 & 0  & e^{-i\omega_0} 
	\end{pmatrix}.
\end{split}
\end{align}
Likewise, the matrix $C_{2x,\text{Co}}$ based on the $C_{3,\text{Co}}$-diagonal-basis $\{ \phi_{a0}, \phi_{b0}, \phi_{b+}, \phi_{b-} \}$ in Eq.~\eqref{sm-eq-basis-wf-C3co} is given by
\begin{align}
\begin{split}
	C_{2x,\text{Co}} &\triangleq (\mathcal{U}_{3,\text{Co}})^{\dagger}  [ C_{2x,\text{Co}} ] \mathcal{U}_{3,\text{Co}}
	= (\mathcal{U}_{3,\text{Co}})^{\dagger}  \begin{pmatrix}
		0 & 0  & 1  &  0 \\ 
		0 & 0  & 0  & 1 \\ 
		-1 & 0  & 0  & 0  \\ 
		0 & -1  & 0  & 0 
	\end{pmatrix}\mathcal{U}_{3,\text{Co}} \\
	&= \frac{1}{\sqrt{3}} \left( \begin{array}{cccc}
		0 & 1 &  e^{-i\omega_0} & e^{i\omega_0} \\
		-1 & 0 & -i e^{-i\omega_0} & i e^{i\omega_0} \\
		-e^{i\omega_0} & -i e^{i\omega_0} & i & 0 \\
		-e^{-i\omega_0} & i e^{-i\omega_0} & 0 & -i \\
	\end{array}	\right).
\end{split}
\end{align}

Among these four states in Eq.~\eqref{sm-eq-basis-wf-C3co}, the two eigen-staets, $\phi_{a0}$ and $\phi_{b0}$, have the same eigenvalue of $C_{3,\text{Co}}$, thus are degenerate. To make each state is related to another by time-reversal symmetry (i.e.~complex conjugate), we need to make a linear combination of $\phi_{a0}$ and $\phi_{b0}$ by defining new states as
\begin{align}
\begin{split}
	\phi_{a0}' & = a_1 \phi_{a0} + b_1 \phi_{b0}, \\
	\phi_{b0}' & = ( \phi_{a0}' )^\ast,
\end{split}
\end{align}
where the choice of the two coefficients $a_1$ and $b_1$ are not uniquely determined yet. The normalization and orthogonalization result in $\vert a_1\vert^2 + \vert b_1\vert^2 = 1$ and $a_1^2+b_1^2=0$, thus we have $a_1=\pm ib_1$ and $\vert a_1\vert^2 =\frac{1}{2}$. Within this new basis ($\{ \phi_{a0}', \phi_{b0}', \phi_{b+}, \phi_{b-} \}$), even though the matrix $C_{3,\text{Co}}$ is unchanged, the transformation matrix $\mathcal{U}_{3,\text{Co}}$ becomes 
\begin{align}
	\mathcal{U}_{3,\text{Co}}   \triangleq   \frac{1}{\sqrt{3}}\begin{pmatrix}
		\sqrt{3}a_1 & \sqrt{3}a_1^\ast & 0 & 0 \\
		-b_1 & -b_1^\ast & -e^{i\omega_0} & -e^{-i\omega_0} \\
		b_1  & b_1^\ast & e^{-i\omega_0} & e^{i\omega_0} \\
		b_1  & b_1^\ast & 1 & 1 
	\end{pmatrix}.
\end{align}
To make the matrix $C_{2x,\text{Co}} \to (\mathcal{U}_{3,\text{Co}} )^{-1} [C_{2x,\text{Co}}] \mathcal{U}_{3,\text{Co}} $ become block-diagonalized, a specific choice of $a_1$ and $b_1=\pm i a_1$ can be mathematically solved by assuming $a_1=\frac{1}{\sqrt{2}} e^{i\theta_a}$ and $b_1=-ia_1$. Here $\theta_a$ can be arbitrary value. Without loss of generality, we choose $\theta_a=-2\pi/3$, so that we have $a_1= \frac{e^{-i \omega _0}}{\sqrt{2}} $  and $b_1=- i\frac{e^{-i \omega _0}}{\sqrt{2}}$. Then, we define the new basis 
\begin{subequations} \label{sm-eq-c3co-new-basis}
\begin{align}
	\phi_{a0}' &= \frac{e^{-i \omega _0}}{\sqrt{2}} \phi_{a0} - i\frac{e^{-i \omega _0}}{\sqrt{2}} \phi_{b0} = \frac{1}{\sqrt{6}} e^{-i \omega_0} \left(  \sqrt{3}, i, -i, -i \right)^T, \\
	\phi_{b0}' &= \frac{e^{i \omega _0}}{\sqrt{2}} \phi_{a0} + i\frac{e^{i \omega _0}}{\sqrt{2}} \phi_{b0} = \frac{1}{\sqrt{6}} e^{i \omega_0} \left(  \sqrt{3}, -i, i, i \right)^T  = (\phi_{a0}')^\ast,  \\
	\phi_{b+} &=\frac{1}{\sqrt{3}}\left( 0,  -e^{i\omega_0}, e^{-i\omega_0} ,1  \right)^T, \\
	\phi_{b-}  &=\frac{1}{\sqrt{3}}\left( 0,  -e^{-i\omega_0}, e^{i\omega_0} ,1  \right)^T = (\phi_{b+})^\ast, 
\end{align}
\end{subequations}
Based on this new basis, the matrix $C_{2x,\text{Co}}$ becomes
\begin{align} \label{sm-eq-C2x-four-Co-basis}
	C_{2x,\text{Co}} = \left(\begin{array}{cccc}
		-\frac{i}{\sqrt{3}} & 0 & \sqrt{\frac{2}{3}} & 0 \\
		0 & \frac{i}{\sqrt{3}} & 0 & \sqrt{\frac{2}{3}} \\
		-\sqrt{\frac{2}{3}} & 0 & \frac{i}{\sqrt{3}} & 0 \\
		0 & -\sqrt{\frac{2}{3}} & 0 & -\frac{i}{\sqrt{3}} \\
	\end{array}	\right).
\end{align}
Moreover, the time-reversal symmetry, complex conjugate, gives rise to 
\begin{align}
	\mathcal{T} \{ \phi_{a0}', \phi_{b0}', \phi_{b+}, \phi_{b-} \} = \{ \phi_{b0}', \phi_{a0}', \phi_{b-}, \phi_{b+} \}.
\end{align}

\subsubsection{Construction of the Irreps for the $e_g$-orbitals}
\label{sm-sub-sub-sec-eg}

We now construct the Irreps for the little group $\{C_3, S_{2x}, S_{2y},\mathcal{T} \}$ for the whole Co-$d$-orbital basis,
\begin{align}
	\{ \phi_{a0}', \phi_{b0}', \phi_{b+}, \phi_{b-} \} \otimes \{ d_{xy},  d_{yz},  d_{xz},  d_{3z^2-1}, d_{x^2-y^2} \}.
\end{align}
We first consider the $20\times20$ matrix $C_3=C_{3,\text{Co}}\otimes D_3$ (e.g.,~$C_{3,\text{Co}}$ is given by Eq.~\eqref{sm-eq-c3-diagonal-wf} and $D_3$ is given by Eq.~\eqref{sm-eq-d3-dx-dy-t2g-eg}). Recall that the five eigenvalues of $D_3$ are given by the $t_{2g}-\text{type: } \{1,e^{i\omega_0} , e^{-i\omega_0}\}$, and the $e_g-\text{type: } \{e^{i\omega_0} , e^{-i\omega_0}\}$. In this sub-section, we focus on the $e_g$-orbital-subspace. As discussed in Eq.~\eqref{sm-eq-irrep-total}, we have the eigenvalues of $C_3$ as $\{ 1, 1, e^{i\omega_0} , e^{-i\omega_0} \} \otimes \{e^{i\omega_0} , e^{-i\omega_0} \} = A_1 \text{ Irrep} \oplus B_1 \text{ Irrep}$. Then, we need to construct the wavefunctions for these two Irreps. Note that the eigen-states of $C_{3,\text{Co}}$ are given by Eq.~\eqref{sm-eq-c3co-new-basis}. Here, the two corresponding eigen-states of $D_3$ with only $e_g$ orbitals are given by
\begin{subequations} \label{sm-eq-basis-wf-D3}
\begin{align}
	\text{Eigenvalue}-e^{i\omega_0}: \;\; d_{-2}  &= \frac{1}{\sqrt{2}}  \left( -id_{3z^2-1} + d_{x^2-y^2} \right) = \frac{1}{\sqrt{2}} \left( 0,0,0,-i,1 \right)^T, \\
	\text{Eigenvalue}-e^{-i\omega_0}: \;\; d_{+2} &= \frac{1}{\sqrt{2}}  \left( id_{3z^2-1}+ d_{x^2-y^2} \right) = \frac{1}{\sqrt{2}} \left( 0,0,0,i,1 \right)^T = (d_{-2})^\ast.
\end{align}
\end{subequations}
where the left side before the colon is the eigenvalue of the matrix $D_3$ and  the right side corresponds to the eigen-state. Besides, it means $D_3 \vert d_{-2}\rangle = e^{i\omega_0} \vert d_{-2}\rangle$ and $D_3 \vert d_{+2}\rangle = e^{-i\omega_0} \vert d_{+2}\rangle$. Then, the eigen-state of $C_3$ is just given by the direct product of eigen-state of $C_{3,\text{Co}}$ (i.e.,~$\{ \phi_{a0}', \phi_{b0}', \phi_{b+}, \phi_{b-} \}$) in Eq.~\eqref{sm-eq-c3co-new-basis} and that of $D_3$ in Eq.~\eqref{sm-eq-basis-wf-D3}. This gives rise to the following eight eigen-states of $C_3$ involving only the $e_g$-orbitals, 
\begin{subequations} \label{sm-eq-8-state-C3-eg}
\begin{align}
	\text{Eigenvalue}-e^{i\omega_0} &:    \;\; \phi_{a0}'\otimes d_{-2} , \;\; \phi_{b0}'\otimes d_{-2},\;\; \phi_{b-}\otimes d_{+2}, \\
	\text{Eigenvalue}-e^{-i\omega_0} &:  \;\; \phi_{a0}'\otimes d_{+2} , \;\; \phi_{b0}'\otimes d_{+2},\;\; \phi_{b+}\otimes d_{-2}, \\
	\text{Eigenvalue}-1 &:  \;\;  \phi_{b+}\otimes d_{+2},\;\; \phi_{b-}\otimes d_{-2}, 
\end{align}
\end{subequations}
where the left side before the colon is the eigenvalue of the matrix $C_3$ and  the right side corresponds to the eigen-state. Notice that $e^{i2\omega_0}=e^{-i\omega_0}$ has been used for $\omega_0=2\pi/3$. Besides, it means $C_3 \vert \phi_{a0}'\rangle \otimes \vert d_{-2}\rangle = e^{i\omega_0} \vert \phi_{a0}'\rangle \otimes \vert d_{-2}\rangle $, $C_3 \vert \phi_{b0}'\rangle \otimes \vert d_{-2}\rangle = e^{i\omega_0} \vert \phi_{b0}'\rangle \otimes \vert d_{-2}\rangle $, $C_3 \vert \phi_{b-}\rangle \otimes \vert d_{+2}\rangle = e^{i\omega_0} \vert \phi_{b-}\rangle \otimes \vert d_{+2}\rangle $; and $C_3 \vert \phi_{a0}'\rangle \otimes \vert d_{+2}\rangle = e^{-i\omega_0} \vert \phi_{a0}'\rangle \otimes \vert d_{+2}\rangle$, $C_3 \vert \phi_{b0}'\rangle \otimes \vert d_{+2}\rangle = e^{-i\omega_0} \vert \phi_{b0}'\rangle \otimes \vert d_{+2}\rangle$, $C_3 \vert \phi_{b+}\rangle \otimes \vert d_{-2}\rangle = e^{-i\omega_0} \vert \phi_{b+}\rangle \otimes \vert d_{-2}\rangle$; and $C_3 \vert \phi_{b+}\rangle \otimes \vert d_{+2}\rangle = \vert \phi_{b+}\rangle \otimes \vert d_{+2}\rangle$, $C_3 \vert \phi_{b-}\rangle \otimes \vert d_{-2}\rangle = \vert \phi_{b-}\rangle \otimes \vert d_{-2}\rangle$.

We next classify these eight eigen-states of the matrix $C_3$ in Eq.~\eqref{sm-eq-8-state-C3-eg} into two different Irreps, i.e., $A_1$-Irrep and $B_1$-Irrep. Once we find those four states for the $A_1$-Irrep, the remaining four states are for the $B_1$-Irrep. Here, we discuss how to get the $A_1$-Irrep in details. According to Eq.~\eqref{sm-eq-A-B-irrep-C3-eig}, we know these two states, $\phi_{b+}\otimes d_{+2}$ and $\phi_{b-}\otimes d_{-2}$, must belong to the $A_1$-Irrep, because they both have $C_3$-eigenvalues-$1$. Then, we use the matrix $S_{2x}=C_{2x,\text{Co}}\otimes D_x$ (e.g.,~$C_{2x,\text{Co}}$ is given by Eq.~\eqref{sm-eq-C2x-four-Co-basis} and $D_x$ is given by Eq.~\eqref{sm-eq-d3-dx-dy-t2g-eg}) to generate the other two distinct states of the $A_1$-Irrep by doing 
\begin{align}
	S_{2x}  \{ \vert \phi_{b+}\rangle \otimes \vert d_{+2}\rangle, \;\;  \vert \phi_{b-}\rangle \otimes \vert d_{-2}\rangle \}.
\end{align}
Notice that $D_x \vert d_{\pm 2}\rangle = \vert d_{\pm 2}\rangle$. Then, we have 
\begin{subequations}
\begin{align}
	&S_{2x} \vert \phi_{b+}\rangle \otimes \vert d_{+2}\rangle \to \left( C_{2x,\text{Co}} \vert \phi_{b+}\rangle \right) \otimes \vert d_{+2}\rangle = ( -\sqrt{\frac{2}{3}} \vert \phi_{a0}'\rangle + \frac{i}{\sqrt{3}} \vert \phi_{b+}\rangle ) \otimes \vert d_{+2}\rangle, \\
	&S_{2x} \vert \phi_{b-}\rangle \otimes \vert d_{-2}\rangle \to \left( C_{2x,\text{Co}} \vert \phi_{b-}\rangle \right) \otimes \vert d_{-2}\rangle = ( -\sqrt{\frac{2}{3}} \vert \phi_{b0}'\rangle - \frac{i}{\sqrt{3}} \vert \phi_{b-}\rangle ) \otimes \vert d_{-2}\rangle.
\end{align}
\end{subequations}
This is due to the block diagonalized matrix of $C_{2x,\text{Co}}$. Therefore, the $A_1$ Irrep is given by
\begin{align}
	\left\{ \phi_{a0}'\otimes d_{+2} ,\;\;\;\;  \phi_{b+}\otimes d_{+2},\;\;\;\;  
	\phi_{b0}'\otimes d_{-2},\;\;\;\;         \phi_{b-}\otimes d_{-2}    \right\}.
\end{align}
We re-organize the above wavefunction-basis to be $\left\{ \phi_{b-}\otimes d_{-2} , \phi_{b0}'\otimes d_{-2},  \phi_{b+}\otimes d_{+2}, \phi_{a0}'\otimes d_{+2} \right\}$, based on which the irreducible matrix representations for the symmetry operators are given by
\begin{subequations}
\begin{align} \label{sm-eq-mat-rep-A1}
	C_3 &= \left(
	\begin{array}{cccc}
		1 & 0 & 0 & 0 \\
		0 & \exp \left(i \omega _0\right) & 0 & 0 \\
		0 & 0 & 1 & 0 \\
		0 & 0 & 0 & \exp \left(-i \omega _0\right) \\
	\end{array}
	\right),  \;\;\;\;\;\;\;\;
	S_{2x} = \left(
	\begin{array}{cccc}
		-\tfrac{i}{\sqrt{3}} & -\sqrt{\tfrac{2}{3}} & 0 & 0 \\
		\sqrt{\tfrac{2}{3}} & \tfrac{i}{\sqrt{3}} & 0 & 0 \\
		0 & 0 & \tfrac{i}{\sqrt{3}} & -\sqrt{\tfrac{2}{3}} \\
		0 & 0 & \sqrt{\tfrac{2}{3}} & -\tfrac{i}{\sqrt{3}} \\
	\end{array}
	\right),  \\
	S_{2y} &= \left(
	\begin{array}{cccc}
		-\frac{i}{\sqrt{3}} & \frac{1-i \sqrt{3}}{\sqrt{6}} & 0 & 0 \\
		\frac{-1-i \sqrt{3}}{\sqrt{6}} & \frac{i}{\sqrt{3}} & 0 & 0 \\
		0 & 0 & \frac{i}{\sqrt{3}} & \frac{1+i \sqrt{3}}{\sqrt{6}} \\
		0 & 0 & \frac{-1+i \sqrt{3}}{\sqrt{6}} & -\frac{i}{\sqrt{3}} \\
	\end{array} \right), \;\;\;\;\;\;\;\;
	\mathcal{T} =  \left(
	\begin{array}{cccc}
		0 & 0 & 1 & 0 \\
		0 & 0 & 0 & 1 \\
		1 & 0 & 0 & 0 \\
		0 & 1 & 0 & 0 \\
	\end{array}
	\right)\mathcal{K}.
\end{align}
\end{subequations}
This confirms it forms the 4D Irrep ($A_1$-Irrep or $R_1R_3$) of the little group $\{C_3,S_{2x},S_{2y}, \mathcal{T}\}$ at the $R$ point. In addition, one can similarly show the remaining four states in Eq.~\eqref{sm-eq-8-state-C3-eg} give rise to the $B_1$ Irrep,
\begin{align}
	\left\{ \phi_{a0}'\otimes d_{-2} , \;\;\;\;   \phi_{b-}\otimes d_{+2},\;\;\;\;  
	\phi_{b0}'\otimes d_{+2},\;\;\;\; \phi_{b+}\otimes d_{-2}  \right\}.
\end{align}

\subsubsection{Construction of the Irreps for the $t_{2g}$-orbitals}
\label{sm-sub-sub-sec-t2g}

We continue to construct the Irreps for the little group $\{C_3, S_{2x}, S_{2y},\mathcal{T} \}$ for the whole Co-$d$-orbital basis,
\begin{align}
	\{ \phi_{a0}', \phi_{b0}', \phi_{b+}, \phi_{b-} \} \otimes \{ d_{xy},  d_{yz},  d_{xz},  d_{3z^2-1}, d_{x^2-y^2} \}.
\end{align}
In addition to the construction of the Irreps for the $e_g$-orbitals in Sec.~\ref{sm-sub-sub-sec-eg}, in this sub-section, we focus on the construction of the Irreps for the $t_{2g}$-orbital-subspace. Again, we use the eigen-states of the $20\times20$ matrix $C_3=C_{3,\text{Co}}\otimes D_3$ (e.g.,~$C_{3,\text{Co}}$ is given by Eq.~\eqref{sm-eq-c3-diagonal-wf} and $D_3$ is given by Eq.~\eqref{sm-eq-d3-dx-dy-t2g-eg}) to construct the Irreps. As discussed in Eq.~\eqref{sm-eq-irrep-total}, we have the eigenvalues of $C_3$ as $\{ 1, 1, e^{i\omega_0} , e^{-i\omega_0} \} \otimes  \{1,e^{i\omega_0} , e^{-i\omega_0}\}   =  A_2 \text{ Irrep} \oplus A_3 \text{ Irrep} \oplus  B_2 \text{ Irrep}$. Then we need to construct the wavefunctions for these three Irreps. Note that the eigen-states of $C_{3,\text{Co}}$ are given by Eq.~\eqref{sm-eq-c3co-new-basis}. Here, the eigen-states of $D_3$ within only the $t_{2g}$-orbital-subspace are given by
\begin{align} \label{sm-eq-basis-wf-D3-t2g}
	\text{Eigenvalue}-1&: \;\; d_0 = \frac{1}{\sqrt{3}}\left( d_{xy} + d_{yz} + d_{xz} \right) = \frac{1}{\sqrt{3}}(1,1,1,0,0)^T, \\
	\text{Eigenvalue}-e^{i\omega_0} &: \;\; d_{+1} =  \frac{1}{\sqrt{3}}\left( e^{i\omega_0} d_{xy} + e^{-i\omega_0} d_{yz} + d_{xz} \right) = \frac{1}{\sqrt{3}}(e^{i\omega_0},e^{-i\omega_0},1,0,0)^T, \\
	\text{Eigenvalue}-e^{-i\omega_0} &: \;\; d_{-1} =  \frac{1}{\sqrt{3}}\left( e^{-i\omega_0} d_{xy} + e^{i\omega_0}d_{yz} + d_{xz} \right) = \frac{1}{\sqrt{3}}(e^{-i\omega_0},e^{i\omega_0},1,0,0)^T = (d_{+1})^\ast. 
\end{align}
where the left side before the colon is the eigenvalue of the matrix $D_3$ and  the right side corresponds to the eigen-state. Besides, it means $D_3\vert d_0\rangle =\vert d_0\rangle$, $D_3 \vert d_{+1}\rangle = e^{i\omega_0} \vert d_{+1}\rangle$ and $D_3 \vert d_{-1}\rangle = e^{-i\omega_0} \vert d_{-1}\rangle$. Then, the eigen-state of $C_3$ is just given by the direct product of eigen-state of $C_{3,\text{Co}}$ (i.e.,~$\{ \phi_{a0}', \phi_{b0}', \phi_{b+}, \phi_{b-} \}$) in Eq.~\eqref{sm-eq-c3co-new-basis} and that of $D_3$ in Eq.~\eqref{sm-eq-basis-wf-D3-t2g}. This gives rise to the following twelve eigen-states of $C_3$ involving only the $t_{2g}$-orbitals, 
\begin{subequations} \label{sm-eq-c3-12-states-t2g}
\begin{align}
	\text{Eigenvalue}-1&: \;\; \phi_{a0}'\otimes d_{0} , \;\;\; \phi_{b0}'\otimes d_{0} ,  \;\;\; \phi_{b+}\otimes d_{-1},\; \phi_{b-}\otimes d_{+1}, \\
	\text{Eigenvalue}-e^{i\omega_0} &: \;\; \phi_{a0}'\otimes d_{+1},\;  \phi_{b0}'\otimes d_{+1} ,  \;\;\; \phi_{b+}\otimes d_{0},\; \phi_{b-}\otimes d_{-1}, \\
	\text{Eigenvalue}-e^{-i\omega_0} &: \;\; \phi_{a0}'\otimes d_{-1},\; \phi_{b0}'\otimes d_{-1} ,  \; \phi_{b+}\otimes d_{+1},\; \phi_{b-}\otimes d_{0}, 
\end{align}
\end{subequations}
where the left side before the colon is the eigenvalue of the matrix $C_3$ and  the right side corresponds to the eigen-state. Notice that $e^{i2\omega_0}=e^{-i\omega_0}$ has been used for $\omega_0=2\pi/3$. Besides, it means $C_3 \vert\phi_{a0}'\rangle \otimes \vert d_{0}\rangle =  \vert\phi_{a0}'\rangle \otimes \vert d_{0}\rangle$, $C_3 \vert \phi_{b0}' \rangle \otimes \vert d_{0}\rangle =  \vert \phi_{b0}' \rangle \otimes \vert d_{0}\rangle$, $C_3 \vert \phi_{b+} \rangle \otimes \vert d_{-1}\rangle =  \vert \phi_{b+} \rangle \otimes \vert d_{-1}\rangle$, $C_3 \vert \phi_{b-} \rangle \otimes \vert d_{+1}\rangle =  \vert \phi_{b-} \rangle \otimes \vert d_{+1}\rangle$; and $C_3 \vert\phi_{a0}'\rangle \otimes \vert d_{+1}\rangle = e^{i\omega_0} \vert\phi_{a0}'\rangle \otimes \vert d_{+1}\rangle$, $C_3 \vert\phi_{b0}'\rangle \otimes \vert d_{+1}\rangle = e^{i\omega_0} \vert\phi_{b0}'\rangle \otimes \vert d_{+1}\rangle$, $C_3 \vert\phi_{b+}\rangle \otimes \vert d_{0}\rangle = e^{i\omega_0} \vert\phi_{b+}\rangle \otimes \vert d_{0}\rangle$, $C_3 \vert\phi_{b-}\rangle \otimes \vert d_{-1}\rangle = e^{i\omega_0} \vert\phi_{b-}\rangle \otimes \vert d_{-1}\rangle$; and $C_3 \vert\phi_{a0}'\rangle \otimes \vert d_{-1}\rangle = e^{-i\omega_0} \vert\phi_{a0}'\rangle \otimes \vert d_{-1}\rangle$, $C_3 \vert\phi_{b0}'\rangle \otimes \vert d_{-1}\rangle = e^{-i\omega_0} \vert\phi_{b0}'\rangle \otimes \vert d_{-1}\rangle$, $C_3 \vert\phi_{b-}\rangle \otimes \vert d_{0}\rangle = e^{-i\omega_0} \vert\phi_{b-}\rangle \otimes \vert d_{0}\rangle$, $C_3 \vert\phi_{b+}\rangle \otimes \vert d_{+1}\rangle = e^{-i\omega_0} \vert\phi_{b+}\rangle \otimes \vert d_{+1}\rangle$.

We next classify these twelve eigen-states of the matrix $C_3$ in Eq.~\eqref{sm-eq-c3-12-states-t2g} into three different Irreps, i.e., $A_2$-Irrep, $A_3$-Irrep and $B_2$-Irrep. Then, we use the matrix $S_{2x}=C_{2x,\text{Co}}\otimes D_x$ (e.g.,~$C_{2x,\text{Co}}$ is given by Eq.~\eqref{sm-eq-C2x-four-Co-basis} and $D_x$ is given by Eq.~\eqref{sm-eq-d3-dx-dy-t2g-eg}) to classify the twelve eigen-states into two classes. For example,
\begin{subequations}
\begin{align}
	S_{2x} \vert \phi_{a0}' \rangle \otimes \vert d_0 \rangle &= \left( C_{2x,\text{Co}} \vert \phi_{a0}'\rangle \right) \otimes \left(  D_x \vert d_0 \rangle \right) ,  \nonumber \\
	&= \left( -\frac{i}{\sqrt{3}}\vert \phi_{a0}'\rangle  - \sqrt{\frac{2}{3}} \vert \phi_{b+} \rangle \right) \otimes \left( -\frac{1}{3} \vert d_0 \rangle + \frac{2}{3}e^{i\omega_0} \vert d_{+1}\rangle + \frac{2}{3} e^{-i\omega_0} \vert d_{-1}\rangle \right), \\
	S_{2x} \vert \phi_{b+} \rangle \otimes \vert d_{-1} \rangle &= \left( C_{2x,\text{Co}} \vert \phi_{b+} \rangle \right) \otimes \left(  D_x \vert d_{-1} \rangle \right) , \nonumber \\
	&= \left( \sqrt{\frac{2}3{}} \vert \phi_{a0}'\rangle  +i \frac{1}{\sqrt{3}} \vert \phi_{b+} \rangle \right) \otimes \left(  \frac{2}{3}e^{i\omega_0} \vert d_0 \rangle  + \frac{2}{3} e^{-i\omega_0} \vert d_{+1}\rangle  -\frac{1}{3} \vert d_{-1}\rangle \right),
\end{align}
\end{subequations}
which indicates that these six eigen-states, $ \phi_{a0}' \otimes \{ d_0, d_{+1}, d_{-1}  \} $  and $ \phi_{b+} \otimes \{ d_0, d_{+1}, d_{-1}  \}$, are the in the same class. Furthermore, we also notice that time-reversal symmetry (i.e.,~complex conjugate) gives rise to
\begin{subequations} \label{sm-eq-wf-trs-relation}
\begin{align} 
	&\{ \phi_{a0}'\otimes d_{0} ,  \phi_{a0}'\otimes d_{+1} ,  \phi_{a0}'\otimes d_{-1} , \phi_{b+} \otimes d_{0} , \phi_{b+} \otimes d_{+1} , \phi_{b+} \otimes d_{-1}  \}  \\ 
		\stackrel{\text{TR}}{\longmapsto}  \;\;
	&\{ \phi_{b0}'\otimes d_{0} ,  \phi_{b0}'\otimes d_{+1} ,  \phi_{b0}'\otimes d_{-1} , \phi_{b-} \otimes d_{0} , \phi_{b-} \otimes d_{+1} , \phi_{b-} \otimes d_{-1}  \} .
\end{align}
\end{subequations}
This indicates that we only need to find three different 2D complex Irreps (e.g.~$R_1$, $R_2$ and $R_3$) for the six eigen-states of $C_3$ in the first line in Eq.~\eqref{sm-eq-wf-trs-relation}, i.e., 
\begin{align} \label{sm-eq-2d-irrep-t2g}
	\{ \phi_{a0}'\otimes d_{0} ,  \phi_{a0}'\otimes d_{+1} ,  \phi_{a0}'\otimes d_{-1} , 
		\phi_{b+} \otimes d_{0} , \phi_{b+} \otimes d_{+1} , \phi_{b+} \otimes d_{-1} \}  
	\;\Rightarrow\; R_1\oplus R_2 \oplus R_3.
\end{align}
Moreover, by using time-reversal symmetry, the remaining six eigen-states (the second line in Eq.~\eqref{sm-eq-wf-trs-relation}) give rise to
\begin{align}
	\{ \phi_{b0}'\otimes d_{0} ,  \phi_{b0}'\otimes d_{+1} ,  \phi_{b0}'\otimes d_{-1} , 
	\phi_{b-} \otimes d_{0} , \phi_{b-} \otimes d_{+1} , \phi_{b-} \otimes d_{-1}  \} 
	\;\Rightarrow\; R_1^\ast \oplus R_2^\ast \oplus R_3^\ast,
\end{align}
where the 2D Irreps labeled with $R_i^\ast$ with $i=1,2,3$ represent the three different 2D Irreps for the other six eigen-states of $C_3$ in second first line in Eq.~\eqref{sm-eq-wf-trs-relation}. Thus, time-reversal symmetry indicates that the combinations, $R_1\oplus R_1^\ast$, $R_2\oplus R_2^\ast$ and $R_3\oplus R_3^\ast$, are for the three different 4D Irreps. Therefore, we only need to construct those three 2D Irreps (e.g.~$R_1$, $R_2$ and $R_3$) in Eq.~\eqref{sm-eq-2d-irrep-t2g}. For this purpose, the representation matrices for $C_3$ and $S_{2x}$ on the subspace expanded by $\{ \phi_{a0}'\otimes d_{0} ,  \phi_{a0}'\otimes d_{+1} ,  \phi_{a0}'\otimes d_{-1} , 
\phi_{b+} \otimes d_{0} , \phi_{b+} \otimes d_{+1} , \phi_{b+} \otimes d_{-1} \} $ are given by
\begin{subequations} \label{sm-eq-c3-s2x-t2g-6-by-6-mat}
\begin{align}
	C_3 &= \left(
		\begin{array}{cccccc}
			1 & 0 & 0 & 0 & 0 & 0 \\
			0 & e^{i\omega_0} & 0 & 0 & 0 & 0 \\
			0 & 0 & e^{-i\omega_0} & 0 & 0 & 0 \\
			0 & 0 & 0 & e^{i\omega_0} & 0 & 0 \\
			0 & 0 & 0 & 0 & e^{-i\omega_0} & 0 \\
			0 & 0 & 0 & 0 & 0 & 1 \\
		\end{array} \right),  \\
	S_{2x} &= \frac{1}{3\sqrt{3}} \left(
		\begin{array}{cccccc}
			e^{i\frac{\pi}{2}} & 2 e^{i\frac{5\pi}{6}} & 2 e^{i\frac{\pi}{6}}  & -\sqrt{2}	 & 2\sqrt{2} e^{-i\frac{2\pi}{3}} & 2\sqrt{2} e^{i\frac{2\pi}{3}}  \\
			2 e^{i\frac{\pi}{6}} &  e^{i\frac{\pi}{2}} & 2 e^{i\frac{5\pi}{6}} & 2\sqrt{2} e^{i\frac{2\pi}{3}}  & -\sqrt{2}  & 2\sqrt{2} e^{-i\frac{2\pi}{3}}  \\
			2 e^{i\frac{5\pi}{6}} & 2 e^{i\frac{\pi}{6}} &  e^{i\frac{\pi}{2}} & 2\sqrt{2} e^{-i\frac{2\pi}{3}} & 2\sqrt{2} e^{i\frac{2\pi}{3}} & -\sqrt{2}   \\
			\sqrt{2} & 2\sqrt{2} e^{i\frac{\pi}{3}} & 2\sqrt{2} e^{-i\frac{\pi}{3}} & e^{-i\frac{\pi}{2}} & 2 e^{-i\frac{\pi}{6}} & 2 e^{-i\frac{5\pi}{6}} \\
			2\sqrt{2} e^{-i\frac{\pi}{3}} & \sqrt{2} & 2\sqrt{2} e^{i\frac{\pi}{3}} & 2 e^{-i\frac{5\pi}{6}} &  e^{-i\frac{\pi}{2}} & 2 e^{-i\frac{\pi}{6}} \\
			2\sqrt{2} e^{i\frac{\pi}{3}} & 2\sqrt{2} e^{-i\frac{\pi}{3}} & \sqrt{2} & 2 e^{-i\frac{\pi}{6}} & 2 e^{-i\frac{5\pi}{6}} & e^{-i\frac{\pi}{2}} \\
		\end{array} \right) .
\end{align}
\end{subequations}
We further notice that these six eigen-states can be grouped into three groups with different eigenvalues of $C_3$,
\begin{align}
\begin{split}
	\text{Group 1, eigenvalue}-1 &: \;\;\;  \phi_{a0}'\otimes d_{0} \;\; \text{ and } \phi_{b+} \otimes d_{-1}  , \\
	\text{Group 2, eigenvalue}-e^{i\omega_0} &: \;\;\;  \phi_{a0}'\otimes d_{+1} \text{ and } \phi_{b+} \otimes d_{0}, \\
	\text{Group 3, eigenvalue}-e^{-i\omega_0} &: \;\;\;  \phi_{a0}'\otimes d_{-1}  \text{ and }  \phi_{b+} \otimes d_{+1}. 
\end{split}	
\end{align}
Therefore, any linear combination of those two eigen-states belonging to the same group will not change the matrix representation of $C_3$ in Eq.~\eqref{sm-eq-c3-s2x-t2g-6-by-6-mat}. By choosing specific transformations, the 6-by-6 matrix representation of $S_{2x}$ in Eq.~\eqref{sm-eq-c3-s2x-t2g-6-by-6-mat} can be reduced into a direct sum of three 2-by-2 matrices. To show that, we define the following new basis 
\begin{align}
&\begin{cases}
	\Phi_{t_{2g}}^{0,1}  &= i\frac{1}{\sqrt{3}} e^{-i\omega_0 }  \phi_{a0}'\otimes d_0 + \sqrt{\frac{2}{3}} e^{i\omega_0} \phi_{b+} \otimes d_{-1}, \\
	\Phi_{t_{2g}}^{0,2}  &= -i\sqrt{\frac{2}{3}}  e^{i\omega_0 }  \phi_{a0}'\otimes d_0 +\frac{1}{\sqrt{3}} \phi_{b+} \otimes d_{-1}, \\
\end{cases} \\
&\begin{cases}
	\Phi_{t_{2g}}^{+,1}  &=-i\sqrt{\frac{2}{3}}  e^{i\omega_0 }  \phi_{a0}'\otimes d_{+1} +\frac{1}{\sqrt{3}} \phi_{b+} \otimes d_{0}, \\
	\Phi_{t_{2g}}^{+,2}  &= \frac{1}{\sqrt{3}}  \phi_{a0}'\otimes d_{+1}  -i\sqrt{\frac{2}{3}}  e^{-i\omega_0 } \phi_{b+} \otimes d_{0}, \\
\end{cases} \\
&\begin{cases}
	\Phi_{t_{2g}}^{-,1}  &= i\frac{1}{\sqrt{3}} e^{i\omega_0 }   \phi_{a0}'\otimes d_{-1} +\sqrt{\frac{2}{3}} \phi_{b+} \otimes d_{+1}, \\
	\Phi_{t_{2g}}^{-,2}  &=\sqrt{\frac{2}{3}} \phi_{a0}'\otimes d_{-1} +  i\frac{1}{\sqrt{3}} e^{-i\omega_0 }   \phi_{b+} \otimes d_{+1}, \\
\end{cases}
\end{align}
where the indices $0,+,-$ labels the eigenvalues $1, e^{i\omega_0}, e^{-i\omega_0}$ of the matrix $C_3$, and the indices $1,2$ label the number of wave functions. Within the new basis
\begin{align}
		\{ \Phi_{t_{2g}}^{0,1} , \; \Phi_{t_{2g}}^{+,1} , \; \Phi_{t_{2g}}^{0,2}, \; \Phi_{t_{2g}}^{-,1} ,\;  \Phi_{t_{2g}}^{+,2},\;  \Phi_{t_{2g}}^{-,2} \},
\end{align}
the 6-by-6 matrix representation of $S_{2x}$ in Eq.~\eqref{sm-eq-c3-s2x-t2g-6-by-6-mat} becomes
\begin{align}
	S_{2x} &= \begin{pmatrix}
			S_{2x,1} & 0  & 0  \\ 
			0 & S_{2x,2}  & 0  \\ 
			0 & 0  &  S_{2x,3}
		\end{pmatrix} , 
\end{align}
where
\begin{align}
	S_{2x,1} = \begin{pmatrix}
			-\frac{i}{\sqrt{3}} & -\frac{1}{\sqrt{2}} - \frac{i}{\sqrt{6}}  \\ 
			\frac{1}{\sqrt{2}} - \frac{i}{\sqrt{6}} & \frac{i}{\sqrt{3}}
		\end{pmatrix} , 
	S_{2x,2}&= \begin{pmatrix}
			\frac{i}{\sqrt{3}} & -\frac{1}{\sqrt{2}} - \frac{i}{\sqrt{6}}  \\ 
			\frac{1}{\sqrt{2}} - \frac{i}{\sqrt{6}} & -\frac{i}{\sqrt{3}}
		\end{pmatrix} , 
	S_{2x,3}= \begin{pmatrix}
			-\frac{i}{\sqrt{3}} & -\frac{1}{\sqrt{2}} + \frac{i}{\sqrt{6}}  \\ 
			\frac{1}{\sqrt{2}} + \frac{i}{\sqrt{6}} & \frac{i}{\sqrt{3}}
		\end{pmatrix}.
\end{align}
Therefore, the three 2D Irreps generated by $C_3$ and $S_{2x}$ are
\begin{subequations}
\begin{align}
	&A-\text{type: } \{ \Phi_{t_{2g}}^{0,1} , \; \Phi_{t_{2g}}^{+,1}\} \text{ and }  \{ \Phi_{t_{2g}}^{0,2}, \; \Phi_{t_{2g}}^{-,1} \}, \\
	&B-\text{type: }  \{ \Phi_{t_{2g}}^{+,2},\;  \Phi_{t_{2g}}^{-,2} \}.
\end{align}
\end{subequations}
Moreover, the time-reversal symmetry leads to the 4D Irrep,
\begin{subequations}
\begin{align}
	&A_2-\text{Irrep: } \{ \Phi_{t_{2g}}^{0,1} , \; \Phi_{t_{2g}}^{+,1}\} \, \oplus  \{ \Phi_{t_{2g}}^{0,1} , \; \Phi_{t_{2g}}^{+,1}\}^\ast  ,\\
	&A_3-\text{Irrep: }   \{ \Phi_{t_{2g}}^{0,2}, \; \Phi_{t_{2g}}^{-,1} \} \, \oplus \{ \Phi_{t_{2g}}^{0,2}, \; \Phi_{t_{2g}}^{-,1} \} ^\ast , \\
	&B_2-\text{Irrep: }  \{ \Phi_{t_{2g}}^{+,2},\;  \Phi_{t_{2g}}^{-,2} \} \oplus \{ \Phi_{t_{2g}}^{+,2},\;  \Phi_{t_{2g}}^{-,2} \}^\ast.
\end{align}
\end{subequations}
We further check the matrix representations for $C_3$, $S_{2x}$, $S_{2y}$ and $\mathcal{T}$ for both the A$_2$ Irrep and A$_3$ Irrep as
\begin{align}
		C_3 &= \left(
		\begin{array}{cccc}
			1 & 0 & 0 & 0 \\
			0 & \exp \left(i \omega _0\right) & 0 & 0 \\
			0 & 0 & 1 & 0 \\
			0 & 0 & 0 & \exp \left(-i \omega _0\right) \\
		\end{array}	\right),  \;\;\;\;\;\;\;\;
		S_{2x} = \left(
		\begin{array}{cccc}
			-\tfrac{i}{\sqrt{3}} & -\sqrt{\tfrac{2}{3}} & 0 & 0 \\
			\sqrt{\tfrac{2}{3}} & \tfrac{i}{\sqrt{3}} & 0 & 0 \\
			0 & 0 & \tfrac{i}{\sqrt{3}} & -\sqrt{\tfrac{2}{3}} \\
			0 & 0 & \sqrt{\tfrac{2}{3}} & -\tfrac{i}{\sqrt{3}} \\
		\end{array}	\right), \\
		S_{2y} &= \left(
		\begin{array}{cccc}
			-\frac{i}{\sqrt{3}} & \frac{1-i \sqrt{3}}{\sqrt{6}} & 0 & 0 \\
			\frac{-1-i \sqrt{3}}{\sqrt{6}} & \frac{i}{\sqrt{3}} & 0 & 0 \\
			0 & 0 & \frac{i}{\sqrt{3}} & \frac{1+i \sqrt{3}}{\sqrt{6}} \\
			0 & 0 & \frac{-1+i \sqrt{3}}{\sqrt{6}} & -\frac{i}{\sqrt{3}} \\
		\end{array}	\right), \;\;\;\;\;\;\;\;
		\mathcal{T} =  \left(
		\begin{array}{cccc}
			0 & 0 & 1 & 0 \\
			0 & 0 & 0 & 1 \\
			1 & 0 & 0 & 0 \\
			0 & 1 & 0 & 0 \\
		\end{array}	\right)\mathcal{K}.
\end{align}
They are completely the same as the matrix representation of the A$_1$ Irrep in Eq.~\eqref{sm-eq-mat-rep-A1}. Please notice that both $A_2$-Irrep and $A_3$-Irrep are the $R_1R_3$-Irrep, which is the same as the $A_1$-Irrep constructed in Sec.~\ref{sm-sub-sub-sec-eg}, and the $B_2$-Irrep is the $R_2R_2$-Irrep that is the same as the $B_1$-Irrep in Sec.~\ref{sm-sub-sub-sec-eg}.

\section{Approach I for the hierarchy of the quasi-symmetry}	
\label{Appendix-C}

In this section, we use the perturbation theory to identify the hierarchy of quasi-symmetry. This is Approach I mentioned in the main text. Here we explain the necessary details. The linear-$k$-order SOC-free Hamiltonian $\mathcal{H}_1({\bf k})$,
\begin{align} 
	\mathcal{H}_{1}(\mathbf{k}) &=  C_0\sigma_0\tau_0 + 2A_1 (\mathbf{k}\cdot\mathbf{L}) ,
\end{align}
where the angular momentum operators are $L_x = \tfrac{1}{2} \sigma_y\tau_0, \; L_y = \tfrac{1}{2}\sigma_x\tau_y, \; L_z = -\tfrac{1}{2}\sigma_z\tau_y$ defined in Eq.~\eqref{eq-ham0-k-order-Lk}. Notice that $\mathcal{H}_1(\mathbf{k})$ is invariant under the spin SU$_{\text{s}}$(2) symmetry group. Moreover, we notice that an additional hidden SU$_{\text{o}}$(2) symmetry also exists for $\mathcal{H}_1(\mathbf{k})$ in the ``orbital'' space that can be generated by the following operators 
\begin{align} \label{sm-eq-qs-m1m2m3}
	\mathcal{M}_{1,2,3} = \tfrac{1}{2}\{ s_0\sigma_y\tau_z, s_0\sigma_y\tau_x, s_0\sigma_0\tau_y \},
\end{align}
which is shown in the main text (see Eq.~(2)), and $\mathcal{M}_{1,2,3}$ all commute with $\mathcal{H}_1(\mathbf{k})$ and satisfy the commutation relations 
\begin{subequations}
\begin{align}
	[\mathcal{M}_{i}, \mathcal{M}_{j} ] &= i\epsilon_{i,j,k} \mathcal{M}_k, \\ 
	\{\mathcal{M}_{i}, \mathcal{M}_{j}  \} &= \frac{1}{2}\delta_{i,j}.
\end{align}
\end{subequations}
Here $i,j=1,2,3$, $\epsilon_{i,j,k}$ is the 3D Levi-Civita symbol, and $\delta_{i,j}$ is the Kronecker delta function. Thus, we refer to it as the orbital SU$_{\text{o}}$(2) quasi-symmetry group for $\mathcal{H}_1({\bf k})+\mathcal{H}_{\text{soc}}$ because of
\begin{align}
	[\mathcal{M}_i,\mathcal{H}_1({\bf k})+\mathcal{H}_{\text{soc}}]=0, 
\end{align}
where the on-site SOC Hamiltonian is given by Eq.~\ref{eq-soc-ham},
\begin{align}
	\mathcal{H}_{\text{soc}} = 4 \lambda_0 ( \mathbf{S} \cdot \mathbf{L}),
\end{align}
where we use ${\bf S}=\tfrac{1}{2}(s_x, s_y, s_z)$ for the spin-1/2 angular momentum operators. And $[S_i, S_j]=i\epsilon_{ijk}S_k$. In addition, we consider the $k^2$-order Hamiltonian that can be represented in a compact form,
\begin{align} \label{sm-eq-h2-m1m2m3-kk}
	\mathcal{H}_{2}(\mathbf{k}) =\mathcal{H}_{2, \mathcal{M}_1}(\mathbf{k}) +\mathcal{H}_{2, \mathcal{M}_2}(\mathbf{k}) +\mathcal{H}_{2, \mathcal{M}_3}(\mathbf{k}), 
\end{align}
where each part of $\mathcal{H}_{2}(\mathbf{k})$ is given by 
\begin{align}
	\mathcal{H}_{2,\mathcal{M}_i}(\mathbf{k})={\bf g}_i\cdot{\bf J}_i, 
\end{align}
for $i=1,2,3$. Here we define the $k$-dependent vectors as
\begin{align}
	\begin{split}
		{\bf g}_1 ({\bf k}) &= (C_2k_xk_y, -C_3k_xk_z,C_1k_yk_z),   \\
		{\bf g}_2 ({\bf k}) &= (C_3k_xk_y,C_1k_xk_z, -C_2  k_yk_z),   \\
		{\bf g}_3 ({\bf k}) &= (C_1k_xk_y,  C_2k_xk_z, -C_3  k_y k_z). 
	\end{split}
\end{align}
And the corresponding vectors of operators 	
\begin{align}
	\begin{split}
		{\bf J}_1 &= ( \sigma_x  \tau_x, -\sigma_z\tau_x,  \sigma_0 \tau_z),  \\
		{\bf J}_2 &= (\sigma_x\tau_z,\sigma_z\tau_z,\sigma_0\tau_x),   \\
		{\bf J}_3 &= (\sigma_z\tau_0,\sigma_x\tau_0,\sigma_y\tau_y). 
	\end{split}
\end{align}
In addition, we also realize that the $k^2$ terms of $\mathcal{H}_2({\bf k})$ break this orbital SU$_{\text{o}}$(2) quasi-symmetry generated by $\{\mathcal{M}_{1,2,3}\}$ and lead to the splitting of all bands. However, different parts of the entire $k^2$-terms can lead to the reduction from SU$_{\text{o}}$(2) to a orbital U(1). To show that, as we discussed in the main text, we find that
\begin{align}
	[{\bf J}_i,\mathcal{M}_i]=0 \text{ and }  \{{\bf J}_i, \mathcal{M}_j\}=0 \text{ for }  i\neq j,
\end{align}
which implies 
\begin{align}
	[\mathcal{H}_{2,\mathcal{M}_i}({\bf k}),\mathcal{M}_i]=0.
\end{align}
This can be also found in the main text and Sec.~\ref{Appendix-A-2}. Below, we focus on how to identity the quasi-symmetry based on the effective perturbation theory (Approach I). Note that alternative Approach II will be presented in Sec.~\ref{Appendix-D}.

Furthermore, we notice there is a conservation of total angular momentum at fixed ${\bf k}$ for $\mathcal{H}_1({\bf k})+\mathcal{H}_{\text{soc}} = C_0 + 2A_1 ({\bf k}\cdot \mathbf{L}) + 4\lambda_0 ({\bf S}\cdot {\bf L}) $. As we have discussed in Sec.~\ref{Appendix-A-4}, the total angular momentum ${\bf S}+{\bf L}$ is conserved at ${\bf k}=0$. But it does not commute with the Hamiltonian $\mathcal{H}_1({\bf k})+\mathcal{H}_{\text{soc}}$ at the nonzero ${\bf k}$. However, we notice that 
\begin{align} \label{sm-eq-slnk-operator}
	[({\bf S}+{\bf L})\cdot \vec{n}_{\bf k}, \mathcal{H}_1({\bf k})+\mathcal{H}_{\text{soc}} ] =0,
\end{align}
where $\vec{n}_{\bf k} = \frac{{\bf k}}{k}$ is the direction of the momentum ${\bf k}$. And $({\bf S}+{\bf L})\cdot \vec{n}_{\bf k}$ is physically similar to the helicity operator at nonzero ${\bf k}$. Also, Eq.~\eqref{sm-eq-slnk-operator} can be proved as follows,
\begin{itemize}
\item Spin and orbital are independent degree of freedoms, so we have $[{\bf S}, {\bf L}]=0$. Therefore, $[({\bf S}+{\bf L})\cdot \vec{n}_{\bf k}, \mathcal{H}_1({\bf k})] = [{\bf L}\cdot \vec{n}_{\bf k}, C_0 + 2A_1 ({\bf k}\cdot {\bf L})] = 0$.
\item We can define total angular momentum ${\bf L}_{\text{tot}} = {\bf S}+{\bf L}$, so that $\mathcal{H}_{\text{soc}}=2\lambda_0 ({\bf L}_{\text{tot}}^2 - {\bf S}^2 - {\bf L}^2) = 2\lambda_0 ({\bf L}_{\text{tot}}^2 - \frac{3}{4}-\frac{3}{4})$. Then one can check 
\begin{align}
	[{\bf L}_{\text{tot},i} , {\bf L}_{\text{tot},j}] = [S_i, S_j] + [L_i, L_j] = i\epsilon_{ijk}S_k + i\epsilon_{ijk}L_k = i\epsilon_{ijk} {\bf L}_{\text{tot},k}.
\end{align}
Besides, we have $[{\bf L}_{\text{tot}} , {\bf L}_{\text{tot}}^2 ]=0$ (i.e.,~the square of the angular momentum commutes with any of the components), which leads to $[{\bf L}_{\text{tot}}, \mathcal{H}_{\text{soc}}]=0$. Therefore, we have 
\begin{align}
	[{\bf L}_{\text{tot}} \cdot \vec{n}_{{\bf k}} , {\bf L}_{\text{tot}}^2 ] =0.
\end{align}
This proves that ${\bf L}_{\text{tot}} \cdot \vec{n}_{{\bf k}} $ is a symmetry operator of $\mathcal{H}_1({\bf k})+\mathcal{H}_{\text{soc}}$ at any nonzero ${\bf k}$.
\end{itemize}

\subsection{Approach I: the U(1) quasi-symmetry protected nodal planes}
\label{Appendix-C-1}

We solve the eigen-problem of the Hamiltonian $\mathcal{H}_1(\mathbf{k})$ in Eq.~\eqref{eq-ham0-k-order} or Eq.~\eqref{eq-ham0-k-order-Lk}. Due to the full rotation symmetry, we choose the spherical coordinate with the momentum ${\bf k}=(k_x,k_y,k_z) = k(\sin\theta\cos\phi, \sin\theta\sin\phi, \cos\theta)$. The eigen-energies of $\mathcal{H}_1(\mathbf{k})$ have two branches $E_{\pm}=C_0\pm A_1k$ with each branch twofold degeneracy (fourfold if spin degeneracy is involved). In this work, we assume $A_1>0$, and the two degenerate eigen-wave functions $\vert \Psi_{A/B+}(\theta,\phi) \rangle$ of the positive energy branch ($E_+$) and $\vert \Psi_{A/B-}(\theta,\phi) \rangle$ of the negative energy branch ($E_-$) are given by
\begin{align}
	\mathcal{H}_1(\mathbf{k}) \begin{cases}
		 \vert \Psi_{A/B+}(\theta,\phi) \rangle = E_+ \vert \Psi_{A/B+}(\theta,\phi) \rangle, \\
		 \vert \Psi_{A/B-}(\theta,\phi) \rangle  = E_- \vert \Psi_{A/B-}(\theta,\phi) \rangle, 
	\end{cases}
\end{align}
where the index $\pm$ represent the eigenvalues of ${\bf L}\cdot \vec{n}_{{\bf k}}$, and the eigen-states in the spherical coordinator are given by
\begin{subequations}
\begin{align}
\begin{cases} \label{eq-basis-wf-P-model}
	\vert \Psi_{A+}(\theta,\phi) \rangle &= \tfrac{1}{\sqrt{2}} \left( \cos\theta\cos\phi - i\sin\phi, -\cos\theta\sin\phi-i\cos\phi, 0 , \sin\theta\right)^T , \\
	\vert \Psi_{B+}(\theta,\phi) \rangle &= \tfrac{1}{\sqrt{2}} \left( -i\sin\theta\cos\phi, i\sin\theta\sin\phi,1,i\cos\theta\right) ^T.
\end{cases} \\ 
\begin{cases} \label{eq-basis-wf-P-model-22}
	\vert \Psi_{A-}(\theta,\phi) \rangle &= \tfrac{1}{\sqrt{2}} \left( i\sin\theta\cos\phi, -i\sin\theta\sin\phi,1,-i\cos\theta\right)^T, \\
	\vert \Psi_{B-}(\theta,\phi) \rangle &= \tfrac{1}{\sqrt{2}} \left( \cos\theta\cos\phi + i\sin\phi, -\cos\theta\sin\phi+i\cos\phi, 0 , \sin\theta \right)^T.
\end{cases}
\end{align}
\end{subequations}
The solution is not unique, since there exists a twofold degeneracy between $\vert \Psi_{A+}(\theta,\phi) \rangle $ and $\vert \Psi_{B+}(\theta,\phi) \rangle$ at arbitrary ${\bf k}$, protected by the orbital SU(2) symmetry generated by $\{\mathcal{M}_1,\mathcal{M}_2,\mathcal{M}_3\}$ in Eq.~\eqref{sm-eq-qs-m1m2m3}. Moreover, the subscript $A(B)$ can represent the eigenvalues of the quasi-symmetry operator defined as $(\mathcal{M}_1,\mathcal{M}_2,\mathcal{M}_3)\cdot \vec{n}$ with a specific real normalized vector $\vec{n}=(n_1, n_2, n_3)$. For the basis in Eq.~\eqref{eq-basis-wf-P-model} and Eq.~\eqref{eq-basis-wf-P-model-22}, the $\vec{n}$-vector reads
\begin{align}
    \vec{n} = \left(-\sin (\theta ) \cos (\phi ), \sin (\theta ) \sin (\phi ), -\cos (\theta )\right).
\end{align}
And one can check 
\begin{subequations}
\begin{align}
    ({\bf \mathcal{M}}\cdot \vec{n} ) \vert \Psi_{A+}(\theta,\phi) \rangle &= \;\;\, \vert \Psi_{A+}(\theta,\phi) \rangle, \\
    ({\bf \mathcal{M}}\cdot \vec{n} ) \vert \Psi_{A-}(\theta,\phi) \rangle &= \;\;\, \vert \Psi_{A-}(\theta,\phi) \rangle, \\
    ({\bf \mathcal{M}}\cdot \vec{n} ) \vert \Psi_{B+}(\theta,\phi) \rangle &= -\vert \Psi_{B+}(\theta,\phi) \rangle, \\
    ({\bf \mathcal{M}}\cdot \vec{n} ) \vert \Psi_{B-}(\theta,\phi) \rangle &= -\vert \Psi_{B-}(\theta,\phi) \rangle.
\end{align}
\end{subequations}
But $\vec{n}$ can be arbitrary due to this twofold degeneracy (fourfold if spin degeneracy is accounted). This means the choice of eigen-states are not unique. Besides, another set of wavefunctions chosen as eigen-states of $\mathcal{M}_3$ will be shown in Sec.~\ref{appendix-D-1} by fixing $\vec{n}=(0,0,1)$. Furthermore, we can project the angular momentum operator $\mathbf{L}$ into the eigenstate subspace,
\begin{align}\label{eq-angul-mom-L-expt-value-1}
	\langle \Psi_{A\pm} (\theta,\phi)\vert \mathbf{L} \vert \Psi_{A\pm}(\theta,\phi)\rangle 
	&= \langle \Psi_{B\pm} (\theta,\phi)\vert \mathbf{L} \vert \Psi_{B\pm}(\theta,\phi)\rangle =\pm\frac{\mathbf{k}}{2k}, \\
	\langle \Psi_{A\pm} (\theta,\phi)\vert \mathbf{L} \vert \Psi_{B\pm}(\theta,\phi)\rangle 
	&= \langle \Psi_{B\pm} (\theta,\phi)\vert \mathbf{L} \vert \Psi_{A\pm}(\theta,\phi)\rangle = 0. 
	\label{eq-angul-mom-L-expt-value-2}
\end{align}
Here, Eqs.~\eqref{eq-angul-mom-L-expt-value-1} and \eqref{eq-angul-mom-L-expt-value-2} mean the emergent angular momentum operator $\mathbf{L}$ is along the momentum direction after the projection. Besides, with involving the spin degree of freedom, the corresponding four-fold degenerate wave-functions are labelled as
\begin{align}\label{sm-eq-p-model-basis-spinful}
\vert \Psi_{+}\rangle=\{\vert \Psi_{A+\uparrow}(\theta,\phi) \rangle, \vert \Psi_{B+\uparrow}(\theta,\phi) \rangle, \vert \Psi_{A+\downarrow}(\theta,\phi) \rangle, \vert \Psi_{B+\downarrow}(\theta,\phi) \rangle \},  
\end{align}
where
\begin{align}\label{eq-four-basis-spin}
\begin{split}
	\vert \Psi_{A+\uparrow}(\theta,\phi) \rangle &= (1,0)^T \otimes \vert \Psi_{A+}(\theta,\phi) \rangle   , \\
	\vert \Psi_{B+\uparrow}(\theta,\phi) \rangle &= (1,0)^T \otimes \vert \Psi_{B+}(\theta,\phi) \rangle ,\\
	\vert \Psi_{A+\downarrow}(\theta,\phi) \rangle &= (0,1)^T \otimes \vert \Psi_{A+}(\theta,\phi) \rangle, \\
	\vert \Psi_{B+\downarrow}(\theta,\phi) \rangle &=  (0,1)^T \otimes \vert \Psi_{B+}(\theta,\phi)\rangle ,
\end{split}
\end{align}
where $(1,0)^T$ and $(0,1)^T$ label the spin-up and spin-down wave functions, respectively. The above set of wave functions serves as the basis for the projected 4-band perturbation model via the first-order perturbation theory, dubbed as ``the $P$-model''. We treat both the on-site SOC in Eq.~\eqref{eq-soc-ham} and the $k^2$-terms in Eq.~\eqref{eq-ham0-k2-order} as perturbations,
\begin{align}
	\mathcal{H}_{perb} ({\bf k}) &= \mathcal{H}_{\text{soc}} +  s_0 \otimes \mathcal{H}_{2}(\mathbf{k}).
\end{align}
The projected Hamiltonian $\left\langle \Psi_{+}  \vert \mathcal{H}_{perb} \vert \Psi_{+} \right\rangle$ is given by
\begin{align}\label{eq-eff-ham-p-model}
	\mathcal{H}_{P}^{eff(1)}({\bf k}) &= (E_+ +B_1k^2)s_0\omega_0 + \mathcal{H}_{\text{soc},+}^{eff(1)}({\bf k}) + \mathcal{H}_{k^2,+}^{eff(1)}({\bf k}),
\end{align}
which is marked as the $P$-model around $R$-point. And
\begin{subequations} \label{sm-eq-hsoc-hk2-p-model}
\begin{align}
	\mathcal{H}_{\text{soc},+}^{eff(1)}({\bf k}) &= \lambda_0 \left( \lambda_x s_x + \lambda_y s_y + \lambda_z s_z \right)  \otimes \omega_0, \\
	\mathcal{H}_{k^2,+}^{eff(1)}({\bf k}) &= \tilde{C}k^2 s_0 \otimes \left( d_x \omega_x + d_y\omega_y + d_z\omega_z \right).
\end{align}
\end{subequations}
where $\tilde{C}=C_1-C_2+C_3$, and $\omega_{x,y,z}$ are Pauli matrices for the $\{A+,B+\}$ band subspace. The coefficients $\lambda_{x,y,z}$ are defined as
\begin{align}\label{sm-eq-soc-p-model-2}
	(\lambda_x,\lambda_y,\lambda_z)&=(\sin\theta\cos\phi,\sin\theta\sin\phi,\cos\theta)= \tfrac{\mathbf{k}}{\vert \mathbf{k}\vert}, 
\end{align}
On the other hand, the coefficients $d_{x,y,z}$ are given by
\begin{align}\label{sm-eq-dxyz-pmodel}
	\begin{split}
		d_x &= \tfrac{1}{4} \sin\theta\sin(2\theta) \sin (2\phi) (\cos\phi + \sin\phi) , \\
		d_y &= \tfrac{1}{4}\sin\theta\sin(2\theta) \sin (2\phi) ( \sin\theta +\cos\theta (\cos\phi - \sin\phi) ), \\
		d_z &= \tfrac{1}{4}\sin\theta\sin(2\theta) \sin (2\phi) ( \cos\theta - \sin\theta (\cos\phi-\sin\phi) ).
	\end{split}
\end{align}

Furthermore, we use the symmetry to understand the above first-order perturbation Hamiltonian. The basis function in Eq.~\eqref{sm-eq-p-model-basis-spinful} can be labeled by eigen-values of symmetries, 
\begin{subequations} \label{sm-eq-basis-p-model-symmetry}
\begin{align} 
	\vert \Psi_{A+\uparrow}(\theta,\phi) \rangle &= \vert \uparrow\rangle \otimes \vert p = +\tfrac{1}{2},  q=+1\rangle ,  \\ 
	\vert \Psi_{B+\uparrow}(\theta,\phi) \rangle &= \vert \uparrow\rangle  \otimes \vert p = +\tfrac{1}{2}, q=-1\rangle ,  \\ 
	\vert \Psi_{A+\downarrow}(\theta,\phi) \rangle &= \vert \downarrow \rangle  \otimes \vert p = +\tfrac{1}{2}, q=+1\rangle,   \\ 
	\vert \Psi_{B+\downarrow}(\theta,\phi) \rangle &= \vert \downarrow \rangle  \otimes \vert p = +\tfrac{1}{2}, q=-1\rangle,  
\end{align}
\end{subequations}
Here we take $p=\pm 1/2$ as the eigenvalues of ${\bf L}\cdot \vec{n}_{{\bf k}}$ and $q=\pm 1$ as the eigenvalues of $\vec{\mathcal{M}}\cdot\vec{n}$. All these four states are degenerate with eigen-energy of $\mathcal{H}_1$ as $C_0 + A_1k$. The linear-$k$ Hamiltonian has both orbital SU(2) symmetry and spin SU(2) symmetry. Specifically, the orbital SU(2) symmetry generated by $\{\mathcal{M}_{1,2,3}\}$ indicates that the vector $\vec{n}$ is arbitrary. And the spin SU(2) symmetry implies that $\vert \uparrow/\downarrow\rangle$ can be any direction in the spin subspace. Then, to show the origin of the hidden quasi-symmetry of the $P$-model, we individually do the projection for the on-site SOC and $k^2$-order Hamiltonian, 
\begin{itemize}
\item Only do the projection of the on-site SOC Hamiltonian. In this case, the orbital SU(2) symmetry generated by $\{\mathcal{M}_{1,2,3}\}$ preserves, because of $[\mathcal{M}_{i}, \mathcal{H}_1({\bf k})+\mathcal{H}_{\text{soc}}]=0$ with $i=1,2,3$. Thus, we obtain
\begin{align}
	\text{SU}_{\text{s}}(2)\times \text{SU}_{\text{o}}(2) 
\stackrel{\left\langle\mathcal{H}_{\text{soc}}\right\rangle_{\mathcal{H}_1}}{\xhookrightarrow{\quad\quad\quad\quad\quad}} \text{U}_{\text{s}}(1) \times \text{SU}_{\text{o}}(2),
\end{align}
where the spin U(1) symmetry generator depends on ${\bf k}$. To understand this, we recall the conservation of $({\bf S}+{\bf L})\cdot \vec{n}_{{\bf k}}$ for $\mathcal{H}_1({\bf k})+\mathcal{H}_{\text{soc}}=C_0 + 
2A_1{\bf k}\cdot {\bf L} + 4\lambda_0 {\bf S}\cdot {\bf L}$. Then, for a fixed nonzero ${\bf k}$, one can take the eigen-states in Eq.~\eqref{sm-eq-basis-p-model-symmetry} as eigen-states of the helicity operator $({\bf S}+{\bf L})\cdot \vec{n}_{{\bf k}}$ by choosing
\begin{subequations}  \label{sm-eq-basis-symmetry-p-model-2}
\begin{align} 
	\vert \Psi_{A+\uparrow}(\theta,\phi) \rangle &= \vert s = +\tfrac{1}{2} \rangle \otimes \vert p = +\tfrac{1}{2},  q=+1\rangle , \\ 
	\vert \Psi_{B+\uparrow}(\theta,\phi) \rangle &= \vert s = +\tfrac{1}{2} \rangle  \otimes \vert p = +\tfrac{1}{2}, q=-1\rangle, \\ 
	\vert \Psi_{A+\downarrow}(\theta,\phi) \rangle &= \vert s = -\tfrac{1}{2} \rangle  \otimes \vert p = +\tfrac{1}{2},q=+1\rangle, \\ 
	\vert \Psi_{B+\downarrow}(\theta,\phi) \rangle &= \vert s = -\tfrac{1}{2} \rangle  \otimes \vert p = +\tfrac{1}{2}, q=-1\rangle. 
\end{align}
\end{subequations}
Here we use $s=\pm1/2$ as eigenvalues of ${\bf S}\cdot\vec{n}_{{\bf k}}$. Here we use $\vec{n}_{{\bf k}}$, $\vec{n}_{{\bf k}}'$, and $\vec{n}_{{\bf k}}''$ to be the set of 3D orthogonal coordinates at the fixed ${\bf k}$ and $\vec{n}_{{\bf k}}$ is the direction of ${\bf k}$ (i.e.,~$\vec{n}_{{\bf k}}=(\sin\theta \cos\phi, \sin\theta\sin\phi, \cos\theta)$). Namely, $\vert\vec{n}_{{\bf k}}\vert=\vert\vec{n}_{{\bf k}}'\vert=\vert\vec{n}_{{\bf k}}''\vert=1$, and $\vec{n}_{{\bf k}} \cdot \vec{n}_{{\bf k}}' = \vec{n}_{{\bf k}} \cdot \vec{n}_{{\bf k}}'' = \vec{n}_{{\bf k}}' \cdot \vec{n}_{{\bf k}}''=0$. Then, the first-order projection of ${\bf S}\cdot {\bf L} = [({\bf S}\cdot\vec{n}_{{\bf k}}) ({\bf L}\cdot\vec{n}_{{\bf k}}) + ({\bf S}\cdot\vec{n}_{{\bf k}}') ({\bf L}\cdot\vec{n}_{{\bf k}}') + ({\bf S}\cdot\vec{n}_{{\bf k}}'') ({\bf L}\cdot\vec{n}_{{\bf k}}'') ]$ onto these four states in Eq.~\eqref{sm-eq-basis-symmetry-p-model-2} leads to 
\begin{align}
\begin{split}
[ \left\langle\mathcal{H}_{\text{soc}}\right\rangle_{\mathcal{H}_1} ]_{i,j} &= 4\lambda_0 \langle \Psi_i(\theta,\phi) \vert  {\bf S}\cdot {\bf L} \vert \Psi_j(\theta,\phi) \rangle , \\
	&= 4\lambda_0 \langle \Psi_i(\theta,\phi) \vert  ({\bf S}\cdot\vec{n}_{{\bf k}}) ({\bf L}\cdot\vec{n}_{{\bf k}}) + ({\bf S}\cdot\vec{n}_{{\bf k}}') ({\bf L}\cdot\vec{n}_{{\bf k}}') + ({\bf S}\cdot\vec{n}_{{\bf k}}'') ({\bf L}\cdot\vec{n}_{{\bf k}}'')  \vert \Psi_j(\theta,\phi) \rangle , \\
	&=4\lambda_0  \langle \Psi_i(\theta,\phi) \vert ({\bf S}\cdot\vec{n}_{{\bf k}}) ({\bf L}\cdot\vec{n}_{{\bf k}})  \vert \Psi_j(\theta,\phi) \rangle, 
\end{split}
\end{align}
where $i,j=\{ A+\uparrow, B+\uparrow, A+\downarrow, B+\downarrow \}$. Thus, the first-order perturbation for on-site SOC Hamiltonian is 
\begin{align}
\left\langle\mathcal{H}_{\text{soc}}\right\rangle_{\mathcal{H}_1} = \lambda_0\begin{pmatrix}
	 1 & 0 & 0 & 0 \\
	 0 & 1 & 0 & 0 \\
	 0 & 0 & -1 & 0 \\
	 0 & 0 & 0 & -1 
	\end{pmatrix}.
\end{align}
Please notice that $\left\langle\mathcal{H}_{\text{soc}}\right\rangle_{\mathcal{H}_1}$ can be also obtained after diagonalizing $\mathcal{H}_{\text{soc},+}^{eff(1)}({\bf k}) $ in Eq.~\eqref{sm-eq-hsoc-hk2-p-model}. At nonzero ${\bf k}$, we find that the fourfold degenerate states [$C_0+A_1k$ for $\mathcal{H}_1({\bf k})$] are split by the on-site SOC Hamiltonian $\mathcal{H}_{\text{soc}}$ into two states and each state has tow fold degeneracy stemming from the orbital SU(2) symmetry, $E=C_0+A_1k \pm \lambda_0$. And the eigen-states are just Eq.~\eqref{sm-eq-basis-symmetry-p-model-2}. And the spin-polarization along the direction of ${\bf k}$ represents the spin U(1) symmetry. 
\item Only do the projection of the $k^2$-order Hamiltonian. In this case, the spin SU(2) symmetry preserves. Since the $k^2$-order Hamiltonian is spin-independent. Thus, 
\begin{align}
	\text{SU}_{\text{s}}(2)\times \text{SU}_{\text{o}}(2) 
	\stackrel{\left\langle\mathcal{H}_{2}\right\rangle_{\mathcal{H}_1}}{\xhookrightarrow{\quad\quad\quad\quad\quad}} \text{SU}_{\text{s}}(2) \times \text{U}_{\text{o}}(1).
\end{align}
We do not have an elegant picture for the $\left\langle\mathcal{H}_{2}\right\rangle_{\mathcal{H}_1}$. But we know it does not dependent on spin, as shown in Eq.~\eqref{sm-eq-hsoc-hk2-p-model}. The four-fold degeneracy is split into two states, and each state has two fold degeneracy (i.e.~spin degeneracy). And, the two-by-two matrix in the $\{A,B\}$-subspace itself severs as the orbital U(1) symmetry. 
\end{itemize}
Therefore, up to the first order perturbation, we obtain a hierarchy of quasi-symmetry for CoSi mentioned in the main text (see Eq.~(7)),  
\begin{align}
	\text{SU}_{\text{s}}(2)\times \text{SU}_{\text{o}}(2) 
	\stackrel{\left\langle\mathcal{H}_{\text{soc}}+\mathcal{H}_{2}\right\rangle_{\mathcal{H}_1}}{\xhookrightarrow{\quad\quad\quad\quad\quad\quad\quad}} \text{U}_{\text{s}}(1) \times \text{U}_{\text{o}}(1) .
\end{align}
And $[\text{U}_{\text{s}}(1) , \text{U}_{\text{o}}(1)]=0$, as shown below.

\subsection{The analytical properties of the effective perturbation 4-band $P$-model}
\label{Appendix-C-2}

The $P$-model has the self-commuting structure at every ${\bf k}$. Explicitly, 
\begin{align}
\left\lbrack \mathcal{H}_{\text{soc},+}^{eff(1)}({\bf k}_1)  , \mathcal{H}_{k^2,+}^{eff(1)}({\bf k}_2) \right\rbrack = 0, \quad \forall {\bf k}_1 \; \&\; \forall {\bf k}_2 \text{ in the whole momentum space} 
\end{align}
as discussed in the main text, which is known as the projected stabilizer code Hamiltonian and directly leads to the U(1) quasi-symmetry group. Below, we provide some analytical study on the properties of the effective $P$-model. The eigen-energies of the $P$-model in Eq.~\eqref{eq-eff-ham-p-model} are 
\begin{align}\label{sm-eq-p-model-disp}
	\begin{split}
		E_{\alpha\beta}(k,\theta,\phi) = C_0+B_1k^2+A_1k +\alpha\lambda_0 
		+\beta\tfrac{\sqrt{3}}{4}\tilde{C}k^2 \vert \sin2\phi\sin2\theta\sin\theta\vert,
	\end{split}
\end{align}
where $\alpha=\pm$ and $\beta=\pm$. Here $\sin2\phi\sin2\theta\sin\theta = 4k_xk_yk_z/k^3$, indicating that there are two-fold degeneracy on high symmetry planes ($k_x=0$ or $k_y=0$ or $k_z=0$).

For the $\Gamma-R-M$ plane, $\phi=\pi/4$ so $\sin2\phi=1$. The eigen-energies along high symmetry lines are listed as follows:
\begin{itemize}
	\item [1.)] Along the $R-M$ line. \\
	$\theta=0$, so $E_\alpha=C_0+B_1k^2+A_1k +\alpha \lambda_0$. All the bands are two-fold degenerate.
	\item[2.)] Along the $R-Z$ line. \\
	$\theta=\pi/2$, so $E_\alpha=C_0+B_1k^2+A_1k +\alpha \lambda_0$. All the bands are two-fold degenerate.             
\end{itemize}

\begin{figure}[t]
	\centering
	\includegraphics[width=0.9\linewidth]{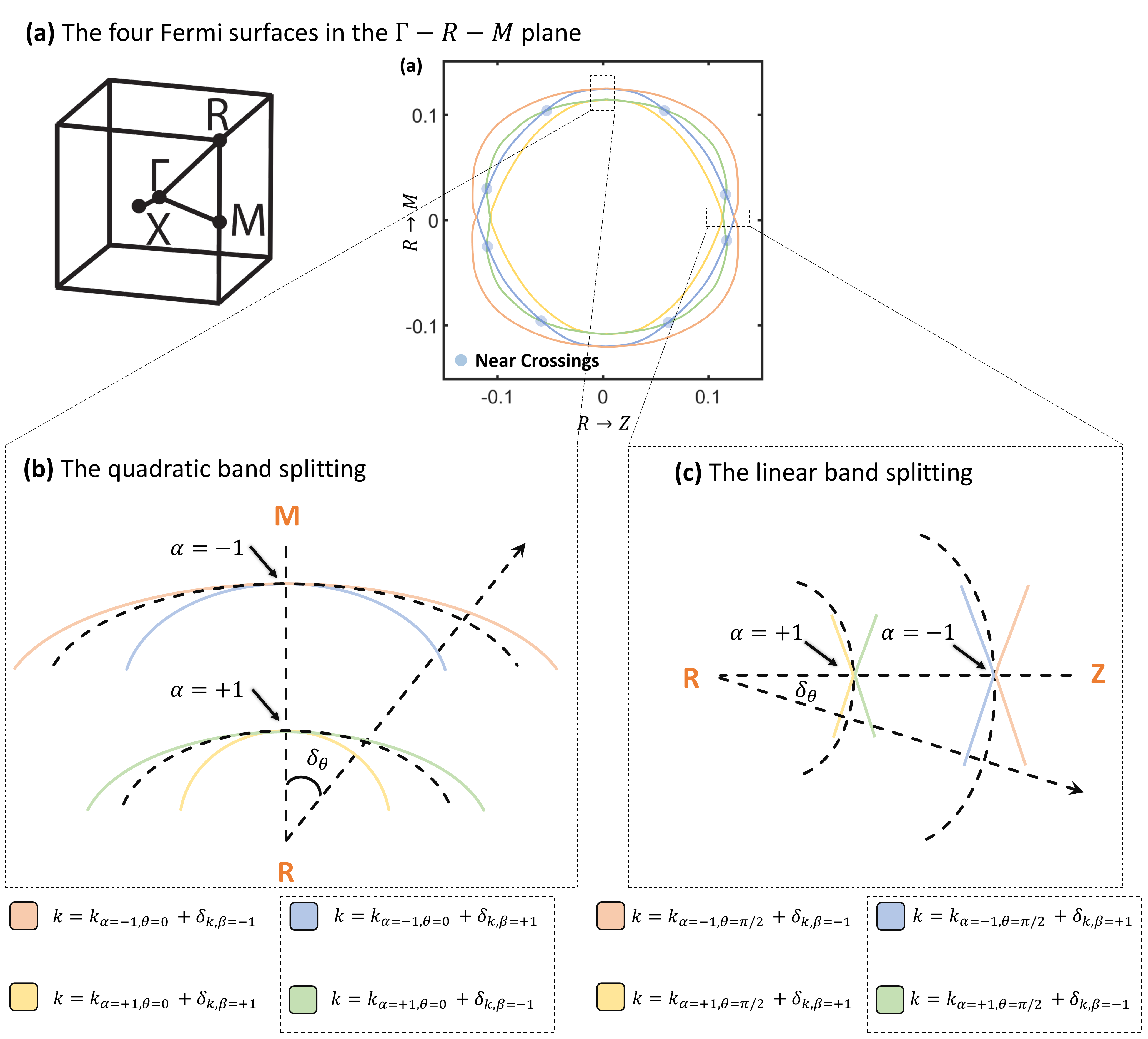}
	\caption{ The four distinct Fermi surfaces of CoSi are numerically calculated, as shown in (a) for the $\Gamma-R-M$ plane. These high-symmetry points are marked in the first Brillouin zone [left panel in (a)]. And these four Fermi surfaces are labeled by four different colors. The middle two have a ``crossing'' behavior with a tiny gap. And the analytical results for the band dispersions (see Eq.~\eqref{sm-eq-p-model-disp}) of the $P$-model along the $R-M$ line in (b) and the $R-Z$ line in (c). The two dashes black lines represents the Fermi surfaces with $\alpha=\pm1$ and $\beta=0$. The $\beta$-term leads to the quadratic band splitting around the $R-M$ line in (a) with $\delta_k\sim\delta_\theta^2$, and the linear band splitting around the $R-Z$ line in (b) with $\delta_k\sim\delta_\theta$. Here, the orange line is for $\alpha=-1,\beta=-1$, the blue line is for $\alpha=-1,\beta=+1$, the green line is for $\alpha=+1,\beta=-1$, and the yellow line is for $\alpha=+1,\beta=+1$. Once the band splitting caused by the $\beta$-term is large enough, the band crossing between the blue line ($\beta=+1$) and the green line ($\beta=-1$) may happen at arbitrary momenta, which are exact and protected by the quasi-symmetry. 
	}	
	\label{sm-fig4}
\end{figure}

We then analyze the perturbation along these two high symmetry lines (i.e.,~the $R-M$ line and the $R-Z$ line) to identify the band crossing types of the Fermi surface (linear or quadratic).  For a given $E_f$, the Fermi surface shape of the momentum $k_\pm$ for the upper four bands is determined by the quadratic equation
\begin{align}\label{eq-k-Ef-high-symmetry-line}
	B_1 k^2 + A_1k + C_\alpha = 0,
\end{align}
where $C_\alpha = \alpha \lambda_0 + C_0 -E_f $ with $\alpha=\pm$. This equation leads to the solution
\begin{align}
	\begin{split}
		k_{\alpha,\theta=0}  = k_{\alpha,\theta=\pi/2} 
		= \frac{1}{2B_1} \left( -A_1 + \sqrt{A_1^2 -4B_1C_\alpha } \right).
	\end{split}
\end{align}
We firstly focus on the $R-M$ line by expanding Eq.~\eqref{sm-eq-p-model-disp} around $\theta=0$. In this case, we have $\theta = 0 + \delta_\theta$ and $k= k_{\alpha,\theta=0}  + \delta_k $. Then, the Eq.~\eqref{eq-k-Ef-high-symmetry-line} should be replaced by 
\begin{align}
	0 =  C_\alpha+B_1k^2+A_1k  +\beta\tfrac{\sqrt{3}}{4}\tilde{C}k^2 \vert \sin2\phi\sin2\theta\sin\theta\vert,
\end{align}
which leads to 
\begin{align}
	\begin{split}
		0= C_\alpha+ B_1 \left( k_{\alpha,\theta=0}  + \delta_k \right)^2 + A_1\left( k_{\alpha,\theta=0}  + \delta_k \right) 
		+ \beta\tfrac{\sqrt{3}}{4}\tilde{C}  \left( k_{\alpha,\theta=0}  + \delta_k \right)^2 \vert \sin2\phi\vert (2\delta_\theta^2).
	\end{split}
\end{align} 
After neglecting the $\delta^2_k$ terms, we find
\begin{align}
	\begin{split}
		\delta_{k,\beta} = -\beta \frac{\sqrt{3}}{2} \frac{\tilde{C}\vert \sin2\phi\vert \delta_\theta^2 (k_{\alpha,\theta=0})^2}{2B_1 k_{\alpha,\theta=0} + A_1 + \beta\sqrt{3}\tilde{C}\vert \sin2\phi\vert\delta_\theta^2  k_{\alpha,\theta=0}}.
	\end{split}
\end{align}
To the $\delta_\theta^2$ order, we have
\begin{align}
	\delta_{k,\beta} = -\beta \frac{\sqrt{3}}{2} \frac{\tilde{C}\vert \sin2\phi\vert  (k_{\alpha,\theta=0})^2}{2B_1 k_{\alpha,\theta=0} + A_1} \times \delta_\theta^2,
\end{align}
which indicates that the exact crossing along the $R-M$ line is quadratic in momentum $k=\sqrt{k_x^2+k_y^2}$, as illustrated in Fig.~\ref{sm-fig4} (a) around the $R-M$ line.

For $\theta=\pi/2$ (i.e.,~the $R-Z$ line), we then expand Eq.~\eqref{sm-eq-p-model-disp} around $\theta = \pi/2 + \delta_\theta$. Because 
\begin{align}
	\begin{split}
		\sin2\theta\sin\theta \to \sin(\pi + 2\delta_\theta) \sin(\pi/2+\delta_\theta) 
		\to -2\delta_\theta + \mathcal{O}(\delta_\theta^3),
	\end{split}
\end{align}
It leads to the solution for $k$ at a fixed $E_f$,
\begin{align}
	\delta_{k,\beta} = -\beta \frac{\sqrt{3}}{2} \frac{\tilde{C}\vert \sin2\phi\vert  (k_{\alpha,\theta=\pi/2})^2}{2B_1 k_{\alpha,\theta=0} + A_1} \times \vert \delta_\theta\vert,
\end{align}
which indicate the exact crossing along $k_z$ is linear, as illustrated in Fig.~\ref{sm-fig4} (b) around the $R-Z$ line.

\begin{figure}[!htbp]
	\centering
	\includegraphics[width=\linewidth]{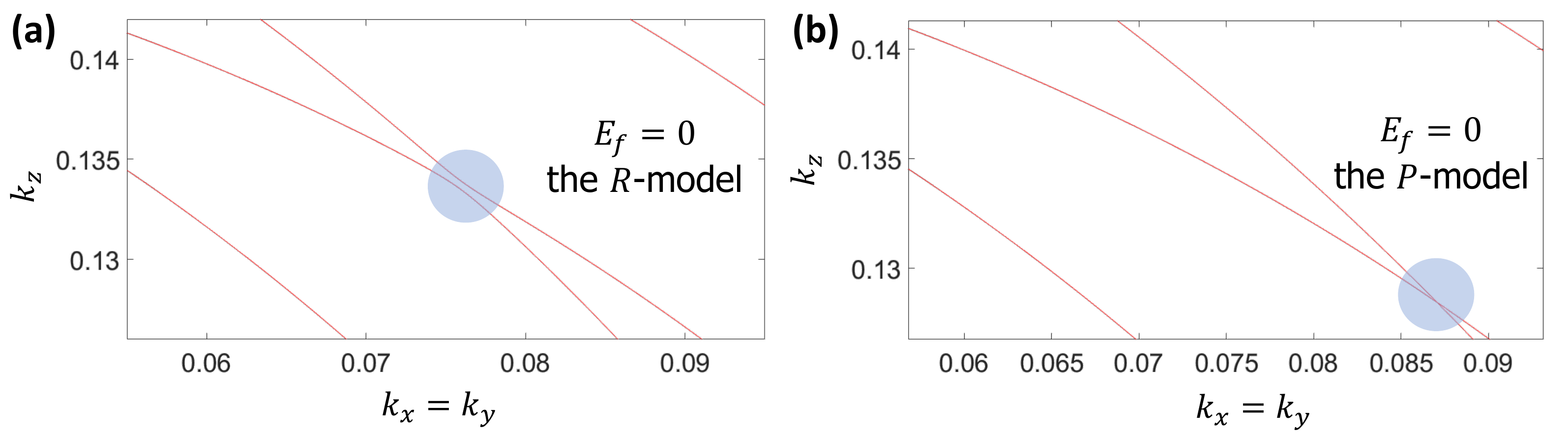}
	\caption{Comparison of the Fermi surfaces with the $R$-model in (a) and the $P$-model in (b) around the quasi nodal points in the $\Gamma-R-M$ plane. (a) shows a tiny gap in the blue circle, and (b) shows the exact quasi-symmetry protected degeneracy.
	}	
	\label{sm-fig5}
\end{figure}

In addition, we discuss the emergent nodal lines on Fermi surfaces for the $P$-model. The comparison of the FSs between the 8-band ${\bf k}\cdot{\bf p}$ effective Hamiltonian, the $R$-model (e.g.~see Eq.~\eqref{eq-R-model-H0-soc-k2-ksoc} in Sec.~\ref{Appendix-A-3}) and the first-order-perturbation 4-band Hamiltonian, the $P$-model (e.g.~see Eq.~\eqref{eq-eff-ham-p-model} in Sec.~\ref{Appendix-C-1}) are shown in Fig.~\ref{sm-fig5}. The $P$-model shows exact degeneracy at non-high-symmetry points. The crossings from two bands can be obtained from the constraint equation
\begin{align}
	E_{\alpha=+1,\beta=-1}=E_{\alpha=-1,\beta=+1} ,
\end{align}
of which the solution 
\begin{align} \label{sm-eq-nodal-plane-solution-1}
	\lambda_0 = \tfrac{\sqrt{3}}{4}\tilde{C}k^2 \vert \sin2\phi\sin2\theta\sin\theta\vert = \sqrt{3}\tilde{C} \frac{\vert k_xk_yk_z\vert}{k}
\end{align}
generally leads to nodal planes. The obtained nodal planes do not intersect with high symmetry planes, e.g. $k_x=0$ or $k_y=0$ or $k_z=0$. At the Fermi energy, we require an additional constraint equation
\begin{align}
	E_f=E_{\alpha=+1,\beta=-1}=E_{\alpha=-1,\beta=+1}. 
\end{align}
By solving the above equation, we find the curve equation for the emergent nodal lines at the Fermi energy
\begin{align}\label{eq-curve-equ-nodal-line}
	\begin{split}
		\sqrt{3}\tilde{C} \frac{\vert k_xk_yk_z\vert}{k^3} = \tfrac{\sqrt{3}}{4}\tilde{C} \vert \sin2\phi\sin2\theta\sin\theta\vert  = \frac{A_1^2\lambda_0}{2(E_f-C_0)^2} \Big{\lbrack} 1 + \frac{2B_1(E_f-C_0)}{A_1^2} + \sqrt{1+\frac{4B_1(E_f-C_0)}{A_1^2}}  \Big{\rbrack}.
	\end{split}
\end{align}
For a fixed Fermi energy $E_f$, solving Eq.~\eqref{eq-curve-equ-nodal-line} generally gives rise to a line solution in the $\theta-\phi$ plane. It represents a nodal line because of Eq.~\eqref{sm-eq-nodal-plane-solution-1}: solving $\theta$ and $\phi$ at fixed $E_f$ from Eq.~\eqref{eq-curve-equ-nodal-line} will fix $k$ simultaneously. However, the existence of such a line solution for Eq.~\eqref{eq-curve-equ-nodal-line} depends on the value of $E_f$. For a critical $E_f$, there is no line solution from the curve equation~\eqref{eq-curve-equ-nodal-line}, instead, we can only get a point solution. To find this minimal $E_f$, we set $\phi=\pi/4$, and notice that the function $\vert\sin2\theta\sin\theta\vert$ reaches its maximum when $\theta\to \arcsin \sqrt{\tfrac{2}{3} }$, we find the minimal $E_f$ as
\begin{align}
	E_{f,c} = \frac{A_1 \sqrt{C_4\lambda _0}+B_1 \lambda _0+C_0 C_4}{C_4}.
\end{align}
Thus, $E_{f,c}$ is the energy for the single nodes with twofold degeneracy on the corresponding nodal planes. Based on this analysis, we realize that the quasi-nodal-line will emerge into single nodes when decreasing the Fermi energy. For $E_f<E_{f,c}$, no solution of Eq.~\eqref{eq-curve-equ-nodal-line} can be obtained anymore. Therefore, one can conclude that each nodal line emerges into a single point after decreasing $E_f$ down to $E_{f,c}$. Moreover, when the nodal plane is split due to high-order perturbations, this point becomes a Weyl point pinned along the (111)-axis. Thus, we conclude that, nodal lines exist only when $E_f \ge E_{f,c}$.

\subsection{The second-order perturbation: gap out the quasi-nodal planes}
\label{Appendix-C-3}

Next we consider the second-order perturbation corrections for the 4-band $P$-model (e.g.~see Eq.~\eqref{eq-eff-ham-p-model} in Sec.~\ref{Appendix-C-1}), which can open a tiny gap for the emergent nodal lines obtained from Eq.~\eqref{eq-curve-equ-nodal-line} at generic momenta. Recall that the first order perturbation Hamiltonian, the $P$-model, is based on the basis in Eq.~\eqref{sm-eq-p-model-basis-spinful}
\begin{align}
	\vert \Psi_{+}\rangle  \triangleq \vert \Psi_{upper}\rangle =\{ (1,0)^T \otimes \vert \Psi_{A+}(\theta,\phi) \rangle, 
	(1,0)^T \otimes \vert \Psi_{B+}(\theta,\phi) \rangle, 
	(0,1)^T \otimes \vert \Psi_{A+}(\theta,\phi) \rangle, 
	(0,1)^T \otimes \vert \Psi_{B+}(\theta,\phi)\rangle \},  
\end{align}
which are all eigen-states of $\mathcal{H}_1$ with the same eigen-energy $E_+=C_0 + A_1 k$. And the spinless wavefunctions $\vert \Psi_{A/B+}(\theta,\phi) \rangle$ are given by Eq.~\eqref{eq-basis-wf-P-model},
\begin{subequations}
\begin{align}
	\vert \Psi_{A+}(\theta,\phi) \rangle &= \tfrac{1}{\sqrt{2}} \left( \cos\theta\cos\phi - i\sin\phi, -\cos\theta\sin\phi-i\cos\phi, 0 , \sin\theta\right)^T , \\
	\vert \Psi_{B+}(\theta,\phi) \rangle &= \tfrac{1}{\sqrt{2}} \left( -i\sin\theta\cos\phi, i\sin\theta\sin\phi,1,i\cos\theta\right) ^T,
\end{align}
\end{subequations}
where $A/B$ represent the eigen-values of $\vec{\mathcal{M}}\cdot \vec{n}$. The second-order perturbation theory has been presented in the supplementary materials in Ref.~\cite{guo_arxiv_2021}. To make sure the completeness of this appendix, we repeat the discussion of the second-order perturbation in this sub-section. Here, we consider the inter-band correction via second-order perturbation. The fourfold degenerate eigen-states of $\mathcal{H}_1$ with lower energy ($E_-=C_0 - A_1k$) are given by
\begin{align}\label{sm-eq-lower-four-bands}
	\vert \Psi_{lower}\rangle= \left\{ (1,0)^T \otimes  \vert \Psi_{B-}(\theta,\phi) \rangle,  (1,0)^T \otimes  \vert \Psi_{A-}(\theta,\phi) \rangle,   (0,1)^T \otimes\vert \Psi_{B-}(\theta,\phi) \rangle,  (0,1)^T \otimes \vert \Psi_{A-}(\theta,\phi) \rangle  \right\},
\end{align}
which are all eigen-states of of $\mathcal{H}_1$ with the same eigen-energy $E_-=C_0 - A_1 k$. And the spinless wavefunctions $\vert \Psi_{A/B-}(\theta,\phi)\rangle$ are given by Eq.~\eqref{eq-basis-wf-P-model-22},
\begin{subequations}
\begin{align}
	\vert \Psi_{A-}(\theta,\phi) \rangle &=
	\tfrac{1}{\sqrt{2}} \left( i\sin\theta\cos\phi, -i\sin\theta\sin\phi,1,-i\cos\theta \right), \\
	\vert \Psi_{B-}(\theta,\phi) \rangle &= \tfrac{1}{\sqrt{2}} \left( \cos\theta\cos\phi + i\sin\phi, -\cos\theta\sin\phi+i\cos\phi, 0 , \sin\theta \right)^T.
\end{align}
\end{subequations}

Therefore, the second-order perturbed Hamiltonian is given by 
\begin{align}\label{sm-eq-2nd-perturbation}
	\Delta\mathcal{H}_{P}^{eff(2)} ({\bf k}) = \frac{1}{\Delta E} \left( \langle \Psi_{upper} \vert (\mathcal{H}_{soc} + \mathcal{H}_{2}(\mathbf{k})) \hat{P}_{lower} ( \mathcal{H}_{soc} + \mathcal{H}_{2}(\mathbf{k}) ) \vert \Psi_{upper}\rangle  \right),
\end{align}
where $\Delta E=E_+-E_-=2A_1k$ is the energy difference between the upper-energy-band the lower-energy-band, and the projection operator $\hat{P}_{lower}=\vert \Psi_{lower} \rangle \langle \Psi_{lower}\vert$ onto the lower four bands in Eq.~\eqref{sm-eq-lower-four-bands}. The mixed terms of $\mathcal{H}_{soc}$ and $\mathcal{H}_{2}(\mathbf{k})$ for the second
order perturbation are given by
\begin{align}\label{sm-eq-ham-p-eff2}
	\Delta\mathcal{H}_{P}^{eff(2)} ({\bf k}) = \frac{1}{\Delta E} \left( \langle \Psi_{upper} \vert \mathcal{H}_{soc}  \vert \Psi_{lower} \rangle \langle \Psi_{lower} \vert \mathcal{H}_{2}(\mathbf{k}) \vert \Psi_{upper} \rangle  \right) + \text{h.c.},
\end{align}
with the matrix elements of $\Delta\mathcal{H}_P^{eff(2)}$
\begin{align}
		\left \lbrack \Delta\mathcal{H}_P^{eff(2)}\right\rbrack _{1,1} &=
		2 \sin ^2(\theta ) \sin (\phi ) \cos (\phi ) \left(\sin ^3(\theta ) \sin (\phi )+\sin (\theta ) \cos ^2(\theta ) \cos (\phi )-\cos ^3(\theta )\right), \\
		\left \lbrack \Delta\mathcal{H}_P^{eff(2)}\right\rbrack _{1,2} &=
		\sin ^2(\theta ) \cos (\phi ) (\sin ^2(\theta ) \sin (2 \phi )+2 i \cos (\theta ) \sin (\phi ) (\cos (\theta ) (\cos (\theta ) \cos (\phi ) \nonumber \\
		&+\sin (\theta ))+\sin (\phi ) (\sin ^2(\theta )+i \cos (\theta ) ) ) ),  \\
		\left \lbrack \Delta\mathcal{H}_P^{eff(2)}\right\rbrack _{1,3} &=
		\tfrac{1}{4} e^{-i \phi } \sin (\theta ) \cos (\theta ) (\sin (2 \phi ) (\sin (3 \theta ) \sin (\phi )+4 \cos ^3(\theta ))-8 \sin (\theta ) \cos ^2(\theta ) \nonumber \\ & \sin (\phi ) \cos ^2(\phi ) \nonumber 
		+8 i \cos (\theta ) \sin ^2(\phi )+(-\tfrac{3}{2}+6 i) \sin (\theta ) \cos (\phi )+(\tfrac{3}{2}+2 i) \sin (\theta ) \\ & \cos (3 \phi )), \\
		~
		\left \lbrack \Delta\mathcal{H}_P^{eff(2)}\right\rbrack _{1,4} &= e^{-i \phi } \sin (\theta ) ((\cos (\theta )-1) (\cos ^2(\theta ) \cos (\phi )\nonumber+i \sin (\theta ) \sin (\phi ) \cos (\phi ) (-\cos ^2(\theta )  \nonumber\\ &+\sin (\theta ) (\cos (\theta )+i) \cos (\phi )) +\sin (\theta ) \cos (\theta ) \sin ^2(\phi ) (-1-i \sin (\theta ) \cos (\phi )))\nonumber \\ &-2 i \cos ^2(\tfrac{\theta }{2}) (-i \cos ^2(\theta ) \cos (\phi ) \cos (2 \phi )+\sin (\phi ) \cos (\phi ) ((\cos ^3(\theta )+\cos (\theta )) \nonumber \\
		& \cos (\phi )-i \sin ^2(\theta ) \cos (\phi )+\sin (\theta ) \cos ^2(\theta )) +\sin (\theta ) \cos (\theta ) \sin ^2(\phi ) (\sin (\theta ) \nonumber \\&\cos (\phi )+i)))
\end{align}
and
\begin{align}
	\left \lbrack \Delta\mathcal{H}_P^{eff(2)}\right\rbrack _{2,2} &=
		-\tfrac{1}{4} \sin ^5(\theta ) \sin (\phi ) \sin (2 \phi ) \nonumber \\&~~~~\left(\cot ^2(\theta ) (-4 \cot (\theta ) \csc (\phi )+4 \cot (\phi )-1)+\csc ^2(\theta )+3\right), \\
		\left \lbrack \Delta\mathcal{H}_P^{eff(2)}\right\rbrack _{2,3} &=
		\tfrac{1}{2} e^{-i \phi } \sin (\theta ) \cos (\theta ) (-4 \sin (\phi ) (\sin (\theta ) \sin (\phi )\nonumber +(2+2 i) \sin ^2(\frac{\theta }{2}) \cos ^2(\phi ))+\nonumber \\
	&\cos ^2(\theta ) \sin (2 \phi ) (-4 i \sin ^2(\tfrac{\theta }{2}) \cos (\phi )+2 i \sin (\theta )+\csc (\phi ))+\cos (\theta ) \csc (\phi ) \nonumber \\
	&(-(2+2 i) \sin ^2(\tfrac{\theta }{2}) \sin ^2(2 \phi ) +\sin ^2(\tfrac{\theta }{2}) \sin (4 \phi )+4 i \sin ^2(\theta ) \sin ^3(\phi ) \cos (\phi )\nonumber \\
	&+\sin (2 \phi ))), \\
		\left \lbrack\Delta\mathcal{H}_P^{eff(2)}\right\rbrack _{2,4} &=
		-\tfrac{1}{8} e^{-i \phi } \sin (\theta ) \cos (\theta ) (8 \cos ^3(\theta ) \sin (2 \phi )-16 \sin (\theta ) \cos ^2(\theta ) \sin (\phi ) \cos ^2(\phi )\nonumber \\&+16 i \cos (\theta ) \sin ^2(\phi ) +\cos (\phi ) (4 \sin (3 \theta ) \sin ^2(\phi )+(-3+12 i) \sin (\theta ))\nonumber \\
	&+(3+4 i) \sin (\theta ) \cos (3 \phi )),
\end{align}
and
\begin{align}
	\left \lbrack \Delta\mathcal{H}_P^{eff(2)}\right\rbrack _{3,3} &=
		-\tfrac{1}{4} \sin ^5(\theta ) \sin (\phi ) \sin (2 \phi ) \nonumber \\
	&\left(\cot ^2(\theta ) (-4 \cot (\theta ) \csc (\phi )+4 \cot (\phi )-1)+\csc ^2(\theta )+3\right), \\
		\left \lbrack \Delta\mathcal{H}_P^{eff(2)}\right\rbrack _{3,4} &=
		-i \sin ^2(\theta ) \sin (\phi ) \cos (\phi ) (2 \cos ^3(\theta ) \cos (\phi )+2 \cos ^2(\theta ) (\sin (\theta )+i \sin (\phi )) \nonumber \\
	&+\sin (\theta ) (\sin (2 \theta ) \sin (\phi )-2 i \sin (\theta ) \cos (\phi ))),
\end{align}
and
\begin{align}
	\left \lbrack \Delta\mathcal{H}_P^{eff(2)}\right\rbrack _{4,4} &=  2 \sin ^2(\theta ) \sin (\phi ) \cos (\phi ) (\sin ^3(\theta ) \sin (\phi )+\sin (\theta ) \cos ^2(\theta ) \cos (\phi )-\cos ^3(\theta )). 
\end{align}
The other parts are related by complex conjugation 
$\left \lbrack \Delta\mathcal{H}_P^{eff(2)}\right\rbrack _{i,j} =  \left \lbrack\Delta\mathcal{H}_P^{eff(2)}\right\rbrack _{j,i}^\ast$.

\subsection{Additional terms due to the linear $k$ SOC}
\label{Appendix-C-4}

In the first-order perturbation for the $P$-model in Sec.~\ref{Appendix-C-1}, we only consider the contribution from on-site SOC Hamiltonian. Here, we further discuss the additional first-order perturbation corrections stemming from the linear-$k$ SOC in Eq.~\eqref{eq-linear-k-soc-ham}, 
\begin{align}
	\begin{split}
		\mathcal{H}_{k,\text{soc}}({\bf k}) &= \lambda_1(k_x s_x + k_y s_y + k_zs_z)\otimes \sigma_0 \tau_0 
		+ \lambda_2 ( k_x s_y\sigma_x\tau_x - k_y s_z\sigma_0\tau_x + k_z s_x\sigma_x\tau_0 ) \\
		&+\lambda_3 ( k_y s_x\sigma_x\tau_x - k_z s_y\sigma_0\tau_x + k_x s_z\sigma_x\tau_0 ) 
		+ \lambda_4(k_x s_y\sigma_x\tau_z - k_ys_z\sigma_y\tau_y -k_zs_x\sigma_z\tau_x) \\
		&+\lambda_5(k_ys_x\sigma_x\tau_z - k_z s_y\sigma_y\tau_y - k_xs_z\sigma_z\tau_x) 
		+\lambda_6(k_x s_y\sigma_z\tau_0 + k_y s_z\sigma_0\tau_z + k_z s_x\sigma_z\tau_z) \\
		&+\lambda_7 (k_ys_x\sigma_z\tau_0 + k_z s_y\sigma_0\tau_z + k_x s_z\sigma_z\tau_z).
	\end{split}
\end{align}
Then, we only need to compute $\langle \Psi_{+} \vert \mathcal{H}_{k,\text{soc}}({\bf k})  \vert \Psi_+\rangle $ based on the basis in Eq.~\eqref{eq-four-basis-spin},
\begin{align}
	\vert \Psi_{+}\rangle  =\{ (1,0)^T \otimes \vert \Psi_{A+}(\theta,\phi) \rangle, 
	(1,0)^T \otimes \vert \Psi_{B+}(\theta,\phi) \rangle, 
	(0,1)^T \otimes \vert \Psi_{A+}(\theta,\phi) \rangle, 
	(0,1)^T \otimes \vert \Psi_{B+}(\theta,\phi)\rangle \},  
\end{align}
which are all eigen-states of $\mathcal{H}_1({\bf k})$ with the same eigen-energy $E_+=C_0 + A_1 k$. And the spinless wavefunctions $\vert \Psi_{A/B+}(\theta,\phi) \rangle$ are given by Eq.~\eqref{eq-basis-wf-P-model}.

First, a term denoted by $\lambda_1$ only renormalizes $\lambda_0$ as $\lambda_0 \to \lambda_0 +\lambda_1 k $.  Second, all the others terms denoted by $\lambda_{2,3,4,5,6,7}$ generally gap the quasi-nodal plane obtained from Eq.~\eqref{sm-eq-nodal-plane-solution-1}, i.e., $\lambda_0 = \tfrac{\sqrt{3}}{4}\tilde{C}k^2 \vert \sin2\phi\sin2\theta\sin\theta\vert$. For an illustration, we consider the $\lambda_2$ term as an example, and the projected matrix is
\begin{align}\label{eq-p-model-linear-ksoc-lambda2}
	\mathcal{H}_{k,\lambda_2,eff}(k,\theta,\phi) = \lambda_2k \left(
	\begin{array}{cccc}
		h_{11} & h_{12} & h_{13} & h_{14} \\
		h_{21} & h_{22} & h_{23} & h_{24} \\
		h_{31} & h_{32} & h_{33} & h_{34} \\
		h_{41} & h_{42} & h_{43} & h_{44} \\
	\end{array} \right),
\end{align}
where the spherical coordinator is used, and 
\begin{align}
	\begin{split}
		h_{11} &= - \sin ^3(\theta ) \sin ^2(\phi ) \cos (\phi ), \\
		h_{12} &=  \frac{1}{2} \sin ^2(\theta ) \sin (2 \phi ) (-\cos (\phi )-i \cos (\theta ) \sin (\phi )), \\
		h_{13} &= - \frac{1}{2}\sin (2\theta ) (\cos (\theta ) \sin (\phi ) +i \sin (\theta ) \cos ^2(\phi ) ), \\
		h_{14} &= \sin (\theta ) \cos (\theta ) (\cos (\theta ) \cos ^2(\phi ) 
		+i \sin (\phi ) (\sin (\theta )+\cos (\phi )) ),\\
		h_{22} &= \sin ^3(\theta ) \sin ^2(\phi ) \cos (\phi ), \\
		h_{23} &= -\sin (\theta ) \cos (\theta ) (\cos (\theta ) \cos ^2(\phi ) 
		-i \sin (\phi ) (\cos (\phi )-\sin (\theta ))) ,\\
		h_{24} &= \frac{1}{2}\sin (2\theta )  \left(\cos (\theta ) \sin (\phi )+i \sin (\theta ) \cos ^2(\phi )\right), \\
		h_{33} &= \sin ^3(\theta ) \sin ^2(\phi ) \cos (\phi ),\\
		h_{34} &= \frac{1}{2}\sin (2\phi ) \sin ^2(\theta )  (\cos (\phi )+i \cos (\theta ) \sin (\phi )),\\
		h_{44} &= -\sin ^3(\theta ) \sin ^2(\phi ) \cos (\phi ). 
	\end{split}
\end{align}
The remaining parts can be obtained by complex conjugation. We find 
\begin{align}
	[\mathcal{M}_{eff}, \mathcal{H}_{k,\lambda_2,eff}({\bf k}) ]\neq 0 \text{ for nonzero } {\bf k},
\end{align}
where the U(1) quasi-symmetry $\mathcal{M}_{eff}$ is given by $\mathcal{H}_{\text{soc},+}^{eff(1)}({\bf k}) $ or $\mathcal{H}_{k^2,+}^{eff(1)}({\bf k})$. This suggests that the nodal planes are generally gapped by involving the linear $k$ SOC terms. But this effect is negligible.

\section{Approach II for the hierarchy of the quasi-symmetry}
\label{Appendix-D}

In this section, we discuss the approach II for the hierarchy of the quasi-symmetry and the perturbation theory based on the solution of $\mathcal{H}_1({\bf k})+\mathcal{H}_{\text{soc}}$ for the $\mathcal{H}_2({\bf k})$. Please note that the algebra of the orbital SU(2) quasi-symmetry defined in Eq.~\eqref{sm-eq-qs-m1m2m3} in Sec.~\ref{appendix-C-0},
\begin{align} 
	\mathcal{M}_{1,2,3} = \tfrac{1}{2}\{ s_0\sigma_y\tau_z, s_0\sigma_y\tau_x, s_0\sigma_0\tau_y \},
\end{align}
which all commute with $\mathcal{H}_1({\bf k})+\mathcal{H}_{\text{soc}}$, but do not commute between themselves. And $[\mathcal{M}_{i}, \mathcal{M}_{j} ] = i \epsilon_{i,j,k} \mathcal{M}_k$. Moreover, we can also defined the rotation for this orbital SU(2) quasi-symmetry group
\begin{align}
\begin{split}
	\mathcal{U}(\theta,\phi) &= e^{i \theta \mathcal{M}_3} e^{i \phi \mathcal{M}_2} \\
	&=   \cos \left(\tfrac{\theta }{2}\right) \cos \left(\tfrac{\phi }{2}\right) s_0\sigma_0\tau_0  + i \sin \left(\tfrac{\theta }{2}\right) \cos \left(\tfrac{\phi }{2}\right) \mathcal{M}_3 
	+ i\cos \left(\tfrac{\theta }{2}\right) \sin \left(\tfrac{\phi }{2}\right) \mathcal{M}_2 + i \sin \left(\tfrac{\theta }{2}\right) \sin \left(\tfrac{\phi }{2}\right) \mathcal{M}_1.
\end{split}
\end{align}
Clearly, $\mathcal{U}(\theta,\phi)$ commutes with $s_0\otimes \mathcal{H}_1(\mathbf{k}) + \mathcal{H}_{\text{soc}}$ for any values of $\theta$ and $\phi$. The $\mathcal{U}(0,\frac{\pi}{2}) $ can rotate $\mathcal{M}_1$ to $\mathcal{M}_3$,
\begin{align}
	\begin{split}
		&\mathcal{U}(0,\frac{\pi}{2})  = e^{i \frac{\pi}{2} \mathcal{M}_2} = \frac{\sqrt{2}}{2} s_0\otimes \left( \sigma_0\tau_0 + i\sigma_y\tau_x \right) \\	
		\Rightarrow  \quad
		&\mathcal{U}(0,\frac{\pi}{2}) \mathcal{M}_1 \mathcal{U}^\dagger(0,\frac{\pi}{2})  = \mathcal{M}_3.
	\end{split}
\end{align}

For better readability of this section, here, we first repeat the linear-$k$ Hamiltonian, on-site SOC Hamiltonian and $k^2$-order Hamiltonian, 
\begin{subequations}
\begin{align} 
	\mathcal{H}_{1}(\mathbf{k}) &=  C_0\sigma_0\tau_0 + 2A_1 (\mathbf{k}\cdot\mathbf{L}) , \\
	\mathcal{H}_{\text{soc}} &= 4\lambda_0 ( \mathbf{S} \cdot \mathbf{L}), \\
	\mathcal{H}_{2}(\mathbf{k}) &=\mathcal{H}_{2, \mathcal{M}_1}(\mathbf{k}) +\mathcal{H}_{2, \mathcal{M}_2}(\mathbf{k}) +\mathcal{H}_{2, \mathcal{M}_3}(\mathbf{k}), 
\end{align}
\end{subequations}
where the spin angular momentum ${\bf S}=\tfrac{1}{2}(s_x, s_y, s_z)$, and the orbital angular momentum operators are ${\bf L}= (\tfrac{1}{2} \sigma_y\tau_0,  \tfrac{1}{2}\sigma_x\tau_y ,  -\tfrac{1}{2}\sigma_z\tau_y )$. And each part of $\mathcal{H}_{2}(\mathbf{k})$ is given by 
\begin{align}
	\mathcal{H}_{2,\mathcal{M}_i}(\mathbf{k})={\bf g}_i({\bf k})\cdot{\bf J}_i,
\end{align}
for $i=1,2,3$. Here we define the $k$-dependent vectors as
\begin{align}
	\begin{split}
		{\bf g}_1 ({\bf k}) &= (C_2k_xk_y, -C_3k_xk_z,C_1k_yk_z),   \\
		{\bf g}_2 ({\bf k}) &= (C_3k_xk_y,C_1k_xk_z, -C_2  k_yk_z),   \\
		{\bf g}_3 ({\bf k}) &= (C_1k_xk_y,  C_2k_xk_z, -C_3  k_y k_z). 
	\end{split}
\end{align}
and the corresponding vectors of operators 	
\begin{align}
	\begin{split}
		{\bf J}_1 &= ( \sigma_x  \tau_x, -\sigma_z\tau_x,  \sigma_0 \tau_z),  \\
		{\bf J}_2 &= (\sigma_x\tau_z,\sigma_z\tau_z,\sigma_0\tau_x),   \\
		{\bf J}_3 &= (\sigma_z\tau_0,\sigma_x\tau_0,\sigma_y\tau_y). 
	\end{split}
\end{align}
In addition, we also realize that the $k^2$ terms of $\mathcal{H}_2({\bf k})$ break this orbital SU$_{\text{o}}$(2) quasi-symmetry generated by $\{\mathcal{M}_{1,2,3}\}$ and lead to the splitting of all bands. However, different parts of the entire $k^2$-terms can lead to the reduction from SU$_{\text{o}}$(2) to a orbital U(1). To show that, as we discussed in the main text, we find that
\begin{align}
	[{\bf J}_i,\mathcal{M}_i]=0 \text{ and }  \{{\bf J}_i, \mathcal{M}_j\}=0 \text{ for }  i\neq j,
\end{align}
which implies 
\begin{align}
	[\mathcal{H}_{2,\mathcal{M}_i}({\bf k}),\mathcal{M}_i]=0.
\end{align}

\subsection{The algebra for the orbital SU(2) quasi-symmetry for the $k^2$-order Hamiltonian}
\label{appendix-C-0}

In this subsection, we discuss the algebra of the orbital SU(2) quasi-symmetry in Eq.~\eqref{sm-eq-qs-m1m2m3}. To show the generality of the breaking of the orbital SU(2) quasi-symmetry down to U(1) quasi-symmetry by the $k^2$-order Hamiltonian. We can rewrite the $\mathcal{H}_2({\bf k})$ into the form
\begin{align}
	\mathcal{H}_{2}(\mathbf{k}) = k_xk_y \left( \vec{C}_{\mathcal{J}} \cdot \vec{\mathcal{J}} \right) 
	+ k_xk_z \left(  \vec{C}_{\mathcal{P}} \cdot \vec{\mathcal{P}} \right) 
	+ k_yk_z \left(  \vec{C}_{\mathcal{Q}} \cdot \vec{\mathcal{Q}} \right),
\end{align}
where the three ${\bf k}$-independent parameter-vectors are 
\begin{align}
	\begin{split}
		\vec{C}_{\mathcal{J}} &=(C_2, -C_3, C_1), \\
		\vec{C}_{\mathcal{P}} &=(-C_3, C_1, C_2), \\
		\vec{C}_{\mathcal{Q}} &=(C_1, -C_2, -C_3), \\		
	\end{split}
\end{align}
and the corresponding operator-vectors are given by
\begin{align}
	\begin{split}
		\vec{\mathcal{J}} &= (\sigma_x\tau_x, -\sigma_x\tau_z, \sigma_z\tau_0), \\ 
		\vec{\mathcal{P}} &= (-\sigma_z\tau_x, \sigma_z\tau_z, \sigma_x\tau_0), \\ 
		\vec{\mathcal{Q}} &= (\sigma_0\tau_z, \sigma_0\tau_x, \sigma_y\tau_y).
	\end{split}
\end{align}
Moreover, we notice that 
\begin{align}
	\begin{cases}
		\; [ \mathcal{J}_a, \mathcal{M}_b ] = i \epsilon_{abc} \mathcal{J}_c, \\
		\; [ \mathcal{P}_a, \mathcal{M}_b ] = i \epsilon_{abc} \mathcal{P}_c, \\
		\; [ \mathcal{Q}_a, \mathcal{M}_b ] = i \epsilon_{abc} \mathcal{Q}_c, 
	\end{cases}	
\end{align}
where $\epsilon_{abc}$ is the three-dimensional Levi-Civita symbol with $a,b,c=1,2,3$. Therefore, for arbitrary real normalized vector $\vec{n}=(n_1,n_2,n_3)$, we have the following commutation relations 
\begin{align}
	[ \vec{\mathcal{J}}\cdot \vec{n},  \vec{\mathcal{M}}\cdot \vec{n} ] = [ \vec{\mathcal{P}}\cdot \vec{n},  \vec{\mathcal{M}}\cdot \vec{n} ]  = [ \vec{\mathcal{Q}}\cdot \vec{n},  \vec{\mathcal{M}}\cdot \vec{n} ] =0.
\end{align}
These can be easily shown, for example, 
\begin{align}
	\begin{split}
		[\vec{\mathcal{J}}\cdot \vec{n},  \vec{\mathcal{M}}\cdot \vec{n} ]  
		&= \sum_{a=1}^{3} \sum_{b=1}^{3} n_a n_b [\mathcal{J}_a, \mathcal{M}_b] = \sum_{a=1}^{3} \sum_{b=1}^{3} n_a n_b \left( i\epsilon_{abc} \mathcal{J}_c \right) \\
		&=\sum_{a=1}^{3} \sum_{b=a+1}^{3} n_a n_b \left( i\epsilon_{abc} \mathcal{J}_c  + i\epsilon_{bac} \mathcal{J}_c  \right) =0.
	\end{split}
\end{align}
Here we have used $\epsilon_{abc} + \epsilon_{bac}=0$. Therefore, we find a general symmetry-breaking case with the orbital SU(2) quasi-symmetry down to the U(1) quasi-symmetry. For any three normalized and orthogonal vector, $\vec{n}$, $\vec{n}'$ and $\vec{n}''$, satisfy $\vert \vec{n}\vert = \vert \vec{n}'\vert =\vert \vec{n}''\vert =1$ and $\vec{n} \perp \vec{n}'$, $\vec{n} \perp \vec{n}''$ and $\vec{n}' \perp \vec{n}''$. Then we can do the projection for the $k^2$-order Hamiltonian,
\begin{align} \label{sm-eq-ham-2-nnn}
	\mathcal{H}_{2} ({\bf k})  = \mathcal{H}_{2,\vec{n}} ({\bf k})  + \mathcal{H}_{2,\vec{n}'} ({\bf k})  + \mathcal{H}_{2,\vec{n}''} ({\bf k}) ,
\end{align}
where we project the parameter-vectors ($\vec{C}_{\mathcal{J}}, \vec{C}_{\mathcal{P}}, \vec{C}_{\mathcal{Q}}$) into the $\{\vec{n}, \vec{n}', \vec{n}''\}$ space, 
\begin{subequations}
	\begin{align}
		\vec{C}_{\mathcal{J}}  = \vec{n} (\vec{C}_{\mathcal{J}}\cdot \vec{n}) +  \vec{n}' (\vec{C}_{\mathcal{J}}\cdot \vec{n}') +  \vec{n}'' (\vec{C}_{\mathcal{J}}\cdot \vec{n}''), \\
		\vec{C}_{\mathcal{P}}  = \vec{n} (\vec{C}_{\mathcal{P}}\cdot \vec{n}) +  \vec{n}' (\vec{C}_{\mathcal{P}}\cdot \vec{n}') +  \vec{n}'' (\vec{C}_{\mathcal{P}}\cdot \vec{n}''), \\
		\vec{C}_{\mathcal{Q}}  = \vec{n} (\vec{C}_{\mathcal{Q}}\cdot \vec{n}) +  \vec{n}' (\vec{C}_{\mathcal{Q}}\cdot \vec{n}') +  \vec{n}'' (\vec{C}_{\mathcal{Q}}\cdot \vec{n}''),
	\end{align}
\end{subequations}
Therefore, the first term in Eq.~\eqref{sm-eq-ham-2-nnn} is given by
\begin{align}
	\mathcal{H}_{2,\vec{n}} ({\bf k}) = k_xk_y (\vec{C}_{\mathcal{J}} \cdot \vec{n}) \times \left( \vec{n} \cdot \vec{\mathcal{J}} \right) 
	+ k_xk_z (\vec{C}_{\mathcal{P}} \cdot \vec{n}) \times \left(  \vec{n} \cdot \vec{\mathcal{P}} \right) 
	+ k_yk_z (\vec{C}_{\mathcal{Q}} \cdot \vec{n}) \left(  \vec{n} \cdot  \vec{\mathcal{Q}} \right),
\end{align}
which commutes with $\mathcal{M}\cdot \vec{n}$,
\begin{align}
	[ \mathcal{M}\cdot \vec{n} , \mathcal{H}_{2,\vec{n}} ({\bf k}) ] =0.
\end{align}
Especially, in the main text, we have mentioned three cases,
\begin{itemize}
	\item $\vec{n}=(1,0,0)$. The $\mathcal{H}_{2,\vec{n}} ({\bf k})$ Hamiltonian is given by $\mathcal{H}_{2,\mathcal{M}_1}({\bf k})$ in Eq.~\eqref{sm-eq-h2-m1m2m3-kk}.
	\item $\vec{n}=(0,1,0)$. The $\mathcal{H}_{2,\vec{n}} ({\bf k})$ Hamiltonian is given by $\mathcal{H}_{2,\mathcal{M}_2}({\bf k})$ in Eq.~\eqref{sm-eq-h2-m1m2m3-kk}.
	\item $\vec{n}=(0,0,1)$. The $\mathcal{H}_{2,\vec{n}} ({\bf k})$ Hamiltonian is given by $\mathcal{H}_{2,\mathcal{M}_3}({\bf k})$ in Eq.~\eqref{sm-eq-h2-m1m2m3-kk}.
\end{itemize}
Moreover, the U(1) quasi-symmetry protected nodal-plane for $\mathcal{H}_{2,\mathcal{M}_1}$ is discussed in the main text. More details will be discussed in the following Sec.~\ref{appendix-D-1}, Sec.~\ref{sec-Appendix-D-3}, and Sec.~\ref{sec-Appendix-D-4}.

\subsection{Analytical solutions by using the U(1) quasi-symmetry}
\label{appendix-D-1}

Furthermore, we show the important role of the orbital SU(2) quasi-symmetry operators (i.e.~$\mathcal{M}_{1,2,3}$) in analytically solving the eigen-state problem. Here, we take $\mathcal{H}_1(\mathbf{k})$ in Eq.~\eqref{eq-ham0-k-order} or Eq.~\eqref{eq-ham0-k-order-Lk} as an example,
\begin{align}
	\mathcal{H}_1(\mathbf{k}) = C_0 +  A_1 (k_x \sigma_y\tau_0 + k_y \sigma_x\tau_y - k_z\sigma_z\tau_y),
\end{align}
which commutes with the orbital SU(2) quasi-symmetry. Here we choose the eigen-states of $\mathcal{H}_1(\mathbf{k})$ to be the common eigen-state of $\mathcal{M}_3=\sigma_0\tau_y$ in Eq.~\eqref{sm-eq-qs-m1m2m3}. It is equivalently to apply a unitary transformation 
\begin{align} \label{sm-eq-trans-unit-m3}
	\mathcal{U} = \frac{1}{\sqrt{2}}s_0\otimes \sigma_0\otimes \begin{pmatrix}
		1	& -i \\ 
		1	& i
	\end{pmatrix},
\end{align}
to the $R$-model, which only leads to a rotation in the $(\tau_x,\tau_y,\tau_z)$ subspace
\begin{align} \label{eq-rotation-unitary-taus}
	\mathcal{U} \tau_0 \mathcal{U}^\dagger = \tau_0, 
	\mathcal{U} \tau_x \mathcal{U}^\dagger = \tau_y, 
	\mathcal{U} \tau_y \mathcal{U}^\dagger = \tau_z, 
	\mathcal{U} \tau_z \mathcal{U}^\dagger = \tau_x. 
\end{align}
Therefore, the quasi-symmetry ${\cal M}_3$ becomes ${\cal U}{\cal M}_3{\cal U}^\dagger = \sigma_0 \tau_z$. Note that the spin Pauli matrix is dropped here. As a result, the linear-$k$ Hamiltonian becomes
\begin{subequations} \label{sm-eq-ham1-m3-solution}
	\begin{align}
		{\cal U}[ \mathcal{H}_1(\mathbf{k}) - C_0] {\cal U}^\dagger &=  \left[ \mathcal{H}_{1,\mathcal{M}_3=+1}(\mathbf{k}) \right]_{2\times2} \oplus  \left[ \mathcal{H}_{1,\mathcal{M}_3=-1}(\mathbf{k}) \right]_{2\times2}  \nonumber \\
		&= \begin{pmatrix}
			\mathcal{H}_{1,\mathcal{M}_3=+1}(\mathbf{k}) & 0 \\
			0 & \mathcal{H}_{1,\mathcal{M}_3=-1}(\mathbf{k})
		\end{pmatrix}, \\
		\mathcal{H}_{1,\mathcal{M}_3=+1}(\mathbf{k}) &= A_1 (k_x \sigma_y + k_y \sigma_x - k_z \sigma_z), \\ 
		\mathcal{H}_{1,\mathcal{M}_3=-1}(\mathbf{k})  &= A_1 (k_x \sigma_y - k_y \sigma_x + k_z \sigma_z).
	\end{align}
\end{subequations}
It is easy to analytically find the eigen-states of the 2-by-2 Hamiltonian $\mathcal{H}_{1,\mathcal{M}_3=\pm1}(\mathbf{k})$, in the spherical coordinate with the momentum ${\bf k}=(k_x,k_y,k_z) = k(\sin\theta\cos\phi, \sin\theta\sin\phi, \cos\theta)$. The eigen-states are given by
\begin{subequations}
\begin{align}
	\mathcal{H}_{1,\mathcal{M}_3=+1}(\mathbf{k})\vert A -  \rangle = -A_1k\vert A -  \rangle , \; \text{ with } \vert A - \rangle = \frac{1}{\sqrt{2} \sqrt{1+\cos\theta}} (i (\cos\theta +1), e^{-i \phi } \sin\theta  )^T, \\
	\mathcal{H}_{1,\mathcal{M}_3=+1}(\mathbf{k})\vert A + \rangle = \;\;\;  A_1k\vert A + \rangle, \; \text{ with } \vert A + \rangle =\frac{1}{\sqrt{2} \sqrt{1-\cos\theta}} (i (\cos\theta - 1), e^{-i \phi } \sin\theta  )^T, 
\end{align}
\end{subequations}
and by substituting the replacement $(k_x, k_y, k_z) \to (k_x, -k_y, -k_z)$ (i.e.,~$(\theta,\phi)\to(\theta+\pi, \pi-\phi)$), then $\cos\theta\to-\cos\theta, \sin\theta\to-\sin\theta, e^{-i\phi}\to -e^{i\phi}$ into the above solution, it results in   
\begin{subequations}
\begin{align}
	\mathcal{H}_{1,\mathcal{M}_3=-1}(\mathbf{k})\vert B -  \rangle = -A_1k\vert B -  \rangle , \; \text{ with } \vert B - \rangle = \frac{1}{\sqrt{2} \sqrt{1-\cos\theta}} (i (-\cos\theta +1), e^{i \phi } \sin\theta  )^T, \\
	\mathcal{H}_{1,\mathcal{M}_3=-1}(\mathbf{k})\vert B + \rangle = \;\;\;  A_1k\vert B + \rangle, \; \text{ with } \vert B + \rangle =\frac{1}{\sqrt{2} \sqrt{1+\cos\theta}} (i (-\cos\theta - 1), e^{i \phi } \sin\theta  )^T,
\end{align}
\end{subequations}
where the subscripts $A(B)$ represent the eigenvalues $+1(-1)$ of the quasi-symmetry $\mathcal{M}_3$. Therefore, the eigen-state of $\mathcal{H}_1(\mathbf{k})$ are given by
\begin{subequations}
\begin{align}
	\mathcal{H}_1(\mathbf{k})\vert \Psi_{A/B+}(\theta,\phi) \rangle = E_+  \vert \Psi_{A/B+}(\theta,\phi) \rangle, \\
	\mathcal{H}_1(\mathbf{k})\vert \Psi_{A/B-}(\theta,\phi) \rangle  = E_- \vert \Psi_{A/B-}(\theta,\phi) \rangle, 
\end{align}
\end{subequations}
where $E_\pm = \pm A_1 k$ and the corresponding eigen-states are given by
\begin{subequations} \label{sm-eq-eig-wf-m3-h1}
	\begin{align}
		\vert \Psi_{A -}(\theta,\phi) \rangle &= \vert A -\rangle \otimes (-\frac{i}{\sqrt{2}}, \frac{1}{\sqrt{2}})^T,  \\
		\vert \Psi_{A +}(\theta,\phi) \rangle &= \vert A +\rangle \otimes (-\frac{i}{\sqrt{2}}, \frac{1}{\sqrt{2}})^T,  \\
		\vert \Psi_{B -}(\theta,\phi) \rangle &= \vert B -\rangle \otimes (\frac{i}{\sqrt{2}}, \frac{1}{\sqrt{2}})^T,  \\
		\vert \Psi_{B +}(\theta,\phi) \rangle &= \vert B +\rangle \otimes (\frac{i}{\sqrt{2}}, \frac{1}{\sqrt{2}})^T.
	\end{align}
\end{subequations}
Due to the presence of the orbital SU(2) symmetry, one can also find the common eigen-states of ${\cal H}_1({\bf k})$ and ${\cal M}\cdot \vec{n}$ for arbitrary real vector $\vec{n}$, which is a rotation acting on the solution in Eq.~\ref{sm-eq-eig-wf-m3-h1}. Thus, the eigen-state solution is not unique due to this twofold degeneracy, protected by the orbital SU(2) symmetry. Note that the degeneracy will be doubled if spin degeneracy is taken into account.

Similarly, we can further analytically solve $\mathcal{H}_1({\bf k})+\mathcal{H}_{\text{soc}}$ by using the U(1) quasi-symmetry generator $\mathcal{M}_3$. The $R$-model in Eq.~\eqref{eq-R-model-H0-soc-k2-ksoc} becomes
\begin{align}\label{eq-rotation-R-model}
	\mathcal{U}	\mathcal{H}_R  \mathcal{U}^\dagger =  \mathcal{U} \left[ s_0\otimes \mathcal{H}_1(\mathbf{k}) + \mathcal{H}_{\text{soc}}	+ s_0\otimes \mathcal{H}_{2}(\mathbf{k}) \right] \mathcal{U}^\dagger. 
\end{align}
The first two parts, $s_0\otimes \mathcal{H}_1({\bf k}) + \mathcal{H}_{\text{soc}}$, preserves the orbital SU(2) symmetry, while some specific terms of the $k^2$-order Hamiltonian can break the SU(2) quasi-symmetry down to U(1). This shows the hierarchy structure of the quasi-symmetry, which will be discussed in details in Sec.~\ref{Appendix-D}. Here we focus on  $s_0\otimes \mathcal{H}_1({\bf k}) + \mathcal{H}_{\text{soc}}$, which becomes block-diagonal after this unitary transformation defined in Eq.~\eqref{sm-eq-trans-unit-m3},
\begin{align}\label{eq-qsR-0+soc-total}
	\begin{split}
		\mathcal{U} \left[ s_0\otimes \mathcal{H}_1({\bf k}) + \mathcal{H}_{\text{soc}} \right] \mathcal{U}^\dagger  
		= C_0 + \begin{pmatrix}
			\mathcal{H}_{A}(\mathbf{k})	&  0 \\ 
			0	& \mathcal{H}_{B}(\mathbf{k})
		\end{pmatrix}, 
	\end{split}
\end{align}
where the index $A(B)$ represent the eigenvalues $+1(-1)$ of quasi-symmetry ${\cal M}_3$, and $\mathcal{H}_{A}(\mathbf{k})$ and $\mathcal{H}_{B}(\mathbf{k})$ are given by
\begin{align}
	\label{eq-qsR-0-soc-H+part}
	\mathcal{H}_{A}(\mathbf{k}) &=  A_1 s_0\otimes \left( k_x\sigma_y + k_y\sigma_x - k_z\sigma_z \right) 
	+ \lambda_0\left( s_x\sigma_y + s_y\sigma_x - s_z\sigma_z \right) , \\
	\label{eq-qsR-0-soc-H-part}
	\mathcal{H}_{B}(\mathbf{k}) &=  A_1 s_0\otimes \left( k_x\sigma_y - k_y\sigma_x + k_z\sigma_z \right) 
	+ \lambda_0\left( s_x\sigma_y - s_y\sigma_x + s_z\sigma_z \right) .
\end{align}
First, one can check that the Hamiltonian can be reduced back to that in Eq.~\eqref{sm-eq-ham1-m3-solution} by setting $\lambda_0=0$. In addition, $\mathcal{H}_{A}(\mathbf{k}) $ and $\mathcal{H}_{B}(\mathbf{k}) $ are related to each other by time-reversal symmetry
\begin{align}
	\mathcal{H}_{B}(\mathbf{k}) = \mathcal{T}[ \mathcal{H}_{A}(-\mathbf{k}) ] \mathcal{T}^\dagger.
\end{align}
Here $\mathcal{T} = i s_y \mathcal{K}$ with $\mathcal{K}$ the complex conjugate. Therefore, we only need to solve the four eigen-states for $\mathcal{H}_{A}(\mathbf{k})$. After straightforward calculation, the four eigen-energies of $\mathcal{H}_{A}(\mathbf{k})$ are given by
\begin{align}
	E({\bf k}) = C_0 + \left\{ \pm A_1k + \lambda_0, \quad \pm\sqrt{A_1^2k^2+4\lambda_0^2}-\lambda_0 \right\}.
\end{align}
Drop the constant $C_0$ for short, the two positive upper bands are 
\begin{subequations}
\begin{align}
	E_{A,+}(\mathbf{k}) &= A_1k + \lambda_0, \\
	E_{A,-} (\mathbf{k}) &=\sqrt{A_1^2k^2+4\lambda_0^2}-\lambda_0, 
\end{align}
\end{subequations}
which both have increasing energy as $k$ increases, and $\pm$ are the band index. Note we have assumed $A_1>0$. And we notice that $E_{A,+}(\mathbf{k}=0) = E_{A,-} (\mathbf{k}=0)$ and $E_{A,+}(\mathbf{k}) > E_{A,-} (\mathbf{k})$ for any nonzero ${\bf k}$. It indicates that the SOC-induced gap between them, $E_{A,+} (\mathbf{k})-E_{A,-} (\mathbf{k})$, approaches to $2\lambda_0$ as $k\to \infty$. By time-reversal symmetry, the four eigen-energies of $\mathcal{H}_{B}(\mathbf{k})$ are the same. Therefore, each state has twofold degeneracy at any nonzero ${\bf k}$. This is due to the presence of the orbital SU(2) symmetry.

Moreover, we can also solve the eigen wavefunctions of $\mathcal{U} \left[ s_0\otimes \mathcal{H}_1({\bf k}) + \mathcal{H}_{\text{soc}} \right] \mathcal{U}^\dagger$. For instance, for the two positive upper bands of $\mathcal{H}_{A}(\mathbf{k}) $, the corresponding wavefunctions are given by,
\begin{align}
	\mathcal{H}_{A}(\mathbf{k}) \begin{cases}
		\vert E_{A,+}(\mathbf{k}) \rangle = E_{A,+}(\mathbf{k}) \vert E_{A,+}(\mathbf{k}) \rangle,  \\
		\vert E_{A,-}(\mathbf{k}) \rangle = E_{A,-} (\mathbf{k}) \vert E_{A,-} (\mathbf{k})\rangle.
	\end{cases}
\end{align}
And the corresponding eigen-wavefunctions are given by
\begin{subequations}
	\begin{align}\label{eq-ham+-eig-wavefunc}
		&\vert E_{A,+}(\mathbf{k})\rangle = \frac{1}{\mathcal{N}_{A,+}(\mathbf{k})} \left( -i,\frac{k_x-ik_y}{k-k_z},\frac{-ik_x+k_y}{k+k_z},1 \right)^T , \\
		&\vert E_{A,-}(\mathbf{k})\rangle = \frac{1}{\mathcal{N}_{A,-}(\mathbf{k})} \Big{(} i(-A_1k_z-\lambda_0+E_{A,-}(\mathbf{k})), 
		A_1(-k_x+ik_y), A_1(-ik_x+k_y), A_1k_z-\lambda_0 + E_{A,-}(\mathbf{k})  \Big{)}^T,
	\end{align}
\end{subequations}
where the normalization factors are
\begin{subequations}
	\begin{align}
		\mathcal{N}_{A,+}(\mathbf{k}) & = 2k/\sqrt{k_x^2+k_y^2},\\
		\mathcal{N}_{A,-}(\mathbf{k}) &= 2\sqrt{(E_{A,-}(\mathbf{k}))^2-\lambda_0^2}.
	\end{align}
\end{subequations}

In addition, we discuss the band index $\pm$ that are actually eigen-values of symmetry. To show that, we emphasize that one can also obtain the common eigen-states for $\mathcal{H}_{1}({\bf k})+\mathcal{H}_{\text{soc}}$ and ${\cal M}\cdot \vec{n}$. As we discussed in Eq.~\eqref{sm-eq-slnk-operator}, there is a helicity-type symmetry operator that commutes with $\mathcal{H}_1({\bf k})+\mathcal{H}_{\text{soc}}$, which indicates the index $\pm$ in the eigen-state solution $\vert E_{A,\pm}(\mathbf{k})\rangle $ are eigen-values of $({\bf S}+{\bf L})\cdot \vec{n}_{{\bf k}}$. Explicitly, one can check that
\begin{subequations}
	\begin{align}
		[({\bf S}+{\bf L})\cdot \vec{n}_{{\bf k}}] \vert E_{A,+}(\mathbf{k})\rangle  &=  \vert E_{A,+}(\mathbf{k})\rangle,\\
		[({\bf S}+{\bf L})\cdot \vec{n}_{{\bf k}}] \vert E_{A,-}(\mathbf{k})\rangle  &=  0.
	\end{align}
\end{subequations}
It represents the eigen-values of the z-component angular moment of the total angular momentum at nonzero ${\bf k}$.

\subsection{Approach II: The U(1) quasi-symmetry protected nodal planes}
\label{sec-Appendix-D-3}

As discussed in Sec.~\ref{appendix-C-0}, parts of the entire $k^2$-order Hamiltonian break the SU(2) quasi-symmetry down to the U(1) quasi-symmetry. For example, we consider	
\begin{align}
	\mathcal{H}_{2,\vec{n}} ({\bf k}) = k_xk_y (\vec{C}_{\mathcal{J}} \cdot \vec{n}) \times \left( \vec{n} \cdot \vec{\mathcal{J}} \right) 
	+ k_xk_z (\vec{C}_{\mathcal{P}} \cdot \vec{n}) \times \left(  \vec{n} \cdot \vec{\mathcal{P}} \right) 
	+ k_yk_z (\vec{C}_{\mathcal{Q}} \cdot \vec{n}) \left(  \vec{n} \cdot  \vec{\mathcal{Q}} \right),
\end{align}
which commutes with $\mathcal{M}\cdot \vec{n}$, 
\begin{align}
	[ \mathcal{M}\cdot \vec{n} , \mathcal{H}_{2,\vec{n}} ({\bf k}) ] =0.
\end{align}
For an illustration, we can choose $\vec{n}=(0,0,1)$ without loss of generality, so that $\mathcal{M}_3$ is the symmetry generator for the remaining U(1) quasi-symmetry. Therefore, we consider the Hamiltonian $\mathcal{H}_{qsR}({\bf k})$ that commute with $\mathcal{M}_3$ as,
\begin{subequations}
\begin{align}
	\mathcal{H}_{qsR}({\bf k}) &= \mathcal{H}_1({\bf k}) + \mathcal{H}_{\text{soc}} + \mathcal{H}_{2,\mathcal{M}_3}(\mathbf{k}), \\
	\mathcal{H}_1({\bf k}) &= C_0\sigma_0\tau_0 + 2A_1 (\mathbf{k}\cdot\mathbf{L}), \\
	\mathcal{H}_{\text{soc}} &= 4\lambda_0 ( \mathbf{S} \cdot \mathbf{L}), \\
	\mathcal{H}_{2,\mathcal{M}_3}(\mathbf{k}) &= {\bf g}_3 \cdot {\bf J}_3 = (C_1k_xk_y,  C_2k_xk_z, -C_3  k_y k_z) \cdot  (\sigma_z\tau_0,\sigma_x\tau_0,\sigma_y\tau_y),
\end{align}
\end{subequations}
where $L_x = \tfrac{1}{2} \sigma_y\tau_0, \; L_y = \tfrac{1}{2}\sigma_x\tau_y, \; L_z = -\tfrac{1}{2}\sigma_z\tau_y$. Notice that $\mathcal{M}_3=\sigma_0\tau_y$ is the quasi-symmetry operator. It is easy to check $[\mathcal{M}_3, \mathcal{H}_{qsR}({\bf k}) ] = 0$. We next solve the common eigen-states of $\mathcal{M}_3$ and $\mathcal{H}_{qsR}({\bf k})$. To do that, we only need to diagonalize the $\tau_y$-term for $\mathcal{H}_{qsR}({\bf k})$. In other words, we apply a unitary transformation 
\begin{align}
	\mathcal{U}_{\mathcal{M}_3} = \frac{1}{\sqrt{2}}s_0\otimes \sigma_0\otimes \begin{pmatrix}
		1	& -i \\ 
		1	& i
	\end{pmatrix},
\end{align}
which leads to the rotating in the $(\tau_x,\tau_y,\tau_z)$ subspace
\begin{align} 
	\mathcal{U}_{\mathcal{M}_3} \tau_0 \mathcal{U}_{\mathcal{M}_3}^\dagger = \tau_0, 
	\mathcal{U}_{\mathcal{M}_3} \tau_x \mathcal{U}_{\mathcal{M}_3}^\dagger = \tau_y, 
	\mathcal{U}_{\mathcal{M}_3} \tau_y \mathcal{U}_{\mathcal{M}_3}^\dagger = \tau_z, 
	\mathcal{U}_{\mathcal{M}_3} \tau_z \mathcal{U}_{\mathcal{M}_3}^\dagger = \tau_x. 
\end{align}
Therefore, after this unitary transformation, the U(1) quasi-symmetry generator $\mathcal{M}_3$ becomes
\begin{align}
	\mathcal{U}_{\mathcal{M}_3} \mathcal{M}_3 \mathcal{U}_{\mathcal{M}_3}^\dagger  \triangleq  \mathcal{M}_{qs} = \begin{pmatrix}
		\sigma_0 & 0  \\  0 & -\sigma_0  
	\end{pmatrix} .
\end{align}
And we take this unitary transformation on the $\mathcal{H}_{qsR}(\mathbf{k})$ Hamiltonian and obtain 
\begin{subequations}
\begin{align}\label{eq-qsR-model-M3-type}
	\mathcal{U}_{\mathcal{M}_3} \mathcal{H}_{qsR}(\mathbf{k})  \mathcal{U}_{\mathcal{M}_3}^\dagger \triangleq \mathcal{H}_{qsR}(\mathbf{k})  = C_0 +B_1k^2 +
		\begin{pmatrix}
			\mathcal{H}_+'(\mathbf{k})	&  0 \\ 
			0	& \mathcal{H}_-'(\mathbf{k})
		\end{pmatrix},  
\end{align}
\end{subequations}
where 
\begin{align}
		\mathcal{H}_+'(\mathbf{k}) &=  s_0\otimes\left\lbrack (A_1k_x-C_3k_yk_z)\sigma_y + (A_1k_y+C_2k_xk_z)\sigma_x -(A_1k_z-C_1k_xk_y)\sigma_z \right\rbrack + \lambda_0\left( s_x\sigma_y + s_y\sigma_x - s_z\sigma_z \right)  , \\
		\mathcal{H}_-'(\mathbf{k}) &=  s_0\otimes\left\lbrack (A_1k_x+C_3k_yk_z)\sigma_y - (A_1k_y-C_2k_xk_z)\sigma_x +(A_1k_z+C_1k_xk_y)\sigma_z \right\rbrack + \lambda_0\left( s_x\sigma_y - s_y\sigma_x + s_z\sigma_z \right)  , 
\end{align}
This means that we have chosen the eigen-states of $\mathcal{H}_{qsR}(\mathbf{k})$ to be the common eigen-state of $\mathcal{M}_3$ or $\mathcal{M}_{qs}$. Due to the presence of the U(1) quasi-symmetry, we dubbed $\mathcal{H}_{qsR}$ as the quasi-symmetric R-model (``qsR''). Therefore, the subscript $\pm$ for $\mathcal{H}_{\pm}'(\mathbf{k})$ represent the different eigenvalues of $\mathcal{M}_3$ or $\mathcal{M}_{qs}$. Besides, $\mathcal{H}_+'(\mathbf{k})$ and $\mathcal{H}_-'(\mathbf{k})$ are related to each other by TR $\mathcal{T}=is_y\mathcal{K}$ with $\mathcal{K}$ complex conjugate,
\begin{align}
		\mathcal{H}_-'(\mathbf{k}) = \mathcal{T} \mathcal{H}_+'(-\mathbf{k}) \mathcal{T}^\dagger.
\end{align}
Next, we compute the eigen-energy of the $qsR$-model. The energy of the upper four bands are given by
\begin{align}
		E_{+,1} (\mathbf{k}) &= C_0+B_1k^2 + \sqrt{A_1^2k^2 + C_1^2k_x^2k_y^2 + C_2^2k_x^2k_z^2 + C_3^2k_y^2k_z^2 - 2A_1\tilde{C}k_xk_yk_z} +\lambda_0,\\ 
		E_{+,2} (\mathbf{k}) &= C_0+B_1k^2 + \sqrt{A_1^2k^2 + C_1^2k_x^2k_y^2 + C_2^2k_x^2k_z^2 + C_3^2k_y^2k_z^2 - 2A_1\tilde{C}k_xk_yk_z + 4\lambda_0^2} -\lambda_0 ,\\ 
		E_{-,1} (\mathbf{k}) &= C_0+B_1k^2 + \sqrt{A_1^2k^2 + C_1^2k_x^2k_y^2 + C_2^2k_x^2k_z^2 + C_3^2k_y^2k_z^2 + 2A_1\tilde{C}k_xk_yk_z} +\lambda_0 ,\\ 
		E_{-,2} (\mathbf{k}) &=  C_0+B_1k^2 + \sqrt{A_1^2k^2 + C_1^2k_x^2k_y^2 + C_2^2k_x^2k_z^2 + C_3^2k_y^2k_z^2 + 2A_1\tilde{C}k_xk_yk_z + 4\lambda_0^2} -\lambda_0 ,
\end{align}
where $\tilde{C}=C_1-C_2+C_3 $. The $qsR$-model breaks the $C_{3,(111)}$ rotation symmetry because of $C_1\neq C_2 \neq C_3$. Moreover, the $\pm$ index for $E_{\pm,i}$ means the eigenvalues of $\mathcal{M}_{qs}$,
\begin{align}
	\begin{split}
		&\mathcal{H}_{qsR}(\mathbf{k})  \vert E_{\pm,i} (\mathbf{k}) \rangle = E_{\pm,i} (\mathbf{k}) \vert E_{\pm,i} (\mathbf{k}) \rangle, \\
		& \mathcal{M}_{qs} \vert E_{\pm,i} (\mathbf{k}) \rangle = \pm \vert E_{\pm,i} (\mathbf{k}) \rangle, 
	\end{split}
\end{align}
where $i=1,2$ is the band index. $E_{+,i}(\mathbf{k}) = E_{-,i}(-\mathbf{k})$ is required by TR symmetry. The quasi-symmetry protected nodal planes are given by
\begin{align}\label{eq-equ-nodal-plane-qsR-model}
	\begin{cases}
		k_xk_yk_z>0, E_{+,1} (\mathbf{k}) = E_{-,2} (\mathbf{k}) , \\
		k_xk_yk_z<0, E_{+,2} (\mathbf{k}) = E_{-,1} (\mathbf{k}).
	\end{cases}	
\end{align}
The crossings are between the bands with different eigenvalue of quasi-symmetry $\mathcal{M}_3$ or $\mathcal{M}_{qs}$, and thus we find nodal planes at generic momenta with the protection from the quasi-symmetry $\mathcal{M}_{qs}$. By further imposing the Fermi energy constraint, there will be nodal lines on the Fermi surface, shown in Fig.~\ref{sm-fig6}.  In Fig.~\ref{sm-fig6}(a), the four Fermi surfaces are plotted in the $\Gamma-R-M$ plane, where two inner FSs intersect with each other and generate the quasi-symmetry protected exact crossings (marked as purple circles). In Fig.~\ref{sm-fig6} (b) and (c),  the solution of Eq.~\eqref{eq-equ-nodal-plane-qsR-model} are explicitly depicted, showing the exact nodal lines on the FSs.

\begin{figure}[!htbp]
	\centering
	\includegraphics[width=\linewidth]{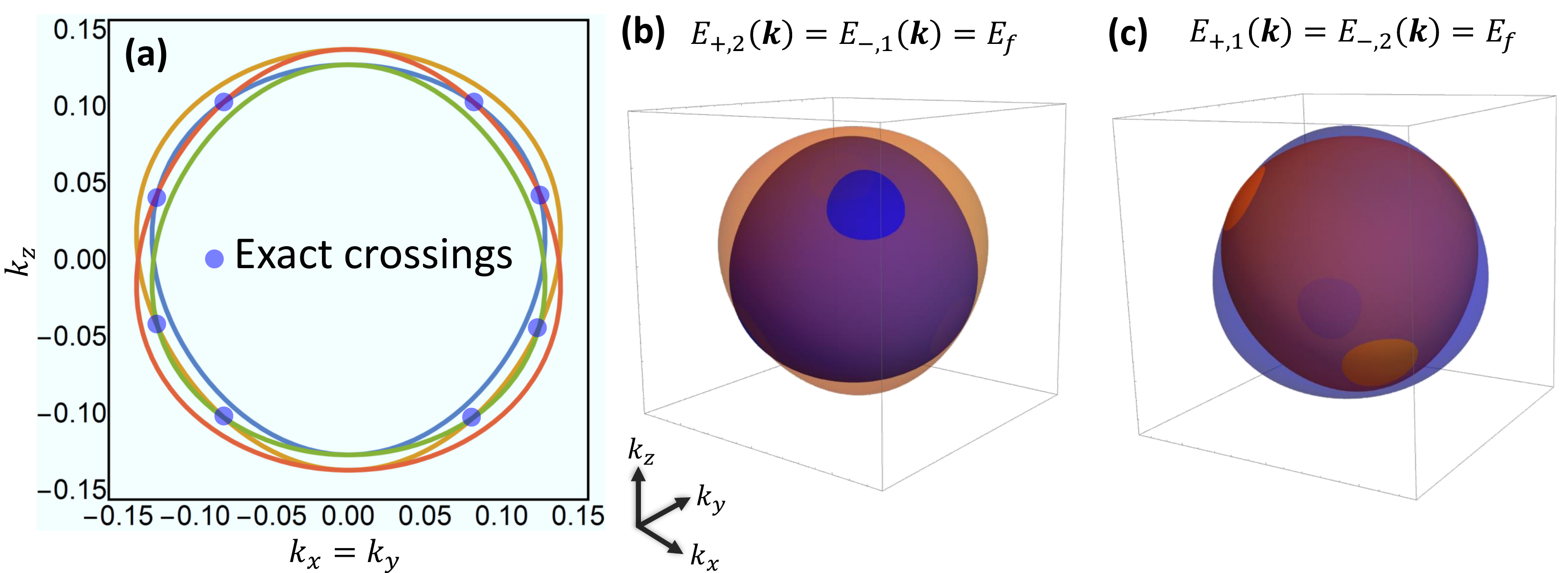}
	\caption{The exact crossings of the 8-band $qsR$-model. In (a), the Fermi surfaces with exact crossing are shown in the $\Gamma-R-M$ plane. In (b) and (c), we show the exact nodal lines on the FSs. These nodal lines are protected by the U(1) quasi-symmetry.
	}	
	\label{sm-fig6}
\end{figure}

Next, let us discuss the terms that break the quasi-symmetry. By including the remaining $k^2$-order Hamiltonian $\mathcal{H}_{2,\mathcal{M}_1}(\mathbf{k})+\mathcal{H}_{2,\mathcal{M}_2}(\mathbf{k})$, the full Hamiltonian is given by
\begin{align}
	\mathcal{H}_{R}({\bf k}) = \mathcal{H}_{qsR}({\bf k}) + \mathcal{U}_{\mathcal{M}_3} [\mathcal{H}_{2,\mathcal{M}_1}(\mathbf{k})+\mathcal{H}_{2,\mathcal{M}_2}(\mathbf{k})]\mathcal{U}_{\mathcal{M}_3}^\dagger,
\end{align}
where the U(1) quasi-symmetry breaking terms are given by
\begin{subequations}
\begin{align}\label{eq-off-diagonal-k2-terms}
	\mathcal{H}_{2,qsb}(\mathbf{k})  &= \mathcal{U}_{\mathcal{M}_3} [\mathcal{H}_{2,\mathcal{M}_1}(\mathbf{k})+\mathcal{H}_{2,\mathcal{M}_2}(\mathbf{k})]\mathcal{U}_{\mathcal{M}_3}^\dagger = \begin{pmatrix}
		0 & s_0\otimes H_{(2)}  \\ 
		s_0\otimes H_{(2)}^\dagger & 0 
	\end{pmatrix}, \\
	H_{(2)} &= C_1(k_yk_z\sigma_0 + k_xk_z\sigma_z) + C_2(-ik_xk_y\sigma_x+ik_yk_z\sigma_0) 
		+ C_3(k_xk_y\sigma_x+ik_xk_z\sigma_z)  \nonumber \\
		&+ (C_1+iC_2)k_yk_z\sigma_0 + (C_1+iC_3)k_xk_z\sigma_z + (-iC_2+C_3)k_xk_y\sigma_x. 
\end{align} 
\end{subequations}
Therefore, we have
\begin{align}
	[\mathcal{M}_{qs} ,\mathcal{H}_{2,qsb}(\mathbf{k})]  = 
	2\begin{pmatrix}
		0	&  h_{(2)}(\mathbf{k}) \\ 
		-h_{(2)}^\dagger(\mathbf{k}) 	& 0
	\end{pmatrix}  
	\neq 0.  
\end{align}
where 
\begin{align}
\begin{split}
	h_{(2)}  = (C_1+iC_2)k_yk_z\sigma_0 + (C_1+iC_3)k_xk_z\sigma_z 
		 + (-iC_2+C_3)k_xk_y\sigma_x. 
\end{split}
\end{align}
For this Hamiltonian, $[\mathcal{M}_{qs} , \mathcal{H}_{2,qsb}(\mathbf{k})] = 0$ only occurs for $\mathbf{k}=(0,0,0)$ (the $R$-point). Therefore, for any Fermi surface that does not cross the $R$-point, the nodal planes will be gapped. This leads to the quasi-symmetry hierarchy mentioned in the main text (see Eq.~(10)), 
\begin{align}
	\text{SU}_{\text{s}}(2)\times \text{SU}_{\text{o}}(2) \stackrel{\mathcal{H}_{\text{soc}}}{\xhookrightarrow{\quad\quad\quad\;}} \text{SU}_{\text{o}}(2) 
	\stackrel{\mathcal{H}_{2,\mathcal{M}_i}}{\xhookrightarrow{\quad\quad\quad\quad\;}} \text{U}_{\text{o}}(1).
\end{align}

More interestingly, as we discussed in Sec.~\ref{appendix-C-0}, we have shown that the choice of $\mathcal{H}_{2,\mathcal{M}_3}$ is just one specific case, instead, for any real vector $\vec{n}$ of $\mathcal{H}_{2,\vec{n}}$ that commute with $\mathcal{M}\cdot \vec{n}$ (see Sec.~\ref{appendix-C-0} for details) can generally lead to the same quasi-symmetry hierarchy results. Because we have
\begin{align}
	[\mathcal{M}\cdot \vec{n}, s_0\otimes \mathcal{H}_1({\bf k}) + \mathcal{H}_{\text{soc}}+ s_0\otimes \mathcal{H}_{2,\vec{n}}({\bf k}) ] =0.
\end{align} 
Thus, one can conclude that this analysis for the reduction from SU(2) quasi-symmetry down to U(1) quasi-symmetry is general, which can help to protect quasi-nodal-plane at generic momenta.

To show this nodal plane ude to the remaining U(1) quasi-symmetry in our approach II more explicitly, we next consider the perturbation to $\mathcal{H}_{2,\vec{n}}({\bf k}) $.  Here we emphasize that one can also obtain the common eigen-states for $\mathcal{H}_{1}({\bf k})+\mathcal{H}_{\text{soc}}$ and ${\cal M}\cdot \vec{n}$. Different from Approach I in Sec.~\ref{Appendix-C}, here we take the eigen-state solution of $s_0\otimes \mathcal{H}_1({\bf k}) + \mathcal{H}_{\text{soc}}$, and do the perturbation for the $k^2$-order Hamiltonian. Below we choose $\vec{n}=(0,0,1)$ and the corresponding U(1) quasi-symmetry opeator is ${\cal M}\cdot \vec{n}={\cal M}_3$. Recall that $\mathcal{H}_{1}({\bf k})+\mathcal{H}_{\text{soc}}$ becomes block-diagonal after this unitary transformation defined in Eq.~\eqref{sm-eq-trans-unit-m3},
\begin{align}
	\begin{split}
		\mathcal{U} \left[ s_0\otimes \mathcal{H}_1({\bf k}) + \mathcal{H}_{\text{soc}} \right] \mathcal{U}^\dagger  
		= C_0 + \begin{pmatrix}
			\mathcal{H}_{A}(\mathbf{k})	&  0 \\ 
			0	& \mathcal{H}_{B}(\mathbf{k})
		\end{pmatrix}, 
	\end{split}
\end{align}
where $\mathcal{H}_{A/B}(\mathbf{k})$ are given by
\begin{align}
	\mathcal{H}_{A}(\mathbf{k}) &=  A_1 s_0\otimes \left( k_x\sigma_y + k_y\sigma_x - k_z\sigma_z \right) 
	+ \lambda_0\left( s_x\sigma_y + s_y\sigma_x - s_z\sigma_z \right) , \\
	\mathcal{H}_{B}(\mathbf{k}) &=  A_1 s_0\otimes \left( k_x\sigma_y - k_y\sigma_x + k_z\sigma_z \right) 
	+ \lambda_0\left( s_x\sigma_y - s_y\sigma_x + s_z\sigma_z \right) .
\end{align}
Again, note that the index $A(B)$ represent the eigenvalues $+1(-1)$ of the quasi-symmetry ${\cal M}_3$. Therefore, after this unitary transformation, the U(1) quasi-symmetry generator $\mathcal{M}_3$ becomes
\begin{align}
	\mathcal{U}_{\mathcal{M}_3} \mathcal{M}_3 \mathcal{U}_{\mathcal{M}_3}^\dagger  \triangleq  \mathcal{M}_{qs} = \begin{pmatrix}
		\sigma_0 & 0  \\  0 & -\sigma_0  
	\end{pmatrix} .
\end{align}
The analytical solution for the two upper bands (i.e.~Eq.~\eqref{eq-ham+-eig-wavefunc}) of $\mathcal{H}_1({\bf k})+\mathcal{H}_{\text{soc}}$ in Sec.~\ref{appendix-D-1} are given by
\begin{subequations}
\begin{align}
	&\vert E_{A,+}(\mathbf{k})\rangle = \frac{1}{\mathcal{N}_{A,+}(\mathbf{k})} \left( -i,\frac{k_x-ik_y}{k-k_z},\frac{-ik_x+k_y}{k+k_z},1 \right)^T , \\
	&\vert E_{A,-}(\mathbf{k})\rangle = \frac{1}{\mathcal{N}_{A,-}(\mathbf{k})} \Big{(} i(-A_1k_z-\lambda_0+E_{A,-}(\mathbf{k})), 
		A_1(-k_x+ik_y), A_1(-ik_x+k_y), A_1k_z-\lambda_0 + E_{A,-}(\mathbf{k})  \Big{)}^T.
\end{align}
\end{subequations}
where the index $\pm$ in the eigen-state solution $\vert E_{A,\pm}(\mathbf{k})\rangle $ are eigen-values of $({\bf S}+{\bf L})\cdot \vec{n}_{{\bf k}}$ for nonzero ${\bf k}$. And the normalization factors are given by $\mathcal{N}_{A,+}(\mathbf{k})  = 2k/\sqrt{k_x^2+k_y^2}$, and $ \mathcal{N}_{A,-}(\mathbf{k}) = 2\sqrt{(E_{A,-}(\mathbf{k}))^2-\lambda_0^2}$. Moreover, note that they are the two positive upper bands with eigen-energies, 
\begin{subequations} \label{sm-eq-EA-pm-1}
\begin{align}
 E_{A,+}(\mathbf{k}) &= A_1k + \lambda_0, \\ 
 E_{A,-} (\mathbf{k}) &= \sqrt{A_1^2k^2+4\lambda_0^2}-\lambda_0,
\end{align} 
\end{subequations}
which have increasing energy as $k$ increases. And, the eigen-states of the $\mathcal{H}_{B}(\mathbf{k})$ block can be related to those of $\mathcal{H}_{A}(\mathbf{k})$ by time-reversal symmetry, 
\begin{subequations}
\begin{align}
	\vert E_{B,+}(\mathbf{k})\rangle &= \mathcal{T} \vert E_{A,+}(-\mathbf{k}) \rangle  ,\\
	\vert E_{B,-}(\mathbf{k})\rangle &=\mathcal{T} \vert E_{A,-}(-\mathbf{k})\rangle , 
\end{align}
\end{subequations}
and
\begin{align}
	\mathcal{H}_{B}(\mathbf{k}) \vert E_{B,\pm}(\mathbf{k})\rangle  = E_{A,\pm}(\mathbf{k})   \vert E_{B,\pm}(\mathbf{k})\rangle,
\end{align}
where $ E_{A,\pm}(-\mathbf{k}) =  E_{A,\pm}(\mathbf{k})$ has been used. Therefore, we have the four bands as a basis,
\begin{align} \label{sm-eq-basis-h1-hsoc-2}
	\{ \vert \Psi_{+} (\mathbf{k}) \rangle \} &=  \Big{\{}  (1,0)^T \otimes \vert E_{A,+}(\mathbf{k})\rangle,   (1,0)^T \otimes \vert E_{A,-}(\mathbf{k})\rangle, 
	(0,1)^T \otimes \vert E_{B,+}(\mathbf{k})\rangle , (0,1)^T \otimes\vert E_{B,-}(\mathbf{k})\rangle \Big{\}}.
\end{align}
In this basis, the $\mathcal{H}_1({\bf k})+\mathcal{H}_{\text{soc}}$ is diagonal,  
\begin{align}
	\begin{split}
		\mathcal{H}_1({\bf k})+\mathcal{H}_{\text{soc}} = C_0+\text{Diag}[&E_{A,+}(\mathbf{k}), E_{A,-}(\mathbf{k}), 
		E_{A,+}(\mathbf{k}) ,E_{A,-}(\mathbf{k}) ].
	\end{split}
\end{align}

After straightforward calculation, we next project the $k^2$-term $\mathcal{H}_{2,\mathcal{M}_3}({\bf k})$ onto the basis in Eq.~\eqref{sm-eq-basis-h1-hsoc-2}, and we arrive at the quasi-symmetry $P$-model ($qsP$-model) as,
\begin{align}
	\mathcal{H}_{qsP}(\mathbf{k}) = \begin{pmatrix}
		E_{A,+}'(\mathbf{k}) + f_1(\mathbf{k}) & d_1(\mathbf{k}) - id_2(\mathbf{k}) & 0 & 0 \\ 
		d_1(\mathbf{k}) + id_2(\mathbf{k})  &   E_{A,-}'(\mathbf{k}) + f_2(\mathbf{k}) & 0  & 0 \\ 
		0 & 0 &  E_{A,+}'(\mathbf{k}) - f_1(\mathbf{k}) & d_1(\mathbf{k}) - id_2(\mathbf{k})   \\ 
		0 & 0 & d_1(\mathbf{k}) + id_2(\mathbf{k})  & E_{A,-}'(\mathbf{k}) - f_2(\mathbf{k})
	\end{pmatrix}, 
\end{align}
where $E_{A,\pm}'(\mathbf{k}) =C_0+B_1k^2+E_{A,\pm}(\mathbf{k})$. For the projected four-band $qsP$-model, we check the protection from quasi-symmetry as
\begin{align}
 [\mathcal{M}_{sq}, \mathcal{H}_{qsP}({\bf k})] = 0.
\end{align}
And all the other components are given by
\begin{subequations} \label{sm-eq-perp-f12-d12}
\begin{align}
	f_1(\mathbf{k}) & = -\tilde{C} \frac{k_xk_yk_z}{k}, \\
	f_2(\mathbf{k}) & = -A_1\tilde{C}\frac{k_xk_yk_z}{\sqrt{A_1^2k^2+4\lambda_0^2}}, \\
	d_1(\mathbf{k}) & = \frac{2(E_{A,+}(\mathbf{k}) -E_{A,-}(\mathbf{k}) )k_xk_y }{\mathcal{N}_{A,+} \mathcal{N}_{A,-} } \left( C_1 + (C_2-C_3)\frac{k_z^2}{k_x^2+k_y^2} \right), \\
	d_2(\mathbf{k}) & = \frac{2(E_{A,+}(\mathbf{k}) -E_{A,-}(\mathbf{k}) )k_zk }{\mathcal{N}_{A,+} \mathcal{N}_{A,-} } \frac{C_2k_x^2 + C_3k_y^2}{k_x^2+k_y^2}.
\end{align}
\end{subequations}

The eigen-energies of the $qsP$-model are given by
\begin{align}
	E_{\alpha,\beta}(\mathbf{k}) &=  \frac{1}{2} \left\lbrack \Delta E_{+}(\alpha\mathbf{k})  +\beta \sqrt{( \Delta E_{-}(\alpha\mathbf{k})  )^2 + E_{d_{12}} }  \right\rbrack ,
\end{align}
with $\alpha,\beta=\pm$. And $\alpha=\pm$ are eigenvalues of $\mathcal{M}_{qs}$ and $\beta=\pm$ are for the band index. 
Here we have defined 
\begin{subequations}
\begin{align}
	E_{d_{12}}(\mathbf{k}) &=4\left[ (d_1(\mathbf{k}) )^2+(d_2(\mathbf{k}) )^2 \right] ,\\
	\Delta E_{\pm}(\mathbf{k})  &= E_{A,+}'(\mathbf{k}) + f_1(\mathbf{k}) \pm ( E_{A,-}'(\mathbf{k}) + f_2(\mathbf{k})  ). 
\end{align}
\end{subequations}
Similar to the discussion for the nodal plane of the $qsR$-model (see Eq.~\eqref{eq-equ-nodal-plane-qsR-model}), the quasi-symmetry protected nodal planes of the $qsP$-model are give by
\begin{align}\label{eq-equ-nodal-plane-qsP-model}
	\begin{cases}
		k_xk_yk_z<0, E_{+,-} (\mathbf{k}) = E_{-,+} (\mathbf{k}) , \\
		k_xk_yk_z>0, E_{+,+} (\mathbf{k}) = E_{-,-} (\mathbf{k}).
	\end{cases}	
\end{align}
The crossings happen for bands with different eigenvalue of quasi-symmetry $\mathcal{M}_{qs}$. We numerically check these exact crossings on the $\Gamma-R-M$ planes, which is consistent with the results in Fig.~\ref{sm-fig6} (a).

\subsection{Perturbation theory for tiny gap of the nodal plane in Approach II}
\label{sec-Appendix-D-4}

The above discussion on the nodal planes in the approach II requires a choice of specific $k^2$-order terms, but in real materials, all the coefficients before the $k^2$-order terms can generally be non-zero and at the same order. Thus, our current approach does not directly explain the near nodal plane seen in real materials. We notice that we treat the SOC terms accurately in our approach II without any approximation, while the existence of near nodal planes in real materials actually require the SOC strength to be much smaller than the Fermi energy. Therefore, we below consider the limit of the SOC strength $\lambda_0\ll A_1 k_F$ ($k_F$ is the Fermi momentum) in our Approach II, taking into account all non-zero $k^2$-order terms.

Following the same procedure of the perturbation projection as in the last section, we project all the $k^2$-order terms $\mathcal{H}_{2,\mathcal{M}_1}(\mathbf{k})+\mathcal{H}_{2,\mathcal{M}_2}(\mathbf{k})+\mathcal{H}_{2,\mathcal{M}_3}(\mathbf{k})$ onto the basis in Eq.~\eqref{sm-eq-basis-h1-hsoc-2}, and get the entire perturbation Hamiltonian, 
\begin{align}\label{eq-qsr-qsb-m3}
	\mathcal{H}_{qsP+qsB}({\bf k}) = \begin{pmatrix}
		E_{A,+}'(\mathbf{k}) + f_1(\mathbf{k}) & d_1(\mathbf{k}) - id_2(\mathbf{k})  &  g_1(\mathbf{k}) &  g_2(\mathbf{k}) \\ 
		d_1(\mathbf{k}) + id_2(\mathbf{k})  &   E_{A,-}'(\mathbf{k}) + f_2(\mathbf{k}) &  g_3(\mathbf{k})  &  g_4(\mathbf{k}) \\ 
		g_1^\ast(\mathbf{k}) &  g_3^\ast(\mathbf{k}) &  E_{A,+}'(\mathbf{k}) - f_1(\mathbf{k}) & d_1(\mathbf{k}) - id_2(\mathbf{k})   \\ 
		g_2^\ast(\mathbf{k}) &  g_4^\ast(\mathbf{k}) & d_1(\mathbf{k}) + id_2(\mathbf{k})  & E_{A,-}'(\mathbf{k}) - f_2(\mathbf{k})
	\end{pmatrix} .
\end{align}
Here the terms $d_1,d_2,f_1,f_2$ have been given by Eq.~\eqref{sm-eq-perp-f12-d12}. In addition, the off-diagonal terms $g_{1,2,3,4}$ are generally to open a gap for the quasi-nodal plane. Because these off-diagonal terms break the quasi-symmetry $\mathcal{M}_{qs}$. And, they are given by 
\begin{subequations}
\begin{align}
	g_1(\mathbf{k}) &= (1-i)\tilde{C}\frac{k_xk_yk_z}{k}, \\
	g_4(\mathbf{k}) &= (i-1)A_1\tilde{C} \frac{k_x k_y k_z} {E_{A,-}(\mathbf{k})+\lambda_0}, \\
	g_{2}(\mathbf{k}) &= \frac{2}{(k_x^2+k_y^2)\mathcal{N}_{A,+}(\mathbf{k}) \mathcal{N}_{A,-}(\mathbf{k}) } \Big{\{}  
	(C_2+iC_3) k_xk_y(k_x^2+k_y^2) (E_{A,-}(\mathbf{k}) -E_{A,+}(\mathbf{k})  )  \nonumber 	\\
	&+ (C_1+iC_3) k_xk_z ( A_1k_xk^2  -ik_yk_z (E_{A,-}(\mathbf{k})-E_{A,+}(\mathbf{k}) ) - k_xk(E_{A,-}(\mathbf{k})-\lambda_0 ) )  \nonumber \\
	&+ (C_2-iC_1) k_yk_z ( A_1k_yk^2 + i k_xk_z(E_{A,-}(\mathbf{k})-E_{A,+}(\mathbf{k}))  - k_yk(E_{A,-}(\mathbf{k})-\lambda_0 ) ) ,\\ 
	g_{3}(\mathbf{k})	&= \frac{2}{(k_x^2+k_y^2)\mathcal{N}_{A,+}(\mathbf{k}) \mathcal{N}_{A,-}(\mathbf{k}) } \Big{\{}  
	-(C_2+iC_3) k_xk_y(k_x^2+k_y^2) ( E_{A,-}(\mathbf{k}) -E_{A,+}(\mathbf{k}) ) \nonumber \\
	&+ (C_1+iC_3) k_xk_z (  A_1k_xk^2 + ik_yk_z(E_{A,-}(\mathbf{k})-E_{A,+}(\mathbf{k}) ) - k_xk(E_{A,-}(\mathbf{k})-\lambda_0 )   ) \nonumber \\
	&+ (C_2-iC_1) k_yk_z (  A_1k_yk^2 - ik_xk_z(E_{A,-}(\mathbf{k})-E_{A,+}(\mathbf{k}) ) - k_yk (E_{A,-}(\mathbf{k})-\lambda_0 )  ) 
	\Big{\}}. 
\end{align}
\end{subequations}
Here $g_3(\mathbf{k}) = -g_2(-\mathbf{k})$. The nodal planes are completely gapped out. Now we use perturbation to explain why the gap is tiny by realizing that the SOC in CoSi is weak enough for doing a perturbation expansion for the coefficients $f_{1,2}({\bf k})$, $d_{1,2}({\bf k})$, and $g_{1,2,3,4}({\bf k})$. By setting $\lambda_0/A_1 k_F\to 0$, we obtain
\begin{subequations}
\begin{align}
	f_1 ({\bf k}) &= f_2 ({\bf k}) =  -\tilde{C} \frac{k_xk_yk_z}{k}, \\
	d_1(\mathbf{k}) & = d_2(\mathbf{k})  = 0 , \\ 
	g_1(\mathbf{k}) &= -g_{2}(\mathbf{k})  =  (1-i)\tilde{C}\frac{k_xk_yk_z}{k}, \\
	g_{2}(\mathbf{k}) &=g_{3}(\mathbf{k})=0.
\end{align}
\end{subequations}
for the zeroth-order terms. Please notice that only the diagonal energies $E_{A,\pm}'$ that are eigen-energies of $s_0\otimes\mathcal{H}_1+\mathcal{H}_{\text{SOC}}$ involve the SOC $\lambda_0$. As a result, the perturbation Hamiltonian in Eq.~\eqref{eq-qsr-qsb-m3} to the zeroth-order in $\lambda_0$ becomes
\begin{subequations}
\begin{align}
	\mathcal{H}^{(0)}_{qsP+qsB}({\bf k})&= \begin{pmatrix}
		E_{A,+}'(\mathbf{k}) + f_1(\mathbf{k}) & 0  &  g_1(\mathbf{k}) &  0\\ 
		0  &   E_{A,-}'(\mathbf{k}) + f_1(\mathbf{k}) &  0 &  -g_1(\mathbf{k}) \\ 
		g_1^\ast(\mathbf{k}) &  0 &  E_{A,+}'(\mathbf{k}) - f_1(\mathbf{k}) & 0  \\ 
		0 &  -g_1^\ast(\mathbf{k}) & 0  & E_{A,-}'(\mathbf{k}) - f_1(\mathbf{k})
	\end{pmatrix} , \\
   &= \begin{pmatrix}
   		E_{A,+}'(\mathbf{k}) + f_1(\mathbf{k}) & g_1(\mathbf{k}) \\
   		g_1^\ast(\mathbf{k}) & E_{A,+}'(\mathbf{k}) - f_1(\mathbf{k}) 
	\end{pmatrix} \oplus 
	\begin{pmatrix}
		E_{A,-}'(\mathbf{k}) + f_1(\mathbf{k})  & -g_1(\mathbf{k}) \\
		-g_1^\ast(\mathbf{k})  & E_{A,-}'(\mathbf{k}) - f_1(\mathbf{k})
	\end{pmatrix} 
\end{align}
\end{subequations}
where the diagonal terms are $E_{A,+}'(\mathbf{k}) =C_0+B_1k^2+A_1k + \lambda_0$ and $E_{A,-}'(\mathbf{k}) =C_0+B_1k^2+\sqrt{A_1^2k^2+4\lambda_0^2}-\lambda_0$. Thus, the eigen-energies are given by
\begin{subequations}
\begin{align}
	E_{1,\pm}(\mathbf{k})  &=  C_0+B_1k^2+A_1k + \lambda_0 \pm \sqrt{\vert f_1({\bf k})\vert^2 + \vert g_1({\bf k})\vert^2  }, \\
	E_{2,\pm}(\mathbf{k})  &=  C_0+B_1k^2+\sqrt{A_1^2k^2+4\lambda_0^2}-\lambda_0  \pm \sqrt{\vert f_1({\bf k})\vert^2 + \vert g_1({\bf k})\vert^2  }, 
\end{align}
\end{subequations}
which leads to the equation for the nodal-plane solution
\begin{align}
	E_{1,-}(\mathbf{k})   = E_{2,+}(\mathbf{k})  
	\quad\Rightarrow \quad 
	A_1 k-\sqrt{A_1^2k^2+4\lambda_0^2} + 2\lambda_0 =  2\sqrt{\vert f_1({\bf k})\vert^2 + \vert g_1({\bf k})\vert^2  }.
\end{align}
In the $k\to \infty $ limit, this equation becomes
\begin{align}
	\lambda_0 =  \sqrt{\vert f_1({\bf k})\vert^2 + \vert g_1({\bf k})\vert^2  } = \sqrt{3} \tilde{C} \left\vert \frac{k_xk_yk_z}{k} \right\vert.
\end{align}
Note that $\tilde{C}>0$ in this work. And it is exactly the same Eq.~\eqref{sm-eq-nodal-plane-solution-1} obtained from the Approach I in Sec.~\ref{Appendix-C}. At small ${\bf k}$, they differs from each other. Moreover, the first-order correction from $\lambda_0$ will open a tiny gap for the nodal planes, which is actually the second-order perturbation theory in Approach I. Based on this analysis, we conclude that the results of Approach II is equivalent from those of Approach I.

\section{The projected two-band model}
\label{Appendix-E}

In this section, we further project the first-order-perturbation 4-band Hamiltonian, the $P$-model (e.g.~see Eq.~\eqref{eq-eff-ham-p-model} in Sec.~\ref{Appendix-C-1}) onto an effective model that only consists of the two bands forming the nodal plane. To do that, we consider the two eigen-states of the $P$-model Hamiltonian in Eq.~\eqref{eq-eff-ham-p-model},
\begin{align}
	\mathcal{H}_{P}^{eff(1)}({\bf k}) &= (E_+ +B_1k^2)s_0\omega_0 + \mathcal{H}_{\text{soc},+}^{eff(1)}({\bf k}) + \mathcal{H}_{k^2,+}^{eff(1)}({\bf k}),
\end{align}
which is marked as the $P$-model around $R$-point. And
\begin{align}
	\begin{split}
		\mathcal{H}_{\text{soc},+}^{eff(1)}({\bf k}) &= \lambda_0 \left( \lambda_x s_x + \lambda_y s_y + \lambda_z s_z \right)  \omega_0, \\
		\mathcal{H}_{k^2,+}^{eff(1)}({\bf k}) &= \tilde{C}k^2 s_0  \left( d_x \omega_x + d_y\omega_y + d_z\omega_z \right).
	\end{split}
\end{align}
where $\tilde{C}=C_1-C_2+C_3$, and $\omega_{x,y,z}$ are Pauli matrices for the $\{A+,B+\}$ band subspace. The eigen-energies are give by Eq.~\eqref{sm-eq-p-model-disp}
\begin{align}
\begin{split}
	E_{\alpha\beta}(k,\theta,\phi) = C_0+B_1k^2+A_1k +\alpha\lambda_0 
		+\beta\tfrac{\sqrt{3}}{4}\tilde{C}k^2 \vert \sin2\phi\sin2\theta\sin\theta\vert,
\end{split}
\end{align}
where $\alpha=\pm$, $\beta=\pm$, and $\sin2\phi\sin2\theta\sin\theta = 4k_xk_yk_z/k^3$. Therefore, these two states $ \{\vert \alpha=+,\beta=-\rangle , \vert \alpha=-,\beta=+ \rangle\}$ that give the eigen-energies $E_{\alpha=+,\beta=-}$ and $E_{\alpha=-,\beta=+}$. The basis eigen-wave function for the two-band model can be derived by solving the four-band $P$ model, $\mathcal{H}_{P}^{eff(1)}$. After straightforward calculation, we find 
\begin{subequations}
\begin{align}
	\vert E_{+,-}\rangle & = \frac{1}{N_{+,-} } 
	(\lambda_z+E_\lambda, \lambda_x + i\lambda_y)^T
	 \otimes (d_z-E_d,d_x+id_y)^T, \\
	\vert E_{-,+}\rangle &= \frac{1}{N_{-,+} }
	(\lambda_z-E_\lambda, \lambda_x + i\lambda_y)^T 
	\otimes (d_z+E_d,d_x+id_y)^T,
\end{align}
\end{subequations}
where we have
\begin{subequations}
\begin{align}
	N_{s,\omega}^2 &=((\lambda_z+s E_\lambda)^2 + \lambda_x^2+\lambda_y^2)  ( (d_z +\omega E_d)^2+d_x^2+d_y^2 ), \\ 
	E_\lambda &= \sqrt{\lambda_x^2 + \lambda_y^2 + \lambda_z^2}, \\
	E_d &= \sqrt{d_x^2 + d_y^2 +d_z^2}, 
\end{align}
\end{subequations}
where we define $\lambda_x = \sin\theta\cos\phi$, $ \lambda_y = \sin\theta\sin\phi$, $\lambda_z =\cos\theta$, $d_x = \cos\theta\sin^2\theta \sin\phi\cos\phi ( \cos\phi + \sin\phi )$, $d_y = \cos\theta\sin^2\theta \sin\phi\cos\phi ( \sin\theta +\cos\theta (\cos\phi - \sin\phi) )$, and $d_z = \sin^2\theta\sin\phi\cos\phi(  \cos^2\theta + \sin\theta\cos\theta (-\cos\phi+\sin\phi) )$.

Within the sub-space formed by these two eigen-states, the effective Hamiltonian $\mathcal{H}_{P}^{eff(1)}$ can be reduced to a two-band model
\begin{align}\label{sm-eq-heff2-two-band}
	\mathcal{H}_{eff} = \epsilon_0 + d_z(\mathbf{k}) \sigma_z,
\end{align}
where $\epsilon_0=C_0+B_1k^2+A_1k $ and $d_z(\mathbf{k}) =\lambda_0  - \tfrac{\sqrt{3}}{4}\tilde{C}k^2 \vert \sin2\phi\sin2\theta\sin\theta\vert=\lambda_0 - \sqrt{3}\tilde{C} \frac{\vert k_xk_yk_z\vert}{k} $. This Hamiltonian commutes with $\sigma_z$, which in turn generates the U(1) quasi-symmetry group for the two-band model. The physical viewpoint of this quasi-symmetry is related to the spin texture of the Fermi surfaces in the momentum space. Due to the SOC term with $\mathbf{k}\cdot\mathbf{s}$ in the $P$-model, the spin texture is in a hedgehog form on each Fermi surface ($\langle \mathbf{s}\rangle \sim \alpha\mathbf{k}$ with $\alpha=\pm$). It indicates that the crossings between these two Fermi surfaces have opposite spin-polarization. Therefore, according to the low-energy effective Hamiltonian, the quasi-symmetry that protects the approximate nodal planes can be viewed as the spin U(1) symmetry along the momentum direction.

Now we discuss the breaking of the quasi-symmetry by second-order perturbation corrections. To see that, we can further project $\mathcal{H}_{P}^{eff(2)}$ into the subspace and the effective Hamiltonian includes two more terms
\begin{align}
	\mathcal{H}_{eff(2)} = \delta d_x(\mathbf{k}) \sigma_x+\delta d_y(\mathbf{k}) \sigma_y, 
\end{align}
which are due to the spin-flipping terms. As a result, the co-dimension of the nodal plane becomes 3 instead of 1, explaining the gap opening. While the detailed expressions for $\delta d_x$ and $\delta d_y$ are complex and not orthogonality, we give an estimate of the typical magnitude of the gap $\Delta_{eff(2)}=2\sqrt{(\delta d_x)^2+(\delta d_y)^2}\sim 0.8 $ meV for $k_F=0.13$ \AA$^{-1}$, being the same order with the DFT estimations~\cite{guo_arxiv_2021}.

Therefore, perturbation theory leads to a general two-band model
\begin{align}\label{eq-two-band-model}
	\mathcal{H}_{2\times 2} =  \delta_{d_x} \sigma_x + \delta_{d_y} \sigma_y + \delta_{d_z} \sigma_z . 
\end{align} 
The gap of this model is given by 
\begin{align}
	\Delta_{k,\text{soc}} = 2\sqrt{ (\delta_{d_x})^2 + (\delta_{d_y})^2 + (\delta_{d_z})^2 }.
\end{align}
If we impose the constraint of the quasi-symmetry $\sigma_z$ that commutes with $\mathcal{H}_{2\times 2}$, 
we can always obtain the constraint equation for the nodal planes by solving $\delta_{d_z}=0$.

Next, let us consider the other terms that break the quasi-symmetry, by taking Eq.~\eqref{eq-p-model-linear-ksoc-lambda2} as an example. The projected two-band model is then numerically calculated directly. The three components are given by
\begin{align}
	\begin{split}
		\delta_{d_x} &= \text{Re}[ \langle E_{+,-} \vert \mathcal{H}_{k,\lambda_2,eff} \vert E_{-,+} \rangle  ], \\
		\delta_{d_y} &= -\text{Im}[ \langle E_{+,-} \vert \mathcal{H}_{k,\lambda_2,eff} \vert E_{-,+} \rangle  ], \\
		\delta_{d_z} &= \frac{1}{2}[ \langle E_{+,-} \vert \mathcal{H}_{k,\lambda_2,eff} \vert E_{+,-} \rangle  
		- \langle E_{-,+}\vert \mathcal{H}_{k,\lambda_2,eff} \vert E_{-,+} \rangle  ], 
	\end{split}
\end{align}
The numerical results are shown in Fig.~\ref{sm-fig7}.  The results only depend on $\theta$ and $\phi$, then, on the $k_r=0.15$ \AA$^{-1}$ sphere with $\lambda_2=0.005$ eV, the numerical calculated $\delta_{d_x}$ and $\delta_{d_y}$ are shown in (a) and (b), respectively. In the whole space, $\delta_{d_x}\neq0$ and $\delta_{d_y}$ expect the Weyl point along the $\Gamma-R$ line, which is shown in Fig.~\ref{sm-fig7} (c) by log-plotting the inverse of the gap $\log\lbrack \frac{1}{\sqrt{\delta_{d_x}^2+\delta_{d_y}^2}}\rbrack$ (i.e.,~the gap vanishes only at the Weyl point). The entire quasi-nodal plane are gapped out by including the general liner-$k$ SOC Hamiltonian.

\begin{figure}[!htbp]
	\centering
	\includegraphics[width=\linewidth]{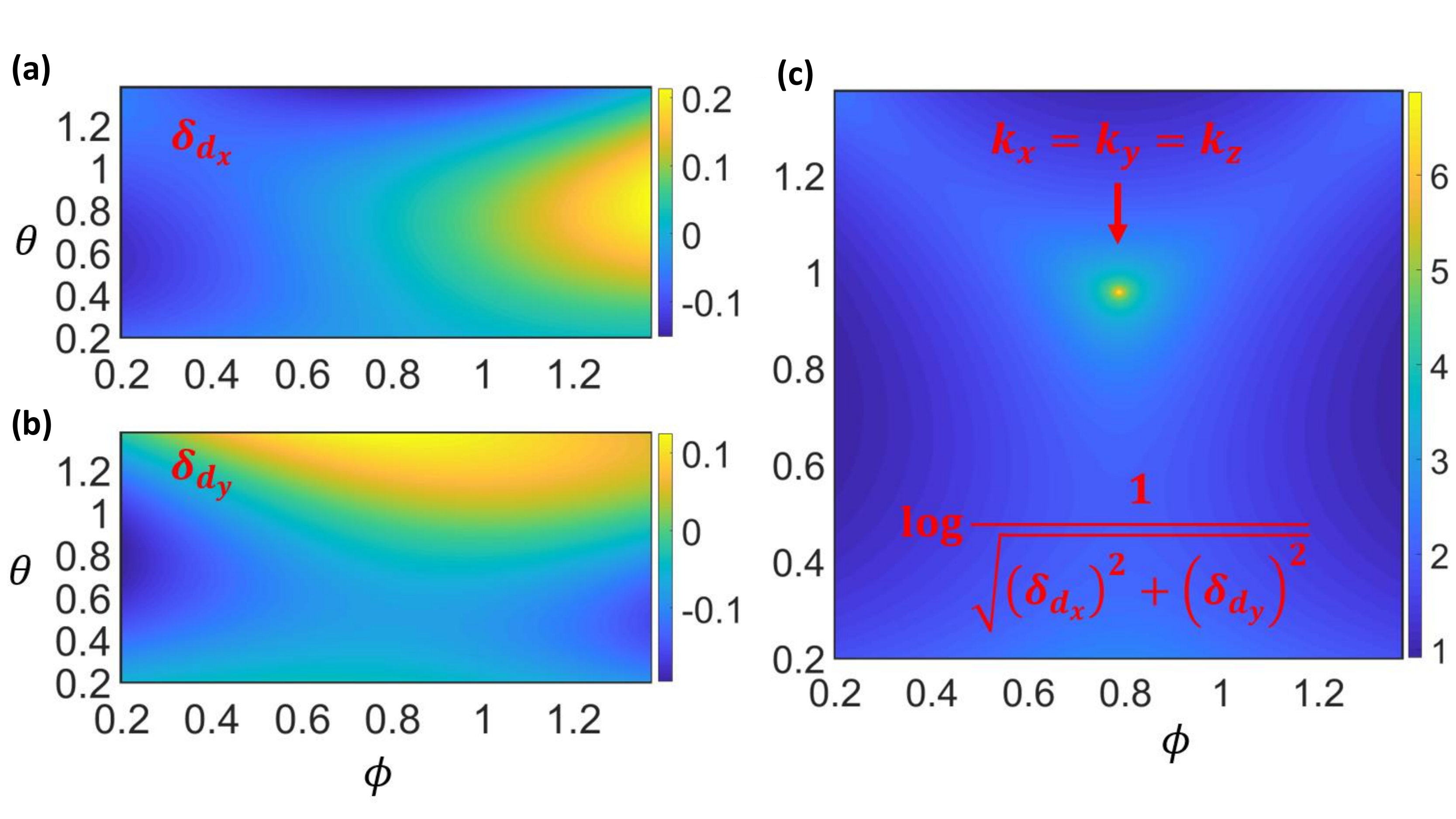}
	\caption{The projected two-band model with $\delta d_x$ in (a), $\delta d_y$ in (b) and the corresponding gap in (c). We find that the gap only closes at the Weyl point. It indicates the quasi-symmetry protected nodal plane is completely gapped out by the linear-k SOC Hamiltonian. Here $k_r=0.15$ \AA$^{-1}$ and $\lambda_2=0.005$ eV.
	}	
	\label{sm-fig7}
\end{figure}

\end{widetext}

\end{document}